\documentclass[lettersize,journal]{IEEEtran}
\usepackage{amsmath,amsfonts}
\usepackage{algorithmic}
\usepackage{array}
\usepackage{textcomp}
\usepackage{stfloats}
\usepackage{url}
\usepackage{verbatim}
\usepackage{graphicx}
\def\BibTeX{{\rm B\kern-.05em{\sc i\kern-.025em b}\kern-.08em
    T\kern-.1667em\lower.7ex\hbox{E}\kern-.125emX}}
\usepackage{balance}

\usepackage{comment}
\usepackage{cite}
\usepackage{amsmath,amssymb,amsfonts}
\usepackage{graphicx}
\usepackage[linesnumbered, ruled,vlined]{algorithm2e}
\usepackage{textcomp}
\usepackage{xcolor}

\usepackage{subfigure}
\usepackage{times}
\usepackage{commath}

\usepackage{multicol}
\usepackage{array}
\usepackage{colortbl}
\usepackage{multirow}
\usepackage{xspace}
\usepackage{hhline}
\usepackage{multirow, tabularx}
\usepackage{makecell}
\usepackage{adjustbox}
\usepackage[colorlinks=true, linkcolor=black, citecolor=black, urlcolor=black]{hyperref}

\usepackage{tikz}
\usepackage{caption}
\usepackage{url}
\usepackage{balance}
\usepackage{indentfirst}
\usepackage{booktabs}
\usepackage{arydshln}
\usepackage{colortbl}
\usepackage{mathtools}

\newcommand*\cced[1]{\tikz[baseline=(char.base)]{
         \node[shape=circle,fill,inner sep=0.8pt] (char) {\textcolor{white}{\footnotesize #1}};}}

\newcommand{\eg}{e.g.,\xspace}
\newcommand{\ie}{{\em i.e.,}\xspace}
\newcommand{\etc}{{\em etc.}\xspace}
\newcommand{\BfPara}[1]{\vspace{1mm}{\noindent\textbf{#1.}}\xspace}
\newcommand{\ours}{$\mathrm{SC_{LSB}}$} %This is my suggestion of the name. It's like Siamese Network, Siamese Twin. Nyang.

\usepackage{balance}
\usepackage{pifont}

\newcommand{\mm}[1]{{\textcolor{black}{#1}}}
\newcommand{\sy}[1]{{\textcolor{black}{#1}}}

\begin{document}
\title{Enhancing Resiliency of Sketch-based Security via LSB Sharing-based Dynamic Late Merging}

\author{
    Seungsam Yang\IEEEauthorrefmark{1}, 
    Seyed Mohammad Mehdi Mirnajafizadeh\IEEEauthorrefmark{1}, 
    Sian Kim\IEEEauthorrefmark{2},
    Rhongho Jang\IEEEauthorrefmark{1},
    DaeHun Nyang\IEEEauthorrefmark{2}\\
    \IEEEauthorblockA{\IEEEauthorrefmark{1}Department of Computer Science, Wayne State University, USA \\}  \IEEEauthorblockA{\IEEEauthorrefmark{2}Cyber Security Department, Ewha Womans University, South Korea \\}
    \thanks{*The first two authors contributed equally to this work.}
    }

\maketitle

\begin{abstract}
With the exponentially growing Internet traffic, sketch data structure with a probabilistic algorithm has been expected to be an alternative solution for non-compromised (non-selective) security monitoring. 
 While facilitating counting within a confined memory space, the sketch's memory efficiency and accuracy were further pushed to their limit through finer-grained and dynamic control of constrained memory space to adapt to the data stream's inherent skewness (i.e., Zipf distribution), namely small counters with extensions. In this paper, we unveil a vulnerable factor of the small counter design by introducing a new sketch-oriented attack, which threatens a stream of state-of-the-art sketches and their security applications. With the root cause analyses, we propose Siamese Counter with enhanced adversarial resiliency and verified feasibility with extensive experimental and theoretical analyses. 
 Under a sketch pollution attack, Siamese Counter delivers 47\% accurate results than a state-of-the-art scheme, and demonstrates up to 82\% more accurate estimation under normal measurement scenarios.
\end{abstract}

\begin{IEEEkeywords}
Stream Data Measurement, Sketch, Dynamic Data Structure, Adversarial Attack, Robustness.\\
\end{IEEEkeywords}
    
    \IEEEpeerreviewmaketitle
%\if
    \def\dsp{\def\baselinestretch{1.00}\large\normalsize}
    \dsp
    \def\ffsp{\def\baselinestretch{1.000}\large\normalsize}
    \ffsp
    \interfootnotelinepenalty=10000
%\fi

\section{Introduction}\label{sec:intro}
Due to the rapidly increasing volume of network traffic, a significant challenge is posed for traffic measurement functions that are crucial for network intrusion detection~\cite{jaqen2021, flowlens, beaucoup2020, mirsky2018kitsune, poseidon, garcia2009anomaly, kim2023count, xing2021ripple, mirnajafizadeh2024enhancing}. 
Therefore, a body of sketch data structures were proposed to overcome computational and memory bottlenecks~\cite{basat2021salsa, gong2017abc, song2020fcm, kim2023count, zhou2018cold, yang2017pyramid, cormode2005improved, charikar2002finding, estan2002new, chen2016countertree, GaoQHD23Splicing, li2022stingy, jang20,nyang16}. 
Sketch technologies offer error-bounded approximation for network traffic measurements within a fixed computation and memory budget, which are particularly beneficial when stream data measurement operations need to meet tight deadlines to support line-rate processing with its minimized footprint (\ie $O(1)$ encoding complexity). Therefore, sketches appear in various advanced security applications in both network~\cite{Xing20NetWarden, namkung2022sketchlib, poseidon, da2020euclid, misa2024leveraging} and cloud settings~\cite{liu2019nitrosketch, huang2017sketchvisor, yang2018elastic}.

Count-Min sketch~\cite{cormode2005improved} is a simple sketch, widely used in practice~\cite{redis, Apachespark, cmjulia}. However, it is often criticized for its bit-wise memory waste due to the full-size counter configuration~\cite{zhou2018cold, yang2017pyramid, kim2023count}. Especially, due to the Zipfian distribution mostly observed in the wild network traffic, the most significant bits (MSBs) of the full-sized (\eg 32-bit) counters are unlikely to be used for mouse flows that dominate the trace.

Unsurprisingly, the memory waste issue motivated a large body of research to utilize a {\it small counter design} (\eg 8-bit counter array instead of 32-bit counter array) to avoid MSB waste of counters and to maximize memory efficiency~\cite{yang2017pyramid,song2020fcm,yang2018elastic, zhou2018cold, chen2016countertree, li2022stingy, GaoQHD23Splicing, kim2023count, basat2021salsa, gong2017abc}. 
Given a fixed memory setting, such a design increases the number of counters significantly; thus, it reduces flow collisions of each counter and increases per-flow accuracy accordingly. 
However, it generates a new issue for larger flows that demand larger counters. The existing strategies to the problem are twofold, namely \textit{static extension} and \textit{dynamic merging}.

The \textit{static} approach, such as Pyramid sketch~\cite{yang2017pyramid}, FCM sketch~\cite{song2020fcm}, Elastic sketch~\cite{yang2018elastic}, Cold Filter~\cite{zhou2018cold}, Counter Tree~\cite{chen2016countertree}, Stingy sketch~\cite{li2022stingy}, and Counter-Splicing sketch~\cite{GaoQHD23Splicing}, employ preserved multi-layer counters with varied counter sizes to adapt to the traffic skewness. This approach allows mouse flows to enjoy massive small counters (e.g., 8-bit) for collision-relaxed measurement, meanwhile elephant flows utilize fewer large counters (e.g., 32-bit) for cost-efficient and interference-free measurement. Nevertheless, the downside of these methods is a failure to adapt to the dynamic shifting of traffic distribution due to their static setting for the amount of small and large counters assuming forthcoming traffic patterns (\ie flow size distribution). Later, the state-of-the-art CountLess sketch~\cite{kim2023count} enhanced the static structure with dynamic operations that flexibly encode mice flows leveraging idle counters across all layers to accommodate shifting distributions for robust traffic measurement.
On the other hand, the \textit{dynamic} approaches, such as ABC~\cite{gong2017abc} and SALSA~\cite{basat2021salsa}, proposed dynamic data structures that allow small counters to grow in size independently by 1) borrowing bits from adjacent counters~\cite{gong2017abc} or 2) merging with contiguous counters upon overflow~\cite{basat2021salsa}; thereby, these dynamic data structures adapt to traffic distribution changes without prior knowledge.

It is worth mentioning that the small counter design with either a static or dynamic strategy forms a clear trade-off between accuracy and overhead. The former simplifies operations but sacrifices accuracy with a fixed amount of small and large counters. The latter maximize memory efficiency via dynamic counter-size extension to prioritize accuracy, but the dynamic operations come with a high overhead. Therefore, static approaches fit better for in-network data plane under critical processing deadline~\cite{yang2018elastic, kim2023count}, whereas the dynamic approaches often appear in cloud settings utilizing general CPU for relatively complex operations~\cite{basat2021salsa, gong2017abc}.

In this work, we shed light on the vulnerabilities inherent from the small counter design by introducing a novel attack, namely a sketch pollution attack, which to our best knowledge is the first attack targeting the sketch itself and assuming adversaries to defeat sketch-based security applications. 
Our analyses indicate that the new attack can efficiently worsen the flow measurement of network traffic, leading to malfunction in security applications with a low cost (see section~\ref{sec::motivation}). 

To guide the sketch design with clear insights, we analyzed the counter-merging behavior in state-of-the-art dynamic data structures upon overflow, leveraging applications in various security domains.
We confirmed that the dynamic strategy delivers a more robust measurement, in general, than static approaches, due to its small counter memory recycling design (\ie memory efficiency).
However, when counter-merging events occur quickly and extensively across the entire data structure, namely sketch pollution attack, the advantage of the small counters no longer exists, imagining many counters are plundered by overflowed neighbors even though they are idle.  
Interestingly, under a certain scenario, the static approaches that employ memory isolation (\ie counter independence) are superior, achieving higher accuracy than dynamic merging methods by retaining more small counters independently from neighboring counters (see section~\ref{sec::motivation}). 
Therefore, we shape our research question as \textit{how to achieve memory efficiency and counter sustainability} by combining advantages of both static and dynamic approaches.

In this paper, we propose a novel dynamic sketch called Siamese Counter with Least Significant Bit Sharing (\ours{}, hereafter). Unlike previous approaches that merge counters instantly upon a counter overflows, we introduce a novel concept called \textit{late merging}, realized by the Least Significant Bit (LSB) sharing of contiguous counters. 
Intuitively, the LSB sharing strategy extends the counting capacity of small counters allowing the sketch to operate more small counters independently as in static approaches.  
Meanwhile, when counters' merging is unavoidable, the late merging strategy minimizes accumulated noise compared to the instant merging approach by discarding more packets from irrelevant flows in the event of counter merging.
By doing so, \ours{} with \textit{late merging} not only maximizes memory efficiency but also achieves better resiliency under the sketch pollution attack that aims to saturate small counters. 
We prove \ours{}'s feasibility through theoretical analysis and extensive experiments, and verified that \ours{} can enhance the sketch's resiliency in various security-related tasks. Our contributions are as follows:
\begin{itemize}
    \item We unveil the vulnerability of state-of-the-art sketches by introducing a novel attack targeting sketches. We estimated the attack impacts using various sketch-based security applications and conducted an in-depth analysis to explain the root causes.%\vspace{-1mm}
    
    \item We propose a novel sketch called \ours{} to enhance adversarial resiliency with a unique sketch design, namely late merging with the least significant bit sharing, inheriting the resiliency from both static and dynamic sketches. %\vspace{-1mm}

    \item We show the feasibility of \ours{} with theoretical proof of error bound and demonstrate that \ours{} outperforms state-of-the-art schemes with extensive experiments under practical security settings.
\end{itemize}

\BfPara{Organizations} The rest of the paper is organized as follows. We first discuss related works, in section~\ref{sec:related}. Then, we motivate our work with a new sketch attack, in section~\ref{sec::motivation}. Next,  we explain \ours{}'s design and theoretical analysis, throughout sections~\ref{sec:main} and
~\ref{sec:analysis}. 
In section~\ref{sec:eval}, we evaluate \ours{}'s performance and adversarial resiliency, followed by discussions and future works of the proposed approach, in section~\ref{sec:disc}. Lastly, we conclude our work, in section~\ref{sec:conclusion}.

\section{Related Works}\label{sec:related}
This section reviews existing sketches in the literature. First, we explain the significance of sketches by discussing recent works. Then, we categorize existing sketch designs that are closely related to our work.

\subsection{Background} A large body of research focuses on integrating the aforementioned sketching algorithms either for specific applications and measurement tasks or by adapting them to particular measurement environments. Here, we highlight the most recent works in each area.
Several works in the domain focus on optimization techniques either by adapting them to constrained hardware devices~\cite{namkung2023sketchovsky, miao2023sketchconf, kim2023count, zhang2021cocosketch, huang2021toward, yang2023sketchint, SketchFeature} or accelerating packet processing in software environments~\cite{liu2019nitrosketch, huang2017sketchvisor, miano2023fast}. These works utilize heuristic and empirical observations~\cite{kim2023count, yang2023sketchint} to identify bottlenecks in resource computation~\cite{huang2017sketchvisor, liu2019nitrosketch, namkung2023sketchovsky} within the given environment and optimize performance accordingly. Additionally, some works~\cite{zhu2019cbfsketch, yang2019diamond, zhu2020oebsa, sheng2021pr} employ ad-hoc designs that combine various data structure types to mitigate collision and improve accuracy. For example, works~\cite{zhu2020oebsa, zhu2019cbfsketch, yang2019diamond} employ Counting Bloom filter alongside the sketch to track counter overflow depth to mitigate collision rate.
Other works focus on application-specific measurements and devising structures optimized for a focused application~\cite{he2023histsketch, shahout2023together,ma2021super, liu2022duet, liu2016one, guo2023sketchpolymer, zhao2021kll, karppa2022hyperlogloglog, dao2022minimizing, JungKJMN22PortCatcher, Jang17, Jang19, sketchlet}. For example, works~\cite{guo2023sketchpolymer, liu2022duet, he2023histsketch, zhao2021kll} leverage sketch design to estimate quantiles of distribution for a given data stream. From a security perspective, recent work~\cite{cheng2024trustsketch} has focused on deploying sketches within the SGX enclave to prevent memory tampering by attackers, which differs from our focus on designing a more resilient sketch within a black-box setting.
In this work, we focus on sketches in a software environment for stream data measurement and security tasks.

\subsection{Structure-Oriented Sketch Design}
\begin{figure}[t]
    \centering
    \subfigure[FCM (static counters)]
    {\includegraphics[width=0.2\textwidth]{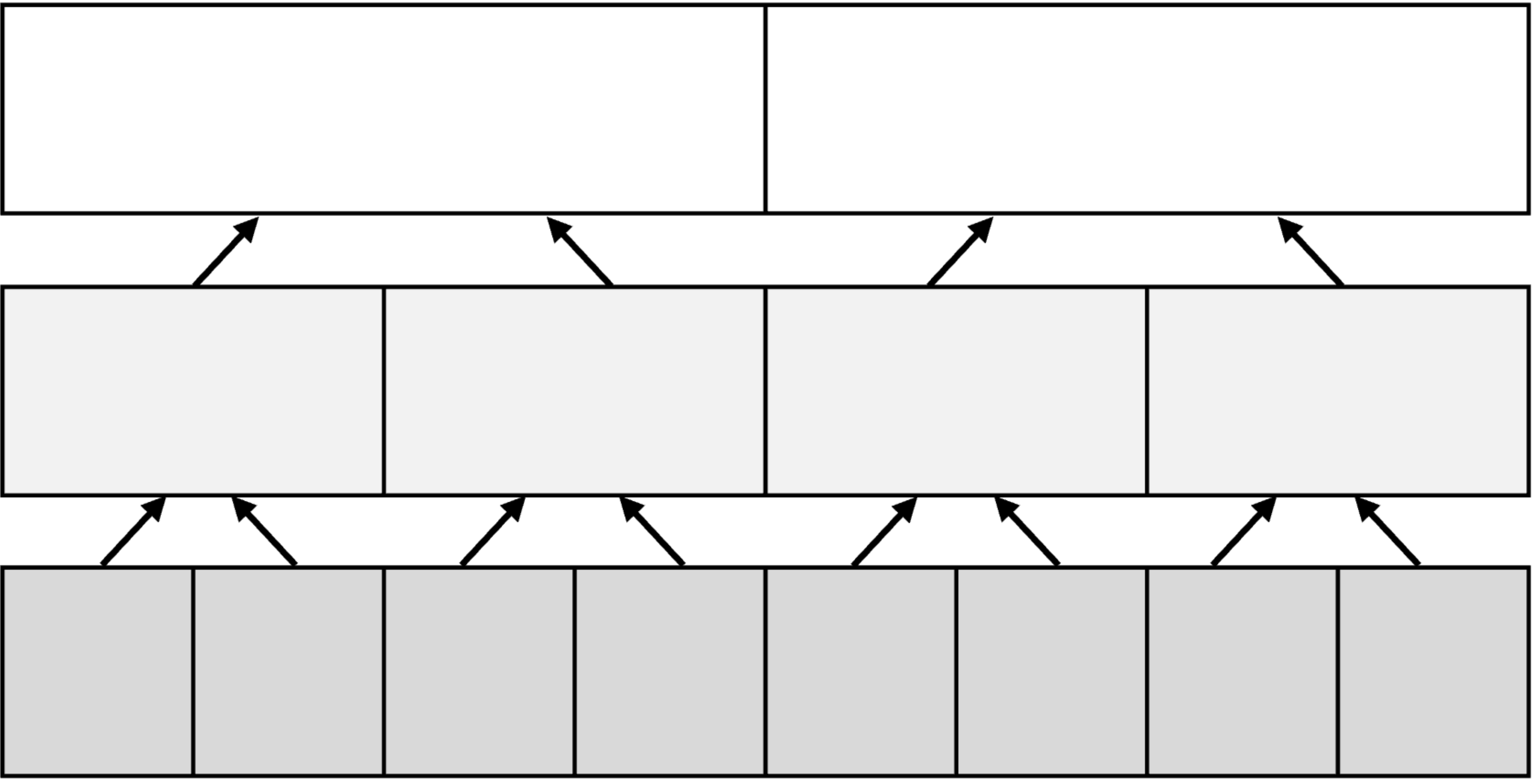}} 
    \subfigure[SALSA (dynamic counters)]
    {\includegraphics[width=0.23\textwidth]{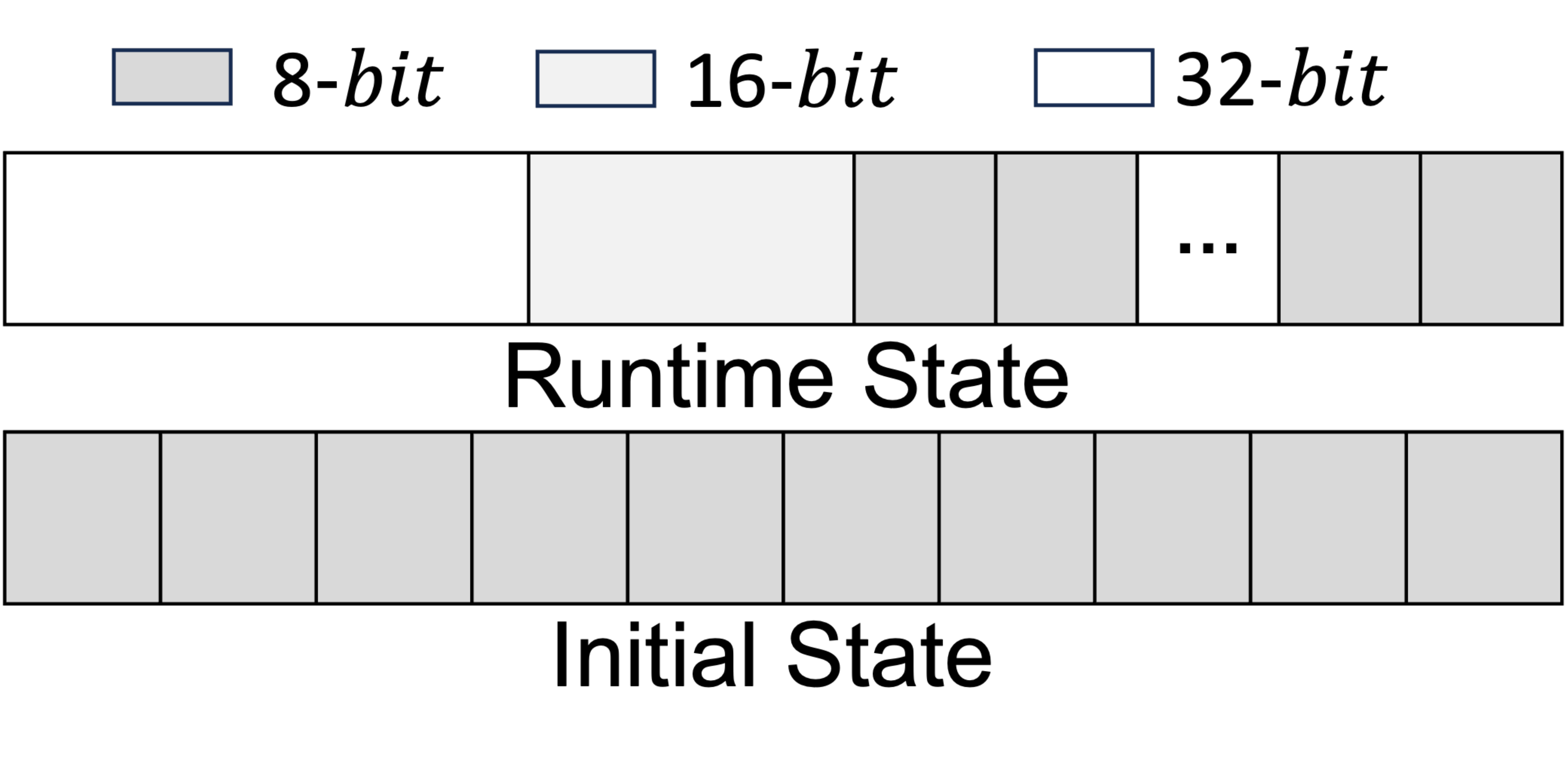}}
    
    \caption{Structure of representative static extension (FCM) and dynamic merging (SALSA).
    }~\label{fig:data_st}
\end{figure}

The initial studies in the sketch domain, including Count-Min~\cite{cormode2005improved} and its variants~\cite{charikar2002finding, estan2002new}, primarily focused on theoretical advancements in noise minimization and relaxed the assumption by maintaining the full-size counters (e.g., 32-bit). However, in practice, an important observation suggested that the Most Significant Bit (MSB) portion of counters is mostly wasted (\ie 75\% of entire memory)~\cite{kim2023count} due to the inherent skewness of the network traffic~\cite{traffic_char2010}, that is the majority of real-world network traffic composed of mice flows (e.g., flow size much smaller than 256). To tackle the issue and maximize the sketch memory usage, a large body of works~\cite{kim2023count, basat2021salsa, gong2017abc, yang2018elastic, li2022stingy, yang2017pyramid, zhou2018cold, chen2016countertree, song2020fcm, GaoQHD23Splicing} utilized ``a small counter design'', which reduces the counter size (\eg 8-bit) for a significantly increased number of counters, as shown in Fig.~\ref{fig:data_st}. Since the sketch allows flows to share counters randomly, the design decreased flow-wise interference (\ie collision reduction) and improved estimation accuracy significantly. For larger flow measurements, two approaches were adopted, namely static counter extension and dynamic counter merging.

\BfPara{Static Counter Extension} 
The static counter extension approach is based on multi-layer sketch designs~\cite{yang2017pyramid, kim2023count, zhou2018cold, yang2018elastic, song2020fcm, li2022stingy, GaoQHD23Splicing, chen2016countertree}, 
which incorporates hierarchical layers to monitor mice and elephant flows separately and independently. Fig.~\ref{fig:data_st} (a) illustrates state-of-the-art work of the static approach, namely FCM~\cite{song2020fcm}. As shown, FCM maintains a tree-like structure composed of layers with varied counter sizes, where the first layer contains a massive number of 8-bit counters, a second layer with fewer medium-sized 16-bit counters, and the last layer with a significantly fewer number of 32-bit counters. 
At the runtime, the flow monitoring is initiated at the first layer for the mice flow measurements. In the event of counter saturation for larger flows, the flow measurement continues at the upper layer with larger counters. 

The design ensures all counters are used fully bit-wise for relaxing flow collisions. However, a notable challenge of such an approach is to pre-define the parameter (\ie the ratio of memory space allocated to each layer) to achieve optimal accuracy. For example, in traffic skewed toward large flows, collisions in the upper layers become more frequent, demanding reduced memory allocation for lower layers and an increase for upper layers.
CountLess~\cite{kim2023count} aimed to improve this layering design by allowing flows to be encoded in all layers by eliminating the flow size confinement, but it still falls in the category of the small counter design.

\BfPara{Dynamic Counter Merging}
Unlike multi-layer sketch design, recent work ABC~\cite{gong2017abc} suggests setting small counters (e.g., 8-bit) across the entire data structure and dynamically enlarging the counter size whenever a small counter is overflowed. ABC~\cite{gong2017abc} implements a bit-borrowing mechanism to extend the counting capacity. Upon counter saturation, it borrows a bit from its adjacent counter and further merges the counters if no more idle bits to borrow from neighbors. The heavy bit-wise state tracking and one-time counter merging later is criticized by Self-Adjusting Lean Streaming Analytics (SALSA)~\cite{basat2021salsa}, which proposes a more general dynamic data structure that simply merges two contiguous counters to double the counter size, and continues the flow counting with a {\it sum} or {\it max} of two previous counter values. Fig.~\ref{fig:data_st} (b) demonstrates the structure SALSA, the representative dynamic counter extension approach. As shown, initially, all counters are small 8-bit counters. During runtime execution,  a counter can be merged with its neighbor depending on the flow size, forming arbitrary counter-size patterns in its data structure. This approach achieves more efficient memory utilization by recycling saturated counters and presents a better ability to adapt to dynamic traffic pattern changes without re-configuring the entire data structure. However, the dynamic counter merging approach simply discards adjacent counter estimations upon merging events. We call this mechanism {\it instant merging}.

\section{Motivation: Adversarial Resiliency of Sketches}~\label{sec::motivation}
Building on the small-counter design to improve estimation accuracy through static or dynamic counter extension, previous works assumed that such extensions would be sufficient for more robust measurement. However, they overlook the possibility of an attacker intentionally saturating these counters, ultimately nullifying the benefits of small-counter design and rolling the behavior back to Count-Min sketch. With this insight, we introduce a novel sketch pollution attack in a black-box setting, enabling an attacker to degrade the effectiveness of the small-counter design. In the following, we systematically analyze the attack impact on both the static counter extension and dynamic counter merging approach demonstrating how adversaries can undermine the foundation of existing sketches.

\BfPara{Threat Model} We assume the attacker is aware of the presence of a sketch for traffic monitoring tasks. This sketch can be deployed in a network router with its primary forward function ~\cite{jaqen2021, Xing20NetWarden} or can be telemetry in a cloud gateway to monitor network traffic entering  hypervisors~\cite{cheng2024trustsketch, liu2019nitrosketch, huang2017sketchvisor}. The attacker's goal is to launch a ``pollution attack'' targeting the sketch function itself. To do so, we assume the attacker is free of injecting/replying crafted packets (headers) into the stream monitored by the sketch, but the query to the sketch is prohibited. Furthermore, we assume a black-box scenario where the attacker has no prior knowledge of the targeted sketch but can obtain the estimated value for the configuration of the array counters ($w$). This can be done by considering the desired accuracy configuration imposed by the administrator to obtain the optimal sketch parameters~\cite{sun2024autosketch, miao2023sketchconf}.

\BfPara{Attack Logistic} Targeting a static sketch with $w$ small counters (\ie 8-bit), an ideal case for an attacker is to select $w$ distinct flows targeting $w$ distinct counters by sending 256 packets for each flow to saturate all small counters. 
Interestingly, due to the unique counter merging behavior of the dynamic sketch, adversaries are allowed to target $w/2$ distinct odd or even counter indexes for the same impact, as shown in Fig.~\ref{fig:atk_logis}. However, obtaining the sketch configuration (e.g., hash algorithm, seed values, memory space, etc.) for a perfect attack is not straightforward.

A desirable way to tackle the issue is to approximate the problem to the coupon collector’s problem~\cite{flajolet1992birthday}, leveraging the sketch's nature that is randomly (uniformly) balancing flows in $w$ counters. 
Thus, the coupon collector’s problem is equivalent to a counter collection problem that determines the expectation of random draws (random selection of attack flows) with a replacement to collect all types of coupons (all distinct counters) at least once. Assuming the hash function is uniform, given $w$ as the number of types of coupons (counters), the expected number of draws (attack flows) $E(X_w)$  needed to collect the full set of coupons (counters) is as follows:
\[
E(X_w) = \sum_{i=1}^{w} \frac{w}{w-i+1} = w \sum_{i=1}^{w} \frac{1}{i} = wH_w,
\]
where $H_w$ is the $w$th harmonic number. Similarly, we can derive $E(X_m)$, the expected number of draws (attack flows) to collect $m$ distinct coupons (counters) out of $w$:
\[
E(X_m) = w(H_w - H_{w-m}).
\]
For example, when a sketch has 155k counters ($w$) and the attacker aims to saturate 50\% of them (\ie $m=w\times0.5$ $\approx$ 78k), the expected number of flows to be generated by the attacker is 108k. 
While it is possible to saturate all counters, it would require approximately 2 million flows, which is impractical. Our experiments show that polluting just 50\% of the counters ($\approx$ 108k flows) is sufficient to increase the error by at least 7x, as shown in Fig.~\ref{fig:atk}.

\subsection{Attack Effectiveness and Sketch Resiliency} 
\BfPara{Setting} With this attack logic, we analyzed the sketch pollution attack using two popular security applications, namely, Heavy Hitter Detection (HHD) and Flow Size Estimation (FSE), which have been widely used for detecting anomalies such as DoS/DDoS attack traffic~\cite{jaqen2021, yang2018elastic, zhou2023efficient, Xing20NetWarden}. The average relative error (ARE) measures estimation error (lower is better), while the F1 score indicates detection accuracy (higher is better). To illustrate,  we selected SALSA~\cite{basat2021salsa}, the representative of dynamic counter merging, and FCM~\cite{song2020fcm} as the representative of static counter extension, as well as Countless~\cite{kim2023count}, which is a special case for the two static and dynamic approaches as it demonstrates a mixed behavior. 
For attack traffic selection, we varied distinct attack flows $m$ from 0 to 108k (equivalent to targeting 0\% -- 50\% of total counters $w$) and send 256 packets for each flow. We note that all schemes utilize the same amount of memory and a similar number of small counters for a fair comparison.

\begin{figure}[t]
    \centering
    \includegraphics[width=0.49\textwidth]{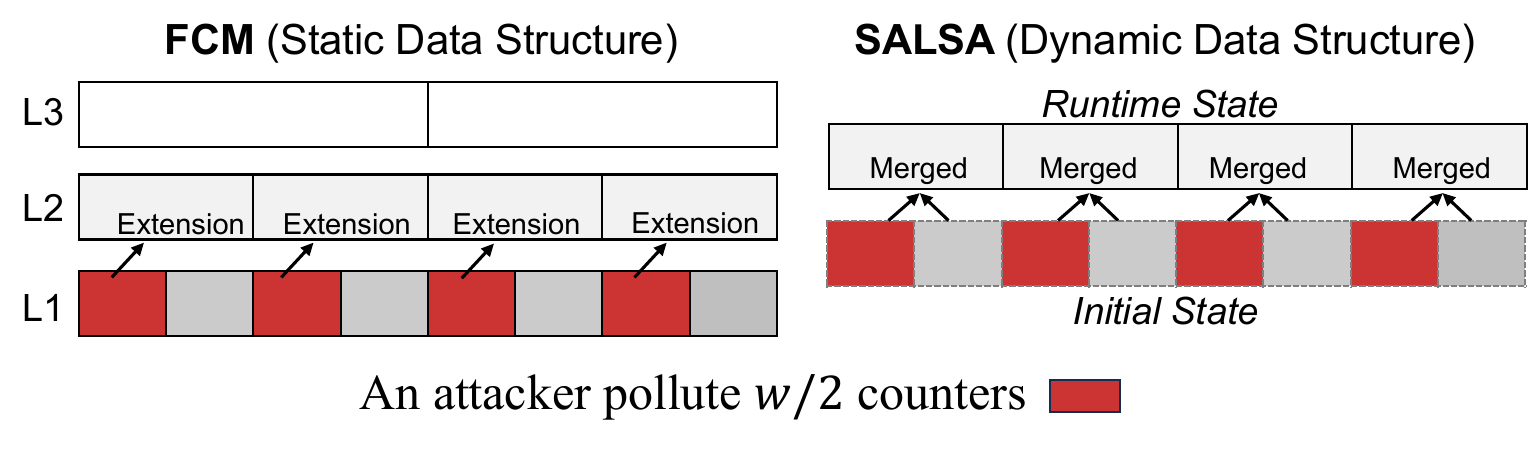}
    \caption{Attack Logistic: saturating counters in the dynamic counter extension SALSA affect idle counters, while in the static counter extension FCM, they operate independently.
    }~\label{fig:atk_logis}
\end{figure}
 
 \begin{figure}[t]
     \centering
     \includegraphics[width=0.28\textwidth]{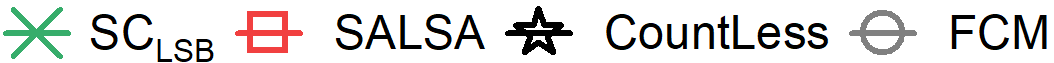}
    \subfigure[Flow size estimation]
     {\includegraphics[width=0.23\textwidth]{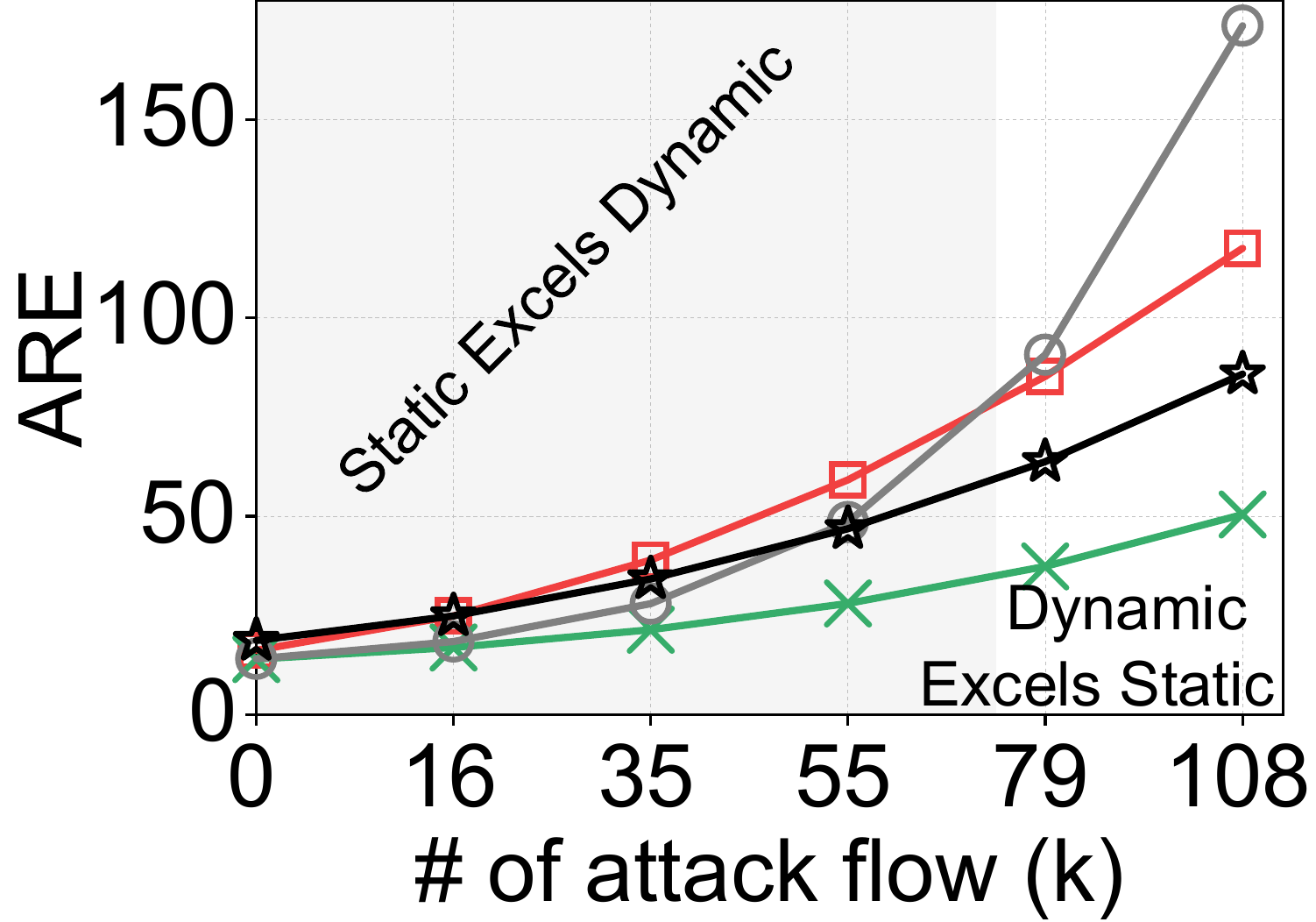}}
     \subfigure[Heavy hitter detection]
    {\includegraphics[width=0.23\textwidth]{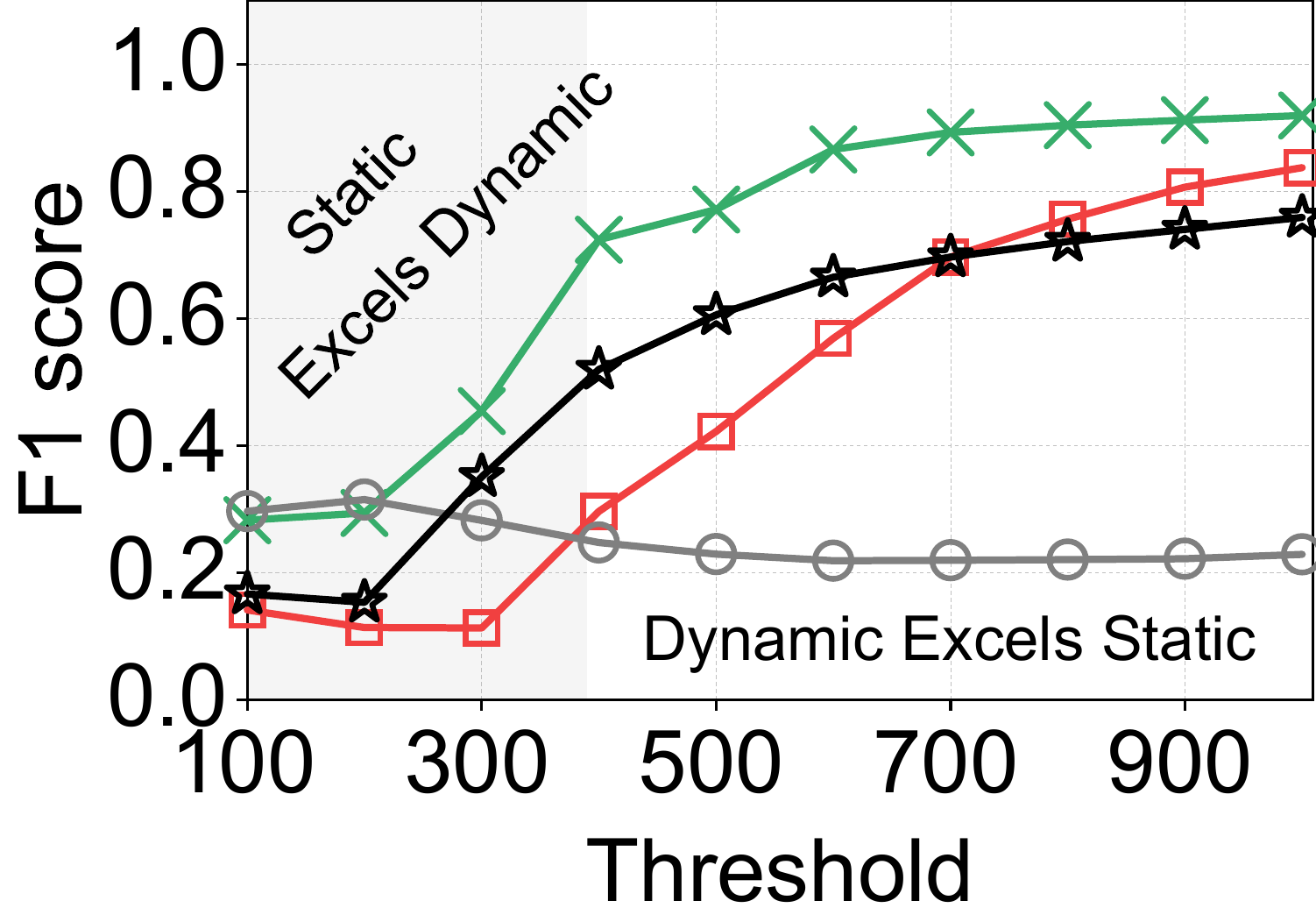}}
     \caption{Adversarial attack on representative dynamic and static sketches: initially, static structure FCM demonstrates better accuracy against attack traffic, but its performance degrades as more counters are targeted. For heavy hitter detection, FCM and CountLess excel at low detection thresholds but underperform compared to dynamic structures when the pre-defined threshold becomes higher.
     }~\label{fig:atk}
\end{figure}

\BfPara{Understanding Counter Merging} As shown in Fig.~\ref{fig:atk_logis}, the small counters in the first layer of FCM operate independently, in which the saturation of one counter does not affect its neighbor. However, when examining the upper layers, the number of counters decreases exponentially. This infers that as the measurement shifts to the upper layers, the collision rate also increases exponentially. Another important observation is that the memory region of the first layer, once saturated, becomes obsolete and is no longer used. 
On the other hand, as shown in Fig.~\ref{fig:atk_logis}, dynamic merging expands the size of the counter via a merging mechanism that reuses the saturated counters. However, the drawback is that once a counter becomes saturated, it plunders the idle adjacent counter, thereby increasing the collision rate. However, when traffic measurement requires larger counter sizes (e.g., 16-bit), the dynamic adaptation recycles the memory space of small counters to meet the demand, thereby it can provide a greater number of large (16-bit) counters compared to static extension, which results in a significant reduction of collision rate.

\BfPara{Performance Under Attack} In light of this, when the sketch pollution attack is light and targets less than 50\% of the counters ($<$50\% of counters merged), we can expect better performance for static counter extension. Conversely, when the pollution attack heavily saturates the small counters across the sketch, we can expect better performance for dynamic merging. This effect has been demonstrated in Fig.~\ref{fig:atk} (a) with the flow size estimation. Furthermore, we can derive similar observations with the heavy hitter detection, which focuses on flows exceeding a pre-defined threshold. Fig.~\ref{fig:atk} (b) demonstrates detection performance when varying the pre-defined threshold, where the attacker targets 30\% of the counters. As shown, static counter extension FCM outperforms SALSA for relatively small thresholds near the merging trigger (\ie 256) by reporting fewer false positives due to the independent counter operation. However, as the threshold increases, dynamic merging consistently outperforms the static extension design, which suffers from high collision rates in the upper layers where the larger heavy flows reside. Notably, this behavior also appears in CountLess~\cite{kim2023count}, a static extension design with dynamic operation. However, its dynamic operation design has resulted in greater resiliency compared to FCM~\cite{song2020fcm}.

\subsection{Motivating Siamese Counter (\ours{})} The interesting observations are that the initial belief in the superiority of dynamic structures is no longer valid under adversarial conditions. When a sketch pollution attack is less aggressive (\ie slow attack), the small counter isolation design in static extension allows independent counter operations and improved accuracy. In contrast, when the sketch pollution attack is aggressive (\ie fast attack), dynamic merging with traffic pattern adaptation outperforms static counter extension.
These findings motivated us to design a novel dynamic sketch, namely Siamese Counter (\ours{}), which introduces a {\it late merging} concept to improve the adversarial resiliency of the dynamic operation. Essentially, \ours{} proposes a least significant bit (LSB) sharing technique to extend the counting capacity of small counter to suppress slow pollution attacks, meanwhile allowing late (dynamic) merging to mitigate fast pollution attacks. Simply put, \ours{} possesses the advantage of static and dynamic to enhance adversarial resiliency for sketch-based security applications.

\section{Siamese Counter (\ours{})} \label{sec:main}
\sy{In this section, we describe the core idea of \ours{}, \textit{late merging with Least Significant Bit (LSB) sharing}. We explain the concept of \ours{} with examples and algorithm.}

\subsection{Design Intuitions}
\begin{figure}[t]
    \centering
    \includegraphics[width=0.48\textwidth]{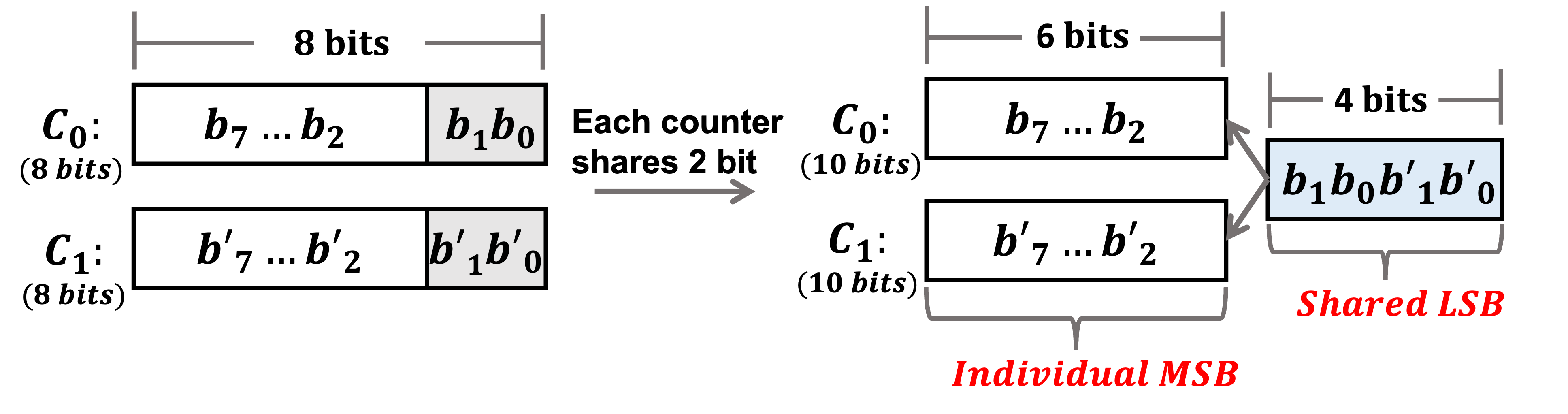}
    \caption{LSB bit sharing between counters $C_0$ and $C_1$: each counter shares 2 bits, resulting in a total of 10 bits per counter, i.e., 4 shared LSBs and the original 6-bit MSB.
    }\label{fig:lsb_sharing}
\end{figure}

\BfPara{Key Idea: Least Significant Bit (LSB) Sharing} To benefit from the long-run sustainability of small counters under dynamic structure sketch design, we introduce the concept of LSB sharing. Figure~\ref{fig:lsb_sharing} demonstrates \ours{}'s LSB sharing concept, where two adjacent counters share $K$-bit LSB upon counter overflow. As shown, each counter $C_0$ and $C_1$ shares two bits from the LSB part. Finally, each one ends with a total of 10 bits, including 4-bit shared LSB and the remaining 6-bit non-shared MSB from the original counter, thereby increasing the counting capacity from 255 to 1023. With this 4x  expansion of counting capacity, \ours{} enables long-time sustainability of small counters. The clear advantage is that in the same adversarial scenario, the attacker must increase the attack budget (\ie 4x more items)  to achieve a similar outcome as with the prior dynamic merging approaches. 
For easy understanding, we refer to the prior merging mechanism as {\it instant merging} throughout the rest of the paper. In SALSA~\cite{basat2021salsa} and SSVS~\cite{melissourgos2023single}, merging occurs upon counter saturation, whereas ABC’s bit borrowing is ineffective due to the unavailability of idle bits for neighboring counters (see section~\ref{sec:eval} for more details), thereby falling in the same category.

\begin{figure}[t]
    \centering
    \hfill
    \includegraphics[width=0.49\textwidth]{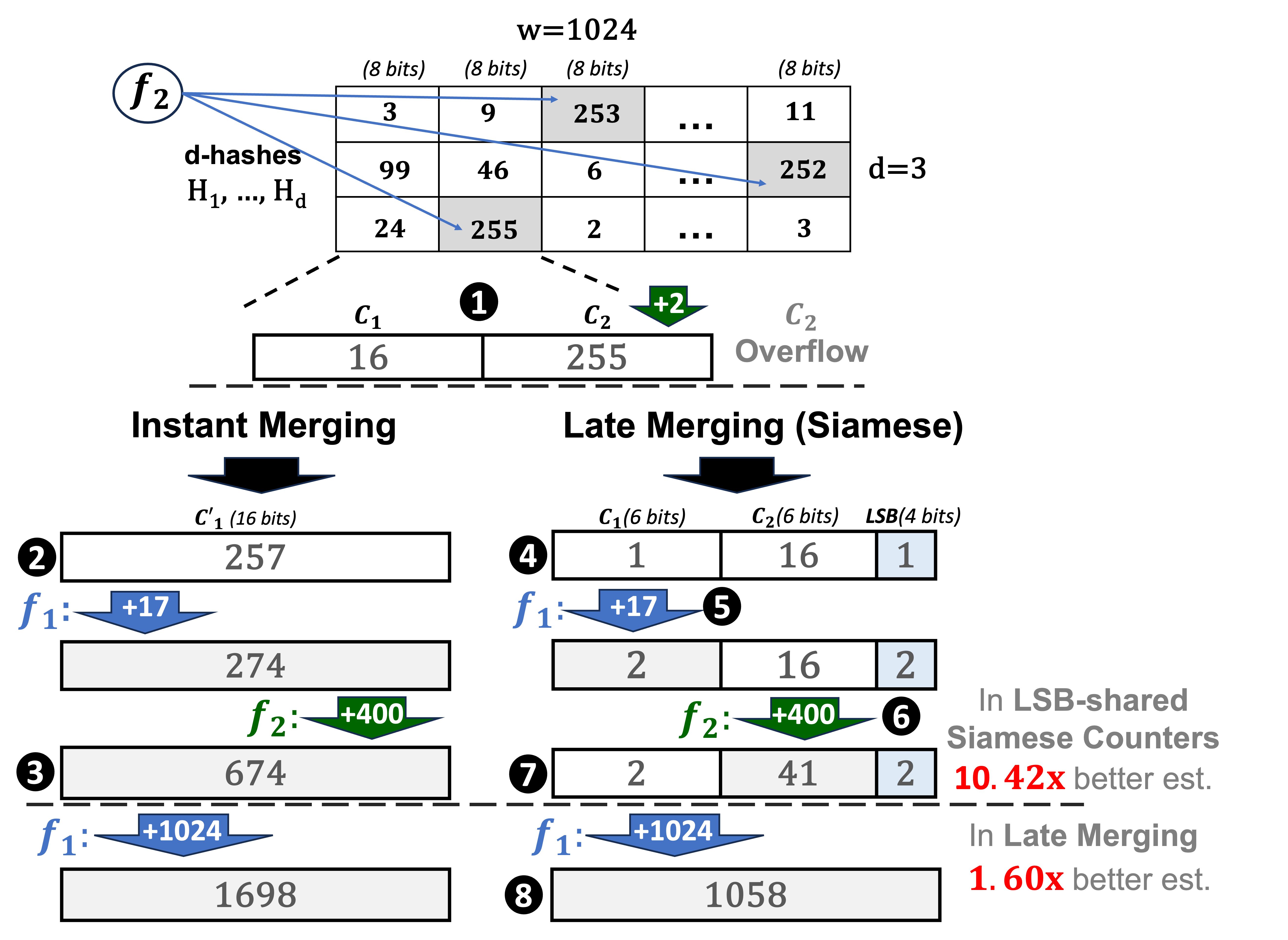}
     \caption{Instant merging vs. late merging: instant merging reduces the number of counters and increases the collision rate, leading to high estimation error. Meanwhile, late merging maintains both the number of counters and the collision rate.
     }
    \label{fig:err_merge}
\end{figure}

\BfPara{Data Structure} Figure~\ref{fig:err_merge} demonstrates the \ours{}'s data structure maintaining $d$ arrays of $w$ 8-bit counters, which is similar with Count-Min sketch~\cite{cormode2005improved} (CMS). In its update process, it runs $d$ independent hash functions using the item's key. Then, each hash value is used as an index for each counter array for item encoding. To query an item's frequency, the same procedures repeat for hash-based location but return the minimum counter value among $d$ arrays, like CMS. 

\BfPara{Workflow with an Example} \mm{Figure~\ref{fig:err_merge} illustrates an example of instant merging and our LSB-shared counters derived from our analysis in section~\ref{sec:eval:in-depth}}. Assume that two items $f_1$ and $f_2$ are encoded in adjacent counters $C_1$ and $C_2$ (\cced 1). According to SALSA, $f_2$ triggers an instant merging of two counters when it overflows the initial 8-bit counter $C_2$ (\cced 2). As a result, forthcoming items of two items will contribute to the merged counter, returning a single estimation of 674 (\cced 3), which results in a 10.42x higher estimation, on average, compared to the ground truth (\ie 33 and 657). 
On the other hand, \ours{} allows LSB sharing in the event of the first counter overflow, and continues the counting with two non-shared MSB counters and one shared LSB counter (\cced 4). With the shared LSB, \ours{} continues to encode both $f_1$ and $f_2$, and employs a \textit{winner-take-all} strategy (see~\ref{sec:eval:winner} for analysis) to update each of the non-shared MSB counters. Simply put, if $f_1$ triggers a saturation of LSB, $C_1$'s MSB value is increased (\cced 5), otherwise, $C_2$'s MSB increases (\cced 6). 
The intuition behind this strategy is that the items coming in random order will saturate the shared LSB proportion to their sizes. As shown in Figure~\ref{fig:err_merge} (\cced 7), \ours{} decode much more accurate values of 34 (=$2\times2^K+2$) and 658 (=$41\times2^K+2$) from independent counters, compared to the ground truth values of 33 and 657, and achieved 10.42x better accuracy than the instant merging.
We note that the enlarged counters by LSB sharing will be merged eventually (\ie late merging) when an item saturates its non-shared MSB counter. However, we stress that the (delayed) late merging of the counters also benefits the final estimation by excluding more noises.  
Figure~\ref{fig:err_merge} (\cced 8) demonstrates the case when the merging is inevitable and both the counter $C_1$ and $C_2$ merge as one. As shown, the final estimation still achieves 1.6x better estimation accuracy than instant merging (\ie 1058 vs. 1698). The proof of winner-take-all approach and late merging are placed in section~\ref{sec:analysis}. 

\subsection{Algorithm}
\BfPara{Encoding} 
For each incoming item $f$, we begin by locating the counter for $f$ ($C_f$) and its adjacent counter ($C_{adj}$) using the $\mathtt{findCounter}$ function (line 3). If $C_f$ shares its LSB with $C_{adj}$ (line 4), we retrieve each $K/2$ LSB from each counter to form a $K$-bit shared LSB counter $C_{LSB_K}$ and increase it by ``1'' (lines 5-6). It is important to note that the LSB concatenation function is for idea explanation. In the real implementation, all $K$ bits are located at the end of two conjunctive (Siamese) counters.  Next, if $C_{LSB_K}$ overflow, the $S-K/2$-bit MSB of $C_f$ is incremented by one, \ie winner-take-all strategy (lines 7-10). Furthermore, if MSB of $C_f$ overflow, we merge $C_f$ and $C_{adj}$ using a max or sum merge method, \ie user choice (lines 11-14).
If $C_f$ and $C_{adj}$ do not share their LSB (\ie independent counters), we simply increase $C_f$ by ``1'' (lines 15-16). However, if $C_f$ overflows, LSB sharing between $C_f$ and $C_{adj}$ will be triggered. That is, each counter contributes $K/2$-bit of their LSBs for sharing $K$ bits ($C_{LSB_K}$) in total (lines 17-18). As a result, each counter has $S+K/2$ bits virtually. At the beginning, $C_{LSB_K}$ is initialized to the maximum value of the two counters' $K$-bit LSBs (line 19). Finally, the non-shared MSB ($(S-K)$ bits) of each counter is re-scaled by dividing the value by $2^{K}$ (lines 20-21). We note that the operation is equivalent to right-shifting the non-shared MSB by $K/2$ bits, thus the total counting capacity is increased by $K/2$ bits shared by neighbor counter $C_{adj}$. 

\BfPara{Decoding} 
Algorithm~\ref{alg:SRC decode} demonstrates the decoding procedure that retrieves the encoded value for item $f$. Similar to the encoding procedure, we begin by locating the counter for $f$ ($C_f$) and its adjacent counter ($C_{adj}$) using the $\mathtt{findCounter}$ function (line 3). If $C_f$ does not share LSB, we return $C_f$ as the estimation value (line 9). Otherwise, if $C_f$ shares its LSB with $C_{adj}$ (line 4), we retrieve each $K/2$ LSB from each counter to form a $K$-bit shared LSB counter $C_{LSB_K}$ (line 5), then using the MSB of $C_f$, the estimation value equals to $2^{K}*C_{MSB} + C_{LSB_K}$ (line 7).

\subsection{Understanding Primitive Functions}
The following are essential building blocks to enable the dynamic operation of Count-Min-based sketches. While the goal \ours{}'s primitive functions align with SALSA~\cite{basat2021salsa}, the internal logic appears different to work with the LSB-sharing. 

\BfPara{Find Counter}
This function is responsible for finding the counter $C_f$ for item $f$ and also its adjacent counter $C_{adj}$ in the array $A[i]$. Therefore, $A[i]$ maintains an even number of counters, and each counter has its designated peer counter $C_{adj}$, with the same size. First, we calculate the location of the small counter among $w$ counters by calculating the hash value of the item's key. Next, we locate $C_f$, which has different states at runtime, namely unmerged, LSB-shared, or merged; thus, the counter size varies depending on the counter state. An auxiliary data structure is employed to track the counter state  (\ie unmerged, LSB-shared, merged). After locating the counter $C_f$, we also identify its adjacent (peer) counter $C_{adj}$ by index $i$ recalculation $i+{(-1)}^{i \mod 2}$, which helps locate a counter's peer counter either on the left or right.

\SetAlgoSkip{}
\begin{algorithm}[t]
\footnotesize
\caption{\ours{} Encoding}~\label{alg:SRC update}
$\textbf{Input:}$ A[d][w]: counter array, K: shared-LSB bit size.

\For{$i=0$ \KwTo $d-1$}{
    $C_f, C_{adj} \gets \mathtt{findCounter}(A[i], f)$\;
    \If{$\texttt{IsShared}(C_f) $}{
         $C_{LSB_K} \gets\ \mathtt{Concat}(\mathtt{lsb}(C_f)_{\frac{K}{2}},
         \mathtt{lsb} (C_{adj})_{\frac{K}{2}})$\;
         $C_{LSB_K} \gets C_{LSB_K} + 1$\;
         \If{$\texttt{IsFull}(C_{LSB_K})$}{
             $S \gets \texttt{sizeof}(C_f)$\;
             $C_{MSB} \gets \mathtt{msb}(C_f)_{S-\frac{K}{2}}$\;
             $C_{MSB} \gets C_{MSB} + 1$\;
             \If{$\texttt{IsFull}(C_{MSB})$}{
                 $\mathtt{Max\_Merge}(C_f, C_{adj})$\;
                 or\\
                 $\mathtt{Sum\_Merge}(C_f, C_{adj})$\;
             }
         }
    }
    
    \Else{
        $C_f \gets C_f + 1$\;
        \If{$\texttt{IsFull}(C_f)$}{
         $C_{LSB_K} \eqqcolon \mathtt{Concat}(\mathtt{lsb}(C_f)_{\frac{K}{2}},
         \mathtt{lsb} (C_{adj})_{\frac{K}{2}})$\;
         $C_{LSB_K} \gets \mathtt{max}(\mathtt{lsb}(C_f)_{K},
         \mathtt{lsb}(C_{adj})_{K})$\;
         $\mathtt{msb}(C_f)_{S-\frac{K}{2}} \gets \mathtt{msb}(C_f)_{S-K}\ /\ 2^K$\;
        $\mathtt{msb}(C_{adj})_{S-\frac{K}{2}} \gets \mathtt{msb}(C_{adj})_{S-K}\ /\ 2^K$\;
        }
    }
    
}
\end{algorithm}

\SetAlgoSkip{}
\begin{algorithm}[t]
\footnotesize
\caption{\ours{} Decoding}~\label{alg:SRC decode}
$\textbf{Input:}$ A[d][w]: counter array, K: shared-LSB bit size.

\For{$i=0$ \KwTo $d-1$}{
    $C_f, C_{adj} \gets \mathtt{findCounter}(A[i], f)$\;
    \If{$\texttt{IsShared}(C_f) $}{
         $C_{LSB_K} \gets\ \mathtt{Concat}(\mathtt{lsb}(C_f)_{\frac{K}{2}},
         \mathtt{lsb} (C_{adj})_{\frac{K}{2}})$\;
         $C_{MSB} \gets \mathtt{msb}(C_f)_{S-\frac{K}{2}}$\;
         \Return $2^{K}\times{C_{MSB}}$ + $C_{LSB_K}$ \;
    }
    \Else{
        \Return $C_f$ 
    }
    
}
\end{algorithm}

\BfPara{Sum vs. Max Merge} 
The LSB-shared counters will eventually overflow and be triggered by counter MSB overflow. Then, two counters will be merged as a single counter to extend the counter capacity further. Two options are given for operators to summarize the counts recorded in two counters, namely max and sum merges (lines 12-14). The sum merge uses a summation of two counter values for the merged counter, with a potential impact of amplified noise by merging all previous items' counts.
On the other hand, max merge discards the smaller count and uses the bigger count for the merged counter. By doing so, the noise from independent items will not accumulate. \ours{} further enhance this property by isolating items longer, thus more noise can be discarded in the event of counter merging (see \ref{sec:error-bound} for detailed proof).  

\BfPara{Counter State Tracking} \ours{} uses an auxiliary data structure that is responsible for tracking counter-merging progress during encoding, which is further used during decoding. We note that the auxiliary data structure is not explicitly presented in the algorithm, as its design is not the focus of this work, as it closely follows SALSA~\cite{basat2021salsa} with slight modifications, while maintaining the same memory requirement of one bit per counter encoding process.
To track the state of counters, \ours{} allocates the memory (\ie bits) according to the possible state of a large counter. Initially, a large counter (\eg 32-bit) contains multiple smaller counters (\eg 8-bit each). Before fully merging, these smaller counters transition into a newly introduced least significant bit (LSB) sharing state, allowing them to extend counting capacity without compromising accuracy. Eventually, as the counters reach their MSB limit, they progressively merge until all smaller counters merge into a single large counter (\eg 32 bits). Following this process, with an initial 8-bit counter size expanding to a 32-bit counter, there are a total of $11$ possible counter states, which can be tracked using only 4 bits (i.e., one bit assigned per 8-bit counter).

\subsection{Time Complexity}
\ours{} performs the same number of hash computations and memory accesses as SALSA, which are the most computationally intensive operations. Thus, the additional computational overhead primarily arises from two factors: 1) the extra increment operation for the MSB counter and 2) the LSB sharing operation, which requires additional bitwise or addition operations. We note that these extra operations occur infrequently enough to keep the amortized overhead low. For instance, when $K=4$, the MSB update triggers only once every 16 counter updates, and the LSB sharing operation occurs at most once every 256. Consequently, on average, our additional operations introduce a negligible overhead per update, maintaining the same overall time complexity as SALSA, as further explored in section~\ref{sec:eval}.

\section{Analysis}\label{sec:analysis}
In this section, we prove the error reduction effect of \ours{} by comparing error bound with state-of-the-art work. Notations are listed in Table~\ref {table: Definitions}.

% \nyang{ To YSS. 1. You have to define X1, and X2 before using them. These are the random variables stored in C-MSB1 and C-MSB2, respectively. Note that X1 and X2 are decoded values. 
% 2. Your claim is that X1 and X2 follow a hypergeometric distribution. So, rewrite in this way.
% 3. First, show the PMF (Probability Mass Function) and then rewrite like this. X1 and X2 follow hypergeometric distribution having following PMF with expectation and variance.}
\subsection{Winner-take-all Analysis}\label{sec:eval:winner}
\noindent\BfPara{Lemma} \textit{Let two adjacent counters $C_1$ and $C_2$ share $K$-LSBs, converting them into $C_{MSB_1}$, $C_{MSB_2}$ and a shared counter $C_\text{LSB}$ to perform LSB sharing, and let $S_1$ and $S_2$ be the number of packets that arrive at counters $C_1$ and $C_2$, respectively. Assuming that all packets are encoded into $C_\text{LSB}$ according to a randomly mixed sequence, each random variable $X_1$ and $X_2$ follows a hypergeometric distribution and has the following probability mass function (PMF), expectation, and variance:}

\begin{align*}
&P(X_1=i) = \frac{\binom{S_1}{i}\binom{S_2}{N-i}}{\binom{S_1+S_2}{N}} = P(X_2=N-i)\\
&E[X_1] = N\left(\frac{S_1}{S_1 + S_2}\right)\text{,} \quad E[X_2] = N\left(\frac{S_2}{S_1 + S_2}\right) \\
&V[X_1] = V[X_2]\\
&\hspace{23pt}= N\left(\frac{S_1}{S_1 + S_2}\right)\left(\frac{S_2}{S_1 + S_2}\right)\left(\frac{S_1 + S_2 - N}{S_1 + S_2 - 1}\right)\text{,}&
\end{align*}
where $N$ is the sum of two random variables $X_1$ and $X_2$, which is equivalent to $\left\lfloor \frac{S_1 + S_2}{2^K} \right\rfloor$.

\noindent\textit{Proof.} \ours{} encodes incoming packets (\ie $S_1+S_2$) into the shared counter $C_{\text{LSB}}$, and follows the winner-take-all strategy to encode $C_{\text{LSB}}$ overflow events to one of non-shared counters $C_{\text{MSB}_1}$ or $C_{\text{MSB}_2}$ based on the last packet. Since $C_{\text{LSB}}$ can count up to $2^K$ packets, the sum of two $C_\text{MSB}s$ (\ie $N$), which is the total number of $C_\text{LSB}$ overflow, can be denoted as $\big\lfloor \frac{S_1+S_2}{2^K} \big\rfloor$. 
Assuming that packets are randomly mixed, it follows the hypergeometric distribution. Assume  $i$ out of $S_1$ packets (to be encoded to $C_1$) triggers $C_\text{LSB}$ overflow among a total $N$ overflow events, then the total number of combinations exact $i$ triggers overflow is $\binom{N}{i}\binom{S_1+S_2-N}{S_1-i}$, where $i$ belongs to the group of $M$ packets that trigger the overflow and $S_1-i$ falls in the group has $S_1+S_2-N$ packets, which do not trigger the overflow events. 
Additionally, since the total number of possible  combinations of all packets (\ie $S_1+S_2$) is $\binom{S_1+S_2}{S_1}$, the probability $P(X_1=i)$ can be denoted as 
\begin{align*}
    P(X_1=i) &=\frac{\binom{N}{i}\binom{S_1+S_2-N}{S_1-i}}{\binom{S_1+S_2}{S_1}} =\frac{\binom{S_1}{i}\binom{S_2}{N-i}}{\binom{S_1+S_2}{N}}\text{.}\label{eq:1}
\end{align*}
Furthermore, the mean and variance can be retrieved based on hypergeometric distribution. Simply put, the above properties prove the winner-takes-all strategy leads non-shared MSB counters (\ie $C_{\text{MSB}_1}$ and $C_{\text{MSB}_2}$) to fairly obtain their shares, on average, even $S_1$ and $S_2$ packets are randomly encoded into $C_{\text{LSB}}$.  $\square$

\begin{table}[t]
\centering
\caption{Parameter definition for analysis.}
    \label{table: Definitions}
\begin{tabular}{|m{0.9cm}|m{6.9cm}|}
\hline
Params & Description \\ \hline
$f$ & The frequency of $f$ \\ \hline
$g$ & The sum of the frequencies of all inputs to $C_1$ except for $f$ \\ \hline
$h$ & The sum of the frequencies of all inputs to $C_2$ \\ \hline
$K$ & The number of shared counter's bits in \ours{} \\ \hline
$X_1, X_2$ & The random variables $C_{MSB_1}$ and $C_{MSB_2}$, respectively\\ \hline
$r$ & The value recorded in the shared LSB of \ours{} \\ \hline
\end{tabular}
\end{table}

\subsection{LSB Sharing vs Merging}\label{sec:error-bound}
In this section, we prove that the performance of \ours{} is better than the state-of-the-art SALSA~\cite{basat2021salsa} through a pairwise comparison of error bound varying counter states between LSB sharing and merging.

\begin{figure}[t]
    \centering
    \includegraphics[width=0.47\textwidth]{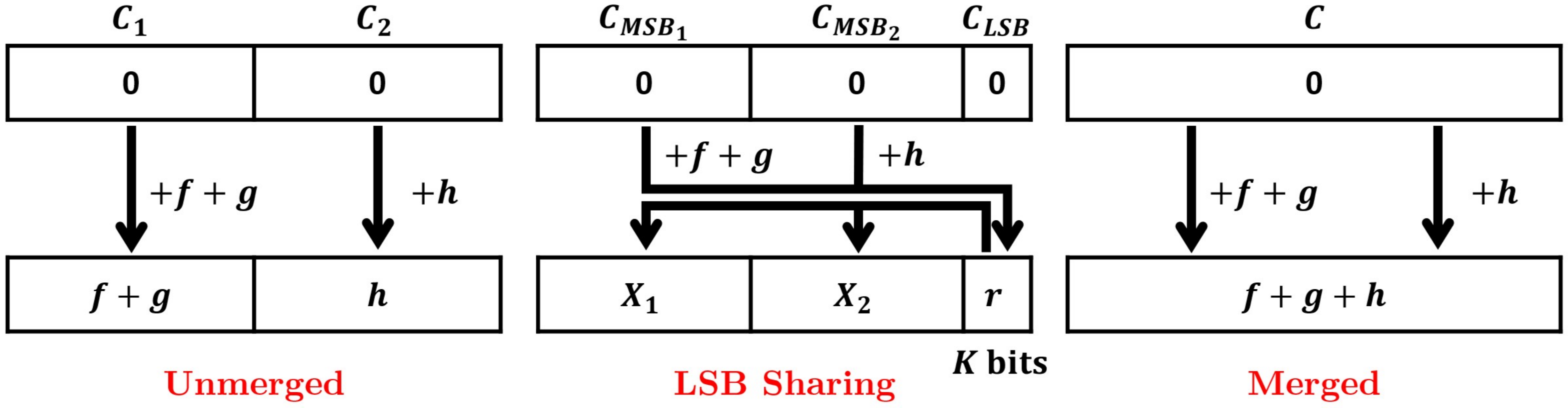}
     \caption{The figure above illustrates the representation of two unmerged counters (Left), LSB sharing (Middle), and Merged counter (Right). While Merging combines all flows into a single counter, LSB sharing retains most flow information separately in individual MSBs.}
     \vspace{2mm}
    \label{fig:LSB_sharing_Merging}
\end{figure}

\noindent\BfPara{Theorem 1}\textit{When SALSA's counters are merged and \ours{} is not yet merged (or unmerged) but uses LSB-shared counters:}
as shown in Fig.~\ref{fig:LSB_sharing_Merging}, let $\hat{f}_{unmerged}$ be the estimation when we count $f$ with two separated counters $C_1$ and $C_2$, $\hat{f}_{LSB}$ be the estimation when we count $f$ with shared LSB, and $\hat{f}_{merged}$ be the estimation when we estimate $f$ with one merged counter, where $\hat{f}_{merged}$ is obtained by merging the two counter $C_1$ and $C_2$.
Then we have the following property:
\begin{equation}
    \hat{f}_{LSB} \leq \hat{f}_{merged}
\label{worst_case}
\end{equation}
\begin{equation}
    \hat{f}_{unmerged} \leq \mathbb{E}[\hat{f}_{LSB}] < \hat{f}_{unmerged} + 2^K\text{.}
\label{avg_case}
\end{equation}

\noindent \sy{\textit{Proof.} In Fig.~\ref{fig:LSB_sharing_Merging}, $C_{MSB_1}$, $C_{MSB_2}$, and $C_{LSB}$ are denoted as $X_1$, $X_2$, and $r$, respectively. As in our lemma, let $N$ be the $X_1+X_2$, which equal to $\big\lfloor \frac{f+g+h}{2^K} \big\rfloor$, then $r$ becomes $f+g+h-2^KN$. Since $\hat{f}_{LSB}$ is $2^KX_1+r$ and $2^K(X_1+X_2)+r$ is $f+g+h$, which is $\hat{f}_{merged}$, we have a property as follows:
\begin{equation}
\label{LSB_Merging}
    \hat{f}_{LSB} \leq \hat{f}_{merged}\text{.} \nonumber
\end{equation}
On the other hand, $E[X_1]=N\big(\frac{f+g}{f+g+h}\big)$ by the lemma. Thus, we also have a property as follows:}
\begin{align}
E[\hat{f}_{LSB}] &= 2^K E[X_1] + r \nonumber \\
&= 2^K N \left(\frac{f+g}{f+g+h}\right) + (f+g+h - 2^K N) \nonumber \\
&= f + g + h \left(1 - \frac{2^K}{f+g+h}N\right)\text{.}
\label{est_avg}
\end{align}

\noindent Since $N$ follows the inequality 
\begin{equation*}
\frac{f+g+h}{2^K}-1 < N \leq \frac{f+g+h}{2^K}\text{,}
\end{equation*}
we get the following inequality from Eq.~(\ref{est_avg}), as
\begin{equation*}
f+g \leq E[\hat{f}_{LSB}] < f+g+\frac{h}{f+g+h}2^K\text{.}
\end{equation*}

Furthermore, since $\hat{f}_{unmerged}$ is $f+g$ and $\frac{h}{f+g+h} \leq 1$, we get the following inequality
\begin{equation*}
    \hat{f}_{unmerged} \leq \mathbb{E}[\hat{f}_{LSB}] < \hat{f}_{unmerged} + 2^K\text{.} \square
\end{equation*}

We note that Eq.~(\ref{worst_case}) guarantees that the error caused by LSB sharing is less than or equal to the error caused by merging in any case. Moreover, Eq.~(\ref{avg_case}) represents that the average error caused by LSB sharing is bounded by the size of the shared counter. Since the drawback of LSB sharing remains minimal as long as $K$ is not too large, LSB sharing enhances the robustness of each counter with insignificant costs. Therefore, this theorem demonstrates that \ours{} achieves higher accuracy than SALSA when utilizing LSB sharing since \ours{} employs LSB sharing where SALSA merges two counters, postponing the merging step further.

\noindent\BfPara{Theorem 2} \textit{When both SALSA's and \ours{}'s counters are max-merged:} \ours{} gives more accurate results compared to SALSA in general.

\noindent \textit{Proof.} Max-merging discards the value of the adjacent counter to be merged with the overflowed counter since the maximum is already larger than the size of the flows using the small counter. LSB sharing can further enhance this merit by maintaining that the two counters are separated for a longer period. Thus, \ours{} gives better results than SALSA, even after \textit{max-merging}. $\square$

\noindent\BfPara{Theorem 3} \textit{When both SALSA's and \ours{}'s counters are sum-merged:} \ours{} and SALSA has the same result.

\noindent \textit{Proof.} When \ours{} merges its counters, the value of the merged counters is defined as $2^K(X_1+X_2)+r$, which is the sum of $C_{MSB_1}$, $C_{MSB_2}$, and $C_{LSB}$, as shown in Fig.~\ref{fig:LSB_sharing_Merging}. Since this sum is $f+g+h$, $\hat{f}_{LSB}$ becomes the same as $\hat{f}_{merged}$. Thus, \ours{} and SALSA have the same value when the sum-merging is used. $\square$

\section{Evaluation} \label{sec:eval}
In this section, we evaluate \ours{} and compare its performance against both dynamic and static structure sketches from the literature across five common security applications, followed by an in-depth analysis of the sketches to demonstrate the benefits of the late merging effect in retaining counter sustainability in the long run.

\subsection{Settings}

\BfPara{Testbed} For all experiments, we used a single core with AMD Ryzen Threadripper PRO 5975WX CPU @1.80 GHz and 512 GB DDR4 3200 MHz RAM on a server.

\begin{figure*}[t]
    \centering
    \subfigure[CAIDA]    {\includegraphics[width=0.19\textwidth]{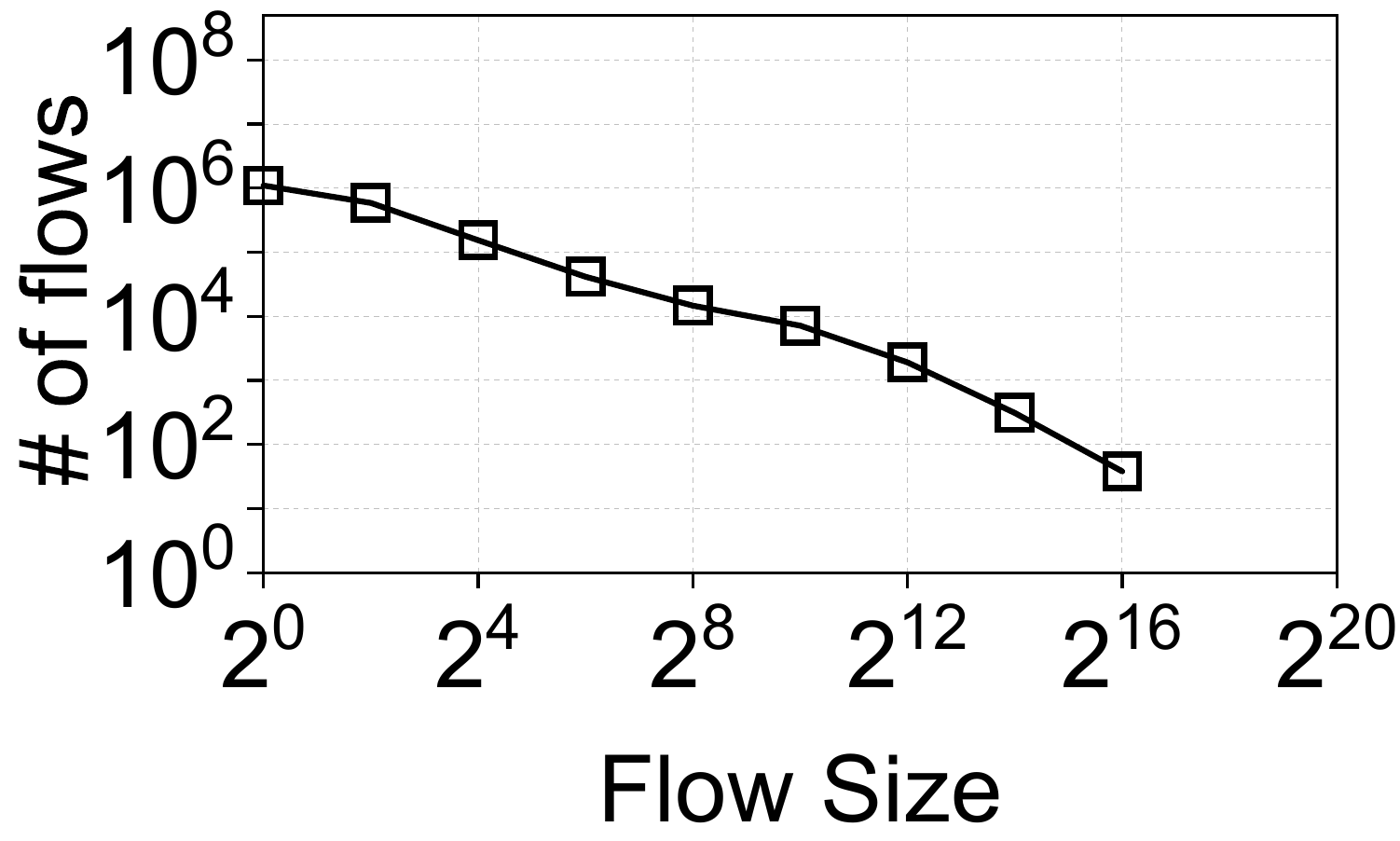}}
    \subfigure[MACCDC]{\includegraphics[width=0.19\textwidth]{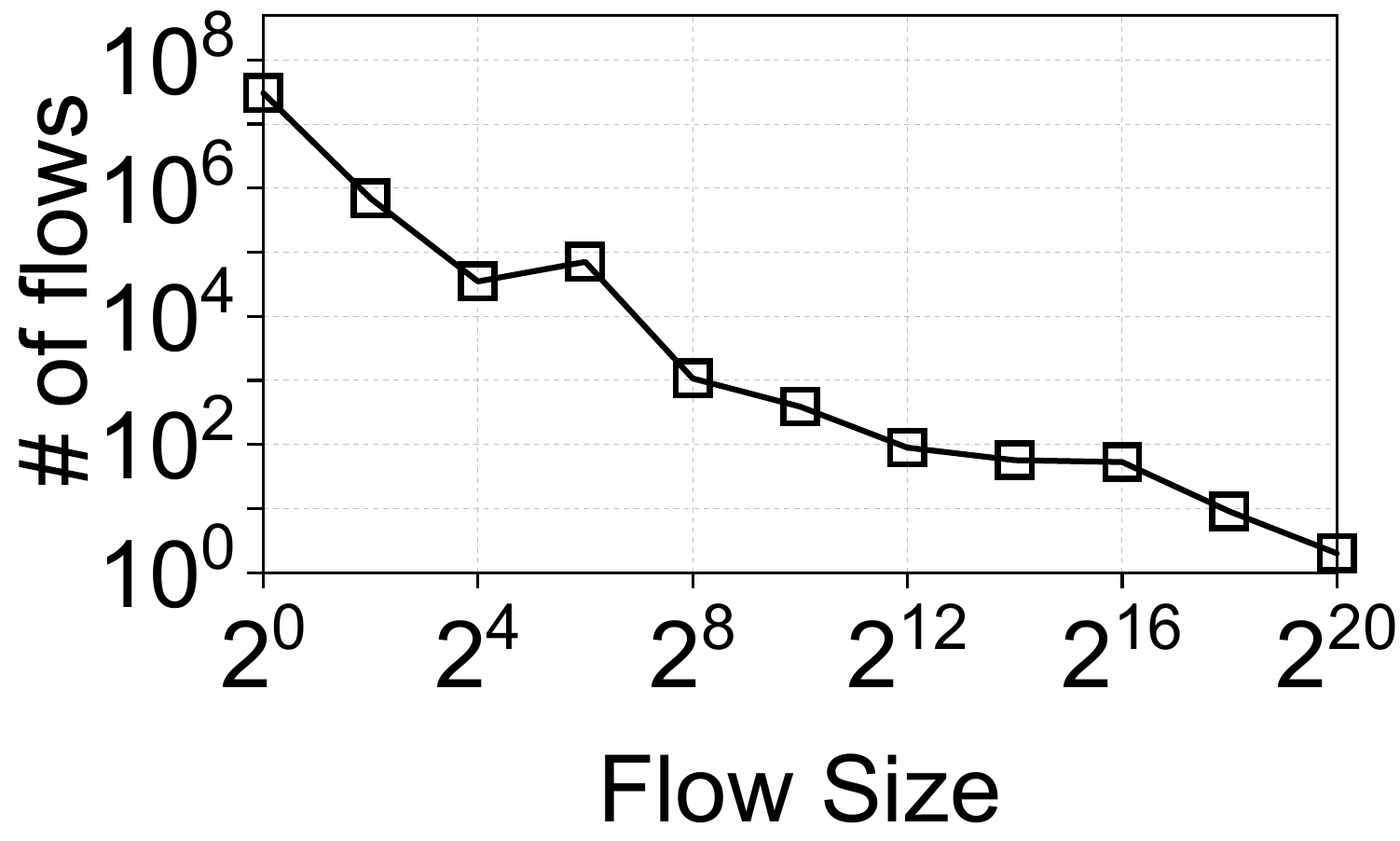}}    
    \subfigure[CIC DDoS]    {\includegraphics[width=0.19\textwidth]{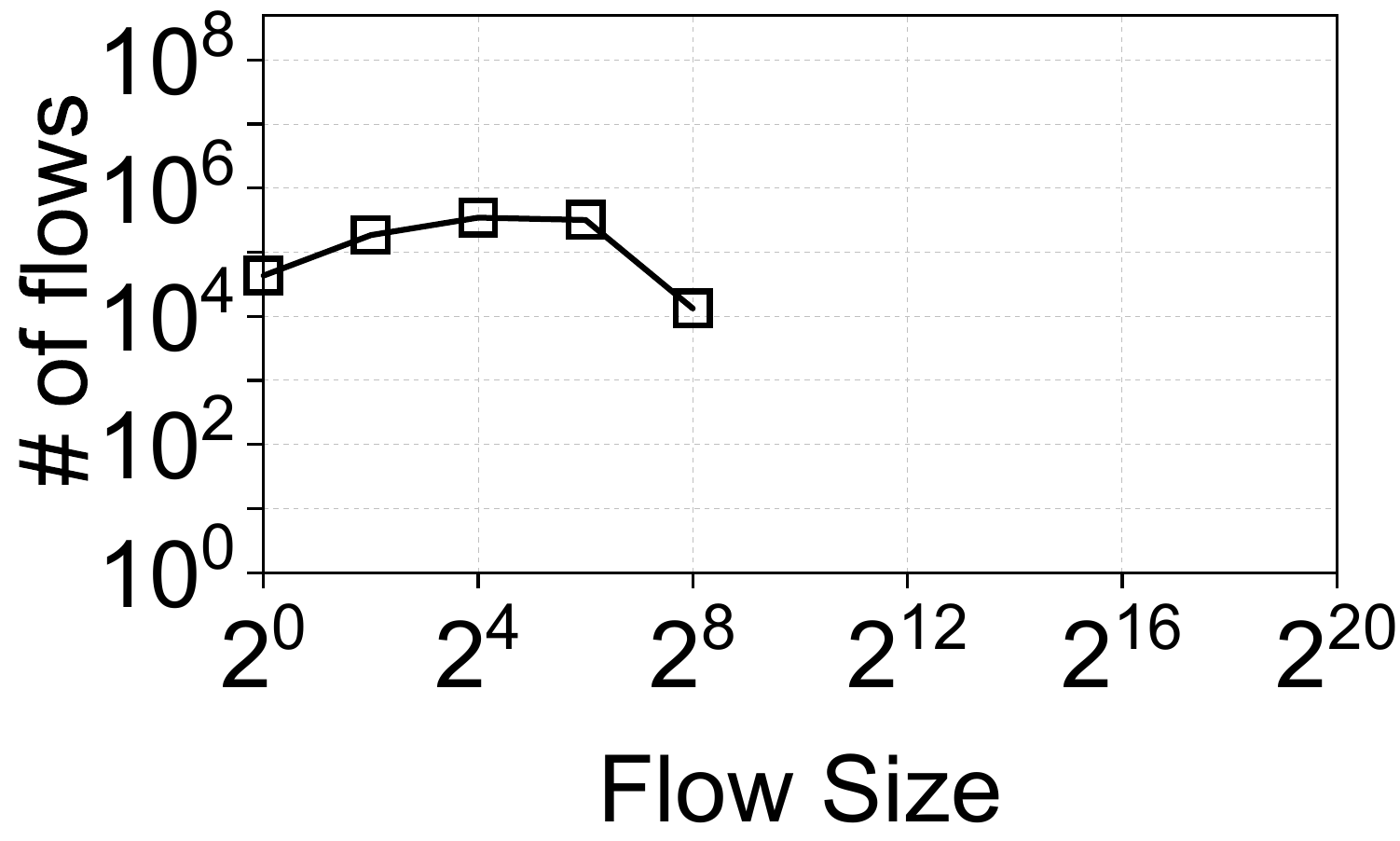}}    
    \subfigure[CIC IDS]    {\includegraphics[width=0.19\textwidth]{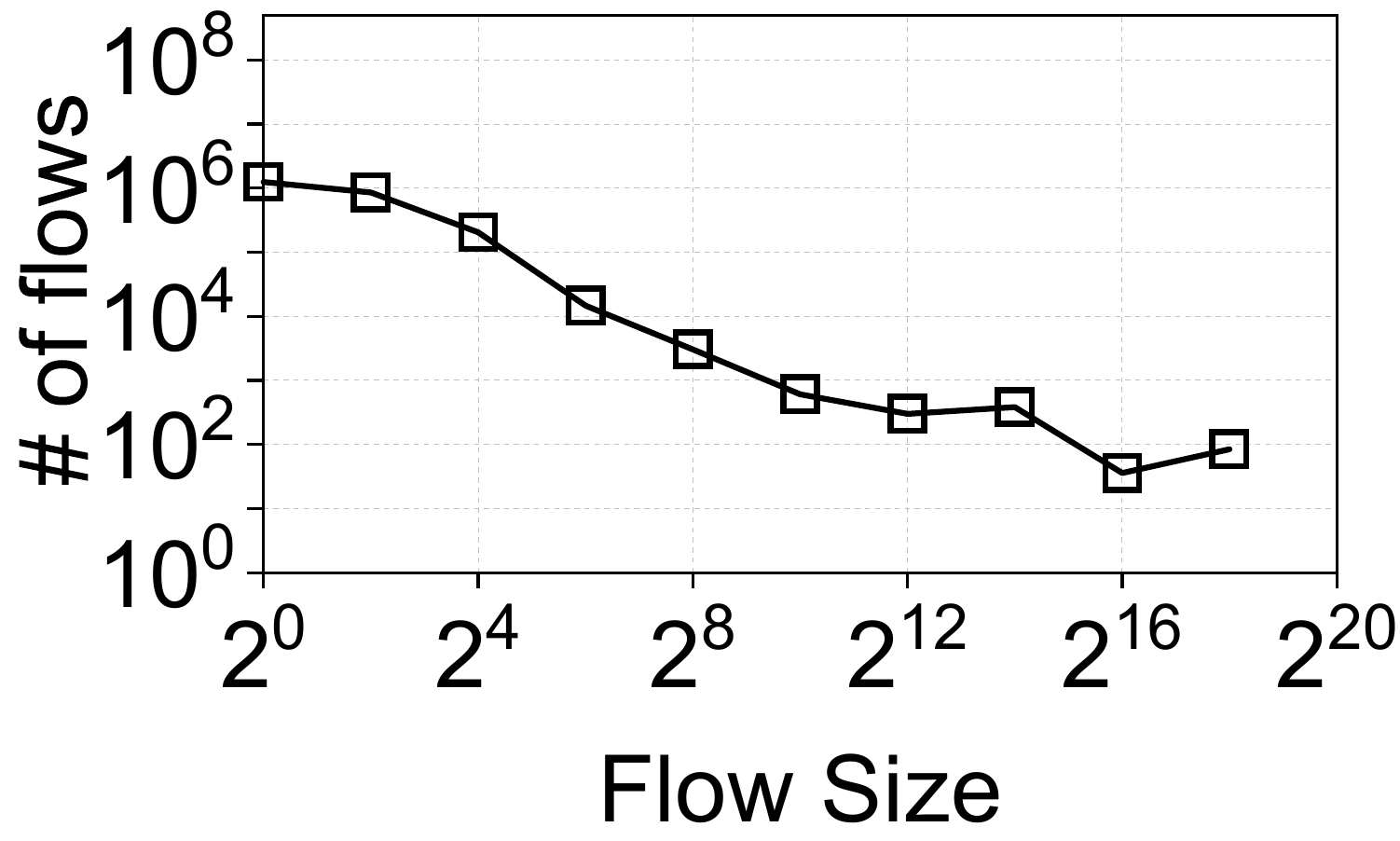}}    
    \subfigure[UNSW DoS]    {\includegraphics[width=0.19\textwidth]{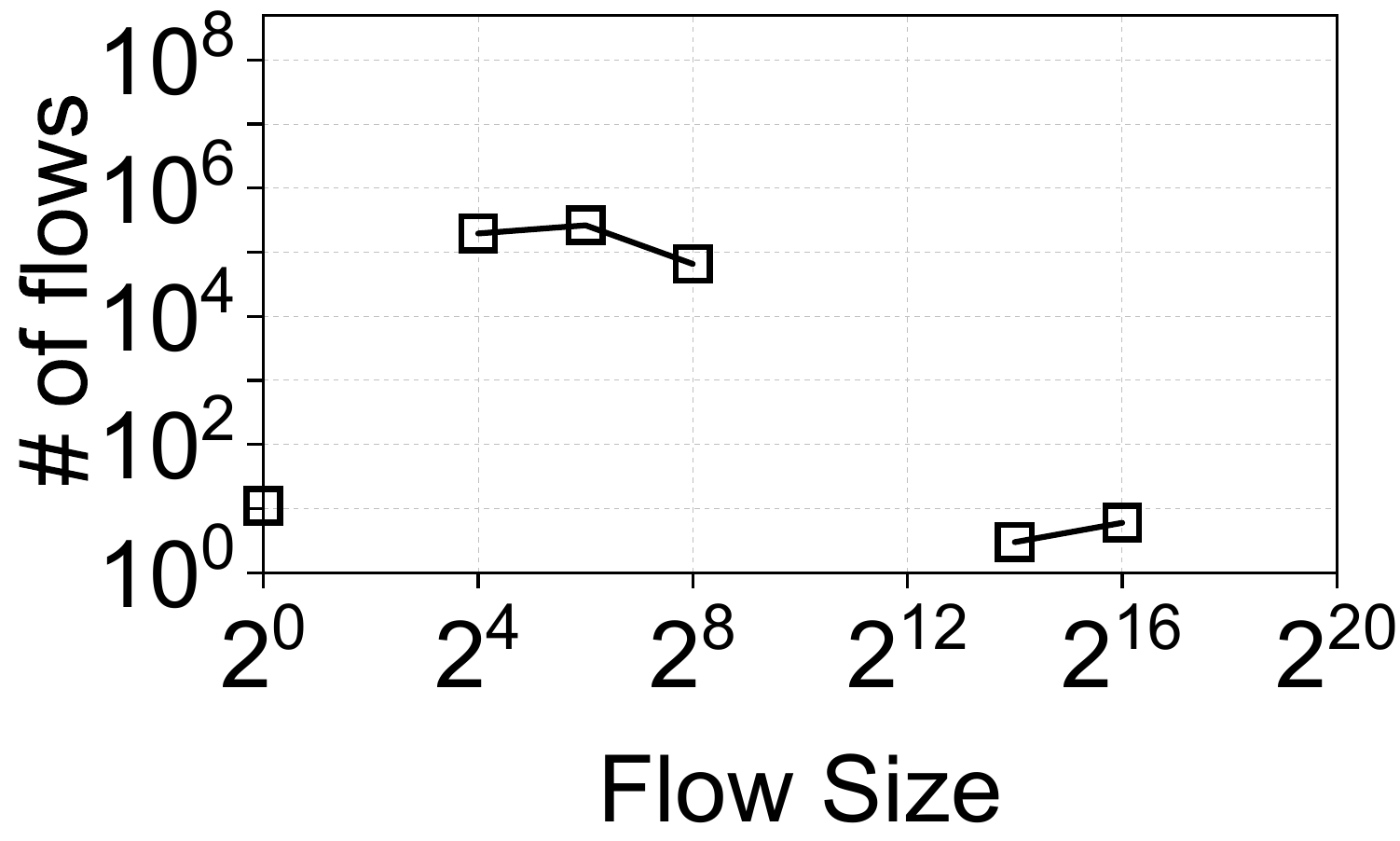}}\vspace{-2mm}
    \caption{Flow size distribution of five real-world datasets for 64 million packets stream. }~\label{fig:Flow_Dist}
\end{figure*}

\BfPara{Dataset} In our experiments, we used five different public real-world traces, and also a synthetic trace, to simulate various traffic patterns and flow size distributions. The synthetic datasets are generated with different Zipf skewness values varied from 0.6 to 1.4. We used five public datasets (see Fig.~\ref{fig:Flow_Dist} for data distribution) comprising one benign trace CAIDA~\cite{CAIDA} and four malicious datasets, such as MACCDC~\cite{maccdc2012}, CIC-DDoS~\cite{sharafaldin2019developing}, CIC-IDS~\cite{sharafaldin2018toward}, and UNSW-DoS~\cite{koroniotis2019towards}.

\BfPara{Parameters}
We compared \ours{} with ABC~\cite{gong2017abc} and SALSA~\cite{basat2021salsa}, which are state-of-the-art schemes employing dynamic sketch approach, as well as CountLess~\cite{kim2023count} and FCM~\cite{song2020fcm}, which are state-of-the-art works leveraging static counter extension approach.
Our analyses indicated that the performance gap is negligible when varying \ours{}'s shared counter sizes ($K$) from 2 to 6 bits, as shown in Fig.~\ref{fig:Synthetic_CMP}. Therefore, we set the size of shared LSB as 4 bit (\ie $K = 4$) for \ours{} in the following experiments. For fairness, all the sketches in our experiments use the same amount of memory initiated with a similar number of small counters (8-bit). Additionally, all schemes use three counter arrays with the same hash functions (BobHash) and operations for indexing.

\subsection{Metrics}
We use the following metrics to evaluate \ours{}'s performance. We denote $N$ as the number of distinct flows, $f_{i}$ as the actual size, and $\hat{f}_{i}$ as the estimated flow size, respectively.

\BfPara{Absolute Relative Error (ARE)} $ARE$ is $\frac{1}{N}\sum_{i=1}^{N}\frac{\lvert f_{i}-\hat{f}_{i} \rvert}{f_{i}}$, which is average of relative error.

\BfPara{Root Mean Square Error (RMSE)} $RMSE$ is defined as $\sqrt{\frac{1}{N}\sum_{i=1}^{N}(f_{i}-\hat{f}_{i})^2}$, which is the standard deviation of error.

\BfPara{F1 Score} F1 score is defined as $\frac{2*PR*RE}{PR+RE}$, where PR represents precision and RE for recall. We employed the F1 score to assess the performance of heavy hitter detection

\BfPara{Weighted Mean Relative Error (WMRE)}
WMRE is defined as $\frac{ \sum_{i=1}^{z} |n_i-\hat{n_i}|}{\sum_{i=1}^{z} (\frac{n_i+\hat{n_i}}{2})}$, where $z$ is the largest flow size, $\hat{n_i}$ is estimated number of flows with size $i$, and $n_i$ is ground truth.

\BfPara{Relative Error (RE)} RE is defined as $|1-\frac{estimated}{actual}|$, which is used to evaluate the accuracy of entropy estimation.

\BfPara{Million Packets Per Second (Mpps)} The number of processed packets per second (\ie throughput) to compare the processing efficiency of sketches.

\subsection{Sketch-based Security Applications under Pollution Attack}
In this section, we analyze the impact of sketch pollution attacks on five well-known security applications by detailing the performance of static extension and dynamic merging approaches.
For each security application, we conducted two sets of experiments. In the first part, we fixed the memory across all schemes to 0.5 MB and used 32 M packets from CAIDA~\cite{CAIDA} trace. We varied the attack flows from 0 to 142 K, which is equivalent to targeting 0\%-60\% of the total counter $w$. In the second part, we fixed the number of attack flows at 108 K and varied the memory from 0.2 MB to 1 MB.

\begin{figure*}[t]
    \centering
    \includegraphics[width=0.45\textwidth]{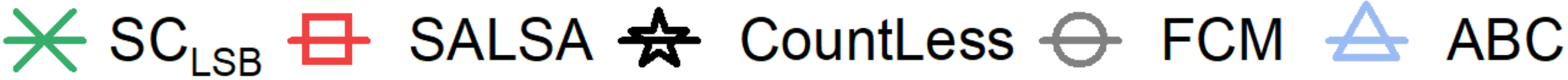} \\
    \subfigure[Flow size estimation]
    {\includegraphics[width=0.19\textwidth]{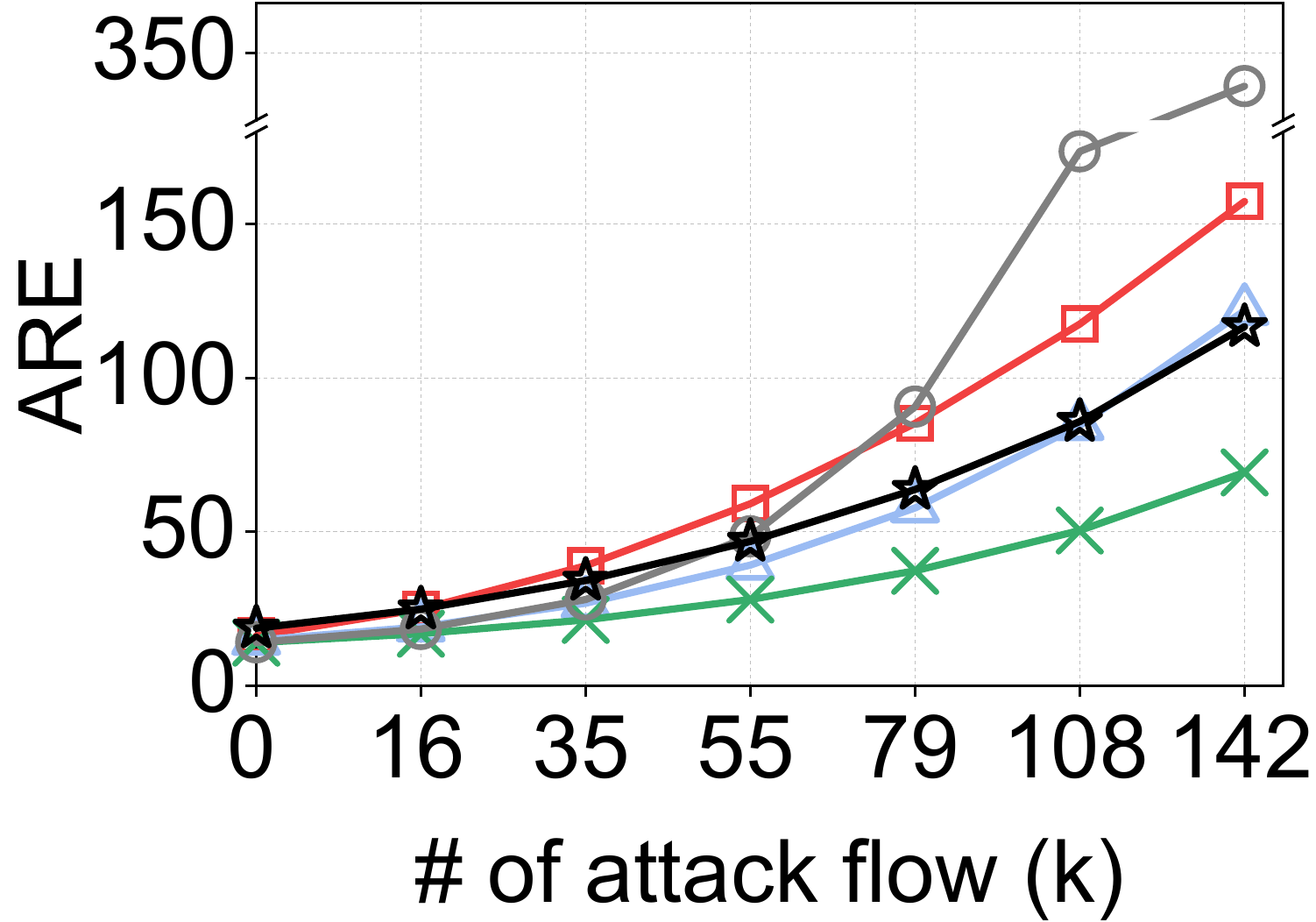}}
       \subfigure[Heavy hitter detection]
    {\includegraphics[width=0.19\textwidth]{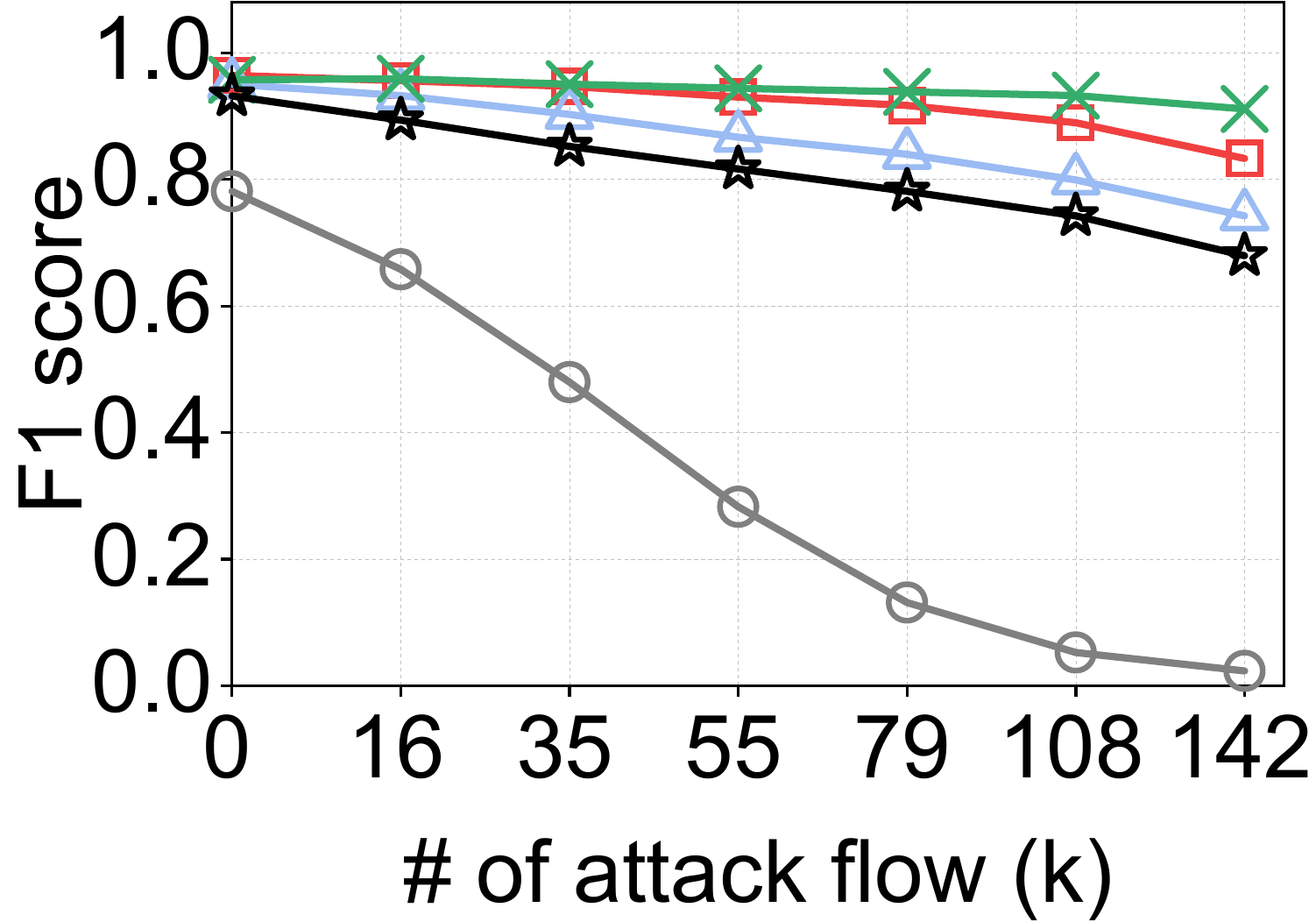}}
    \subfigure[Change detection]
    {\includegraphics[width=0.19\textwidth]{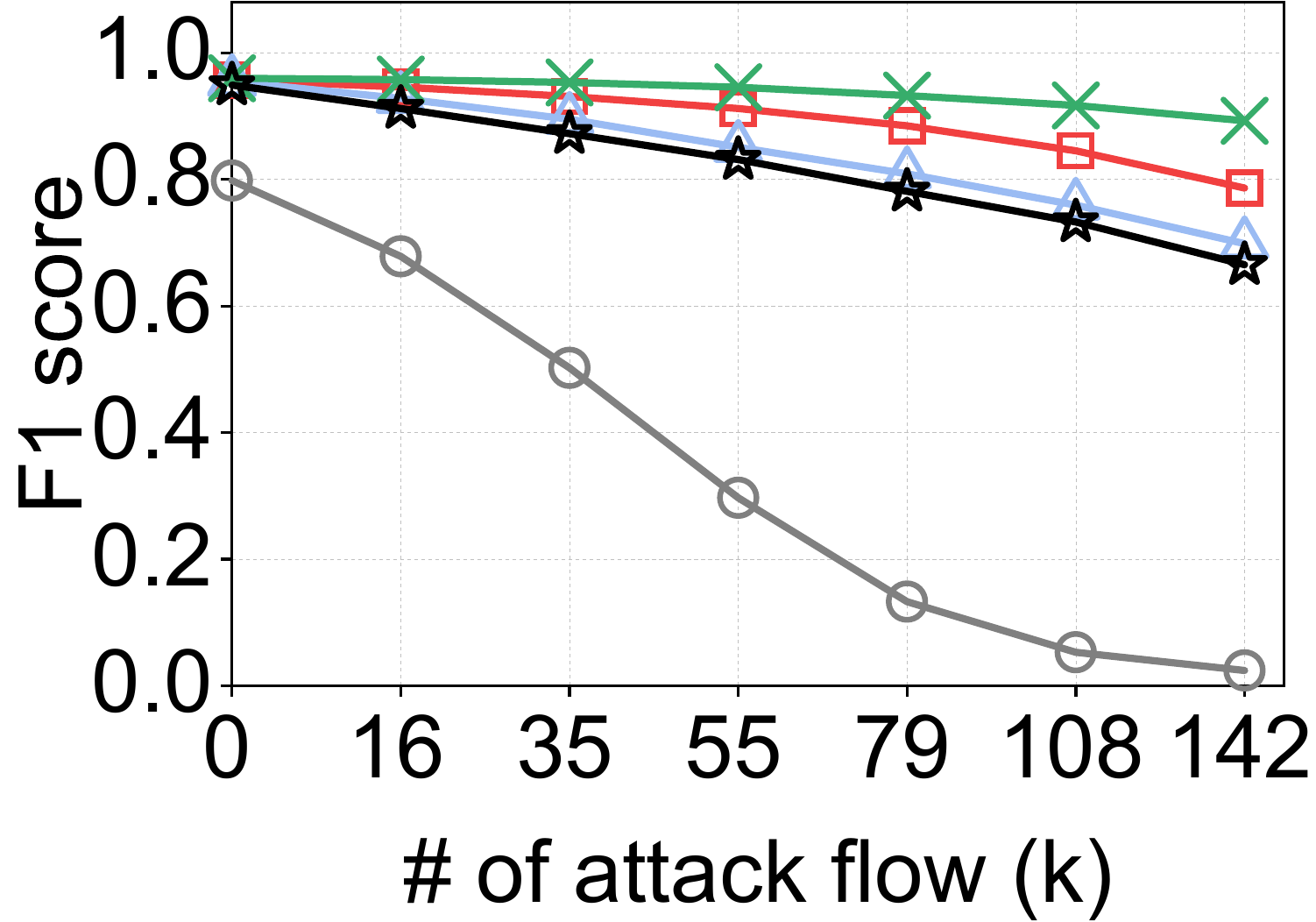}}
    \subfigure[Flow size distribution]
    {\includegraphics[width=0.19\textwidth]{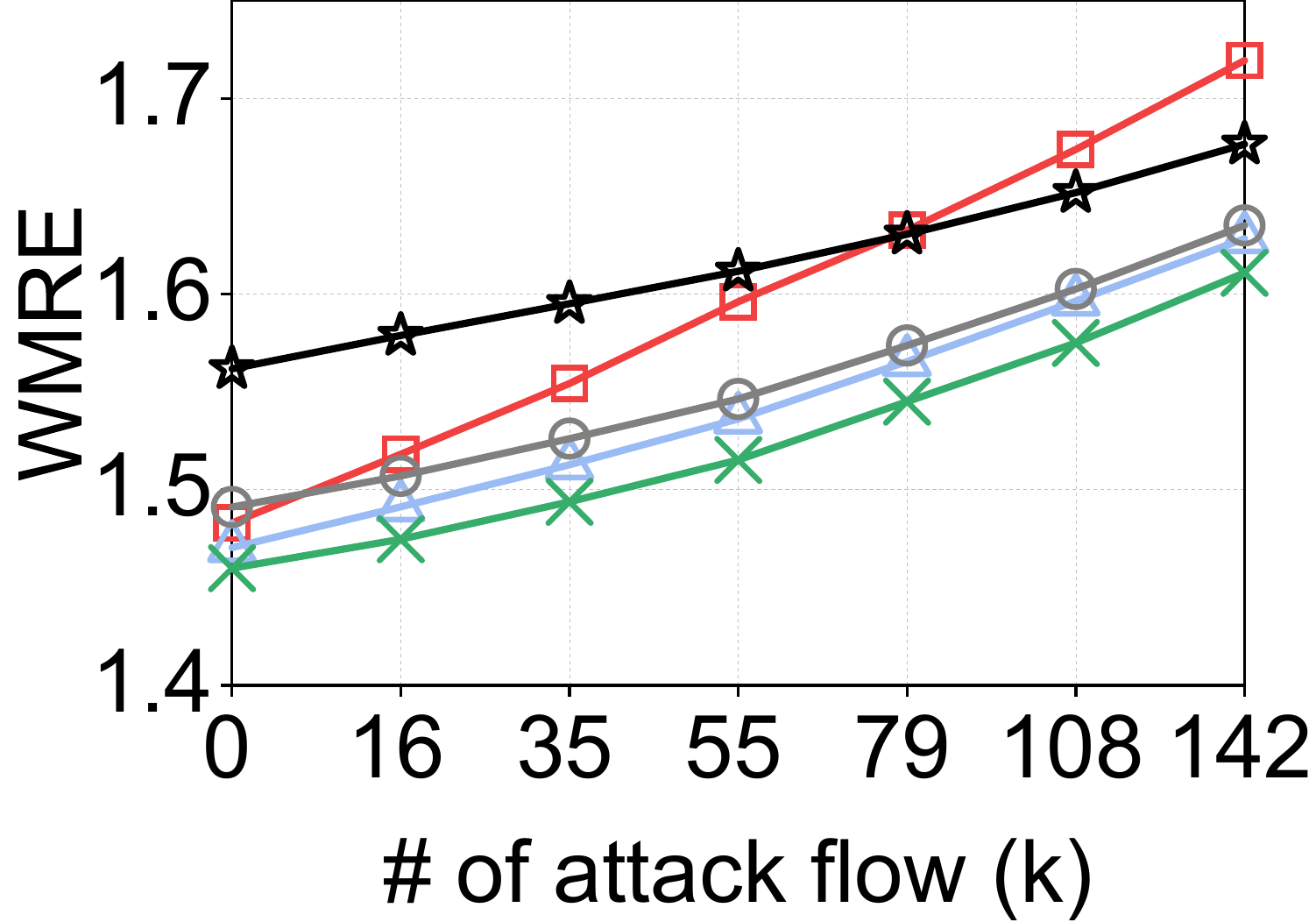}}
    \subfigure[Entropy]
    {\includegraphics[width=0.19\textwidth]{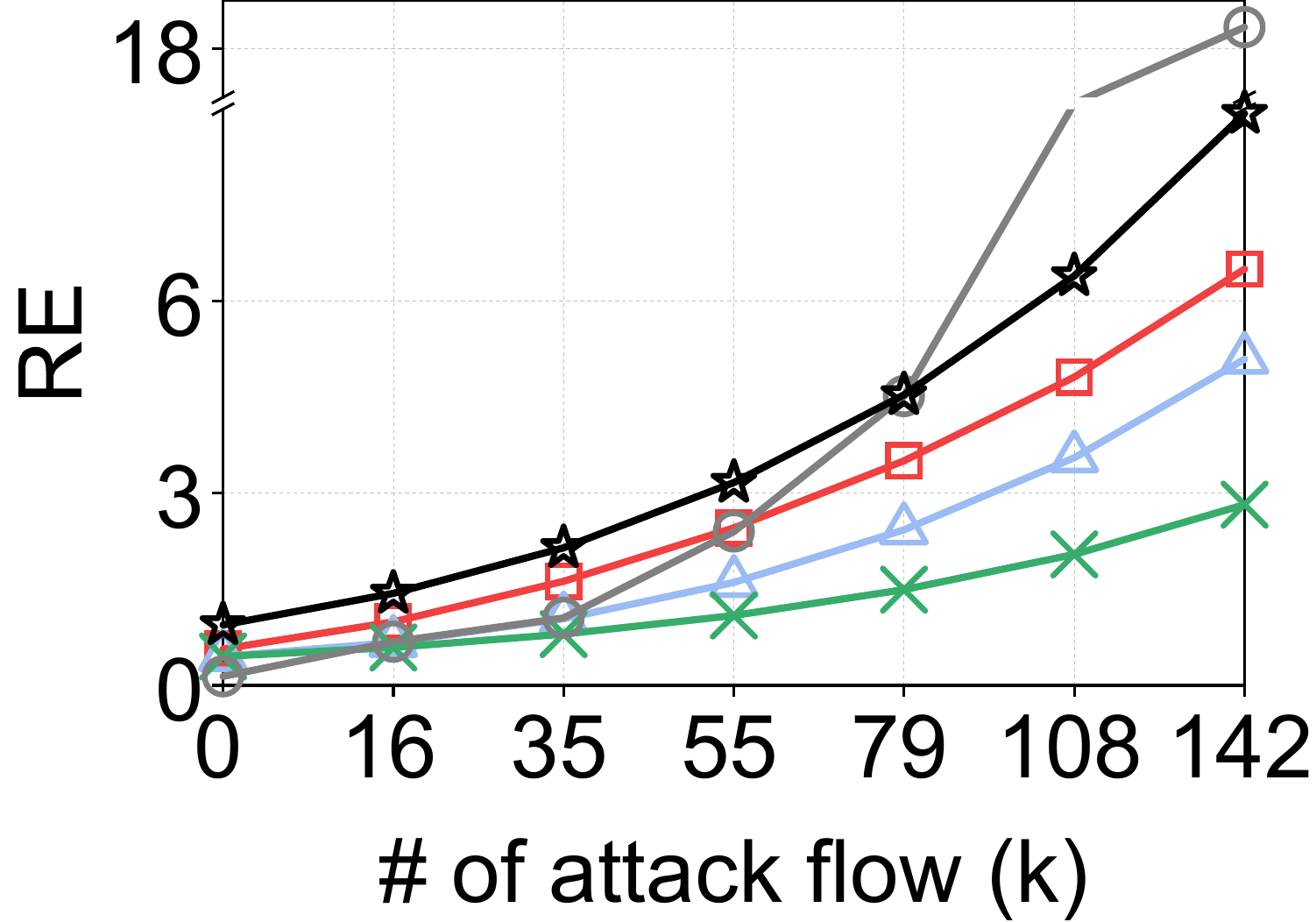}}
    \caption{Sketch-based security application under pollution attack: initially, static structures perform well against attack traffic, but their performance degrades as more counters become targeted.
    }~\label{fig:sec5_eval}\vspace{-4mm}
\end{figure*}

\BfPara{Flow Size Estimation} Flow size estimation is typically used to calculate the size of a flow and can further be extended to estimating the volume of certain information in packets (e.g., flags, protocol, \etc), which is a valuable feature when identifying malicious attack traffic~\cite{jaqen2021}. As shown in Fig.~\ref{fig:sec5_eval} (a), the static counter extension FCM performs well under light pollution attacks (fewer than 79k attack flows, equivalent to targeting approximately 50\% of the counters), outperforming the dynamic sketch SALSA by having 21\% less error on average. However, with more aggressive attack flows, dynamic approaches SALSA and ABC demonstrate 30\% less error on average. Notably, although CountLess benefits from a design that combines dynamic and static features, it eventually exhibits similar behavior under heavy attack traffic (not shown). On the other hand, \ours{}, with its late merging design, demonstrates resiliency by having  11\%, 17\%, 8\%, and 16\% less error when compared to FCM, SALSA, ABC, and CountLess, respectively, under light pollution attacks, while outperforming others by an average of 35\% during more aggressive attacks.

\begin{figure*}[t]
    \centering
    \subfigure[Flow size estimation]
    {\includegraphics[width=0.19\textwidth]{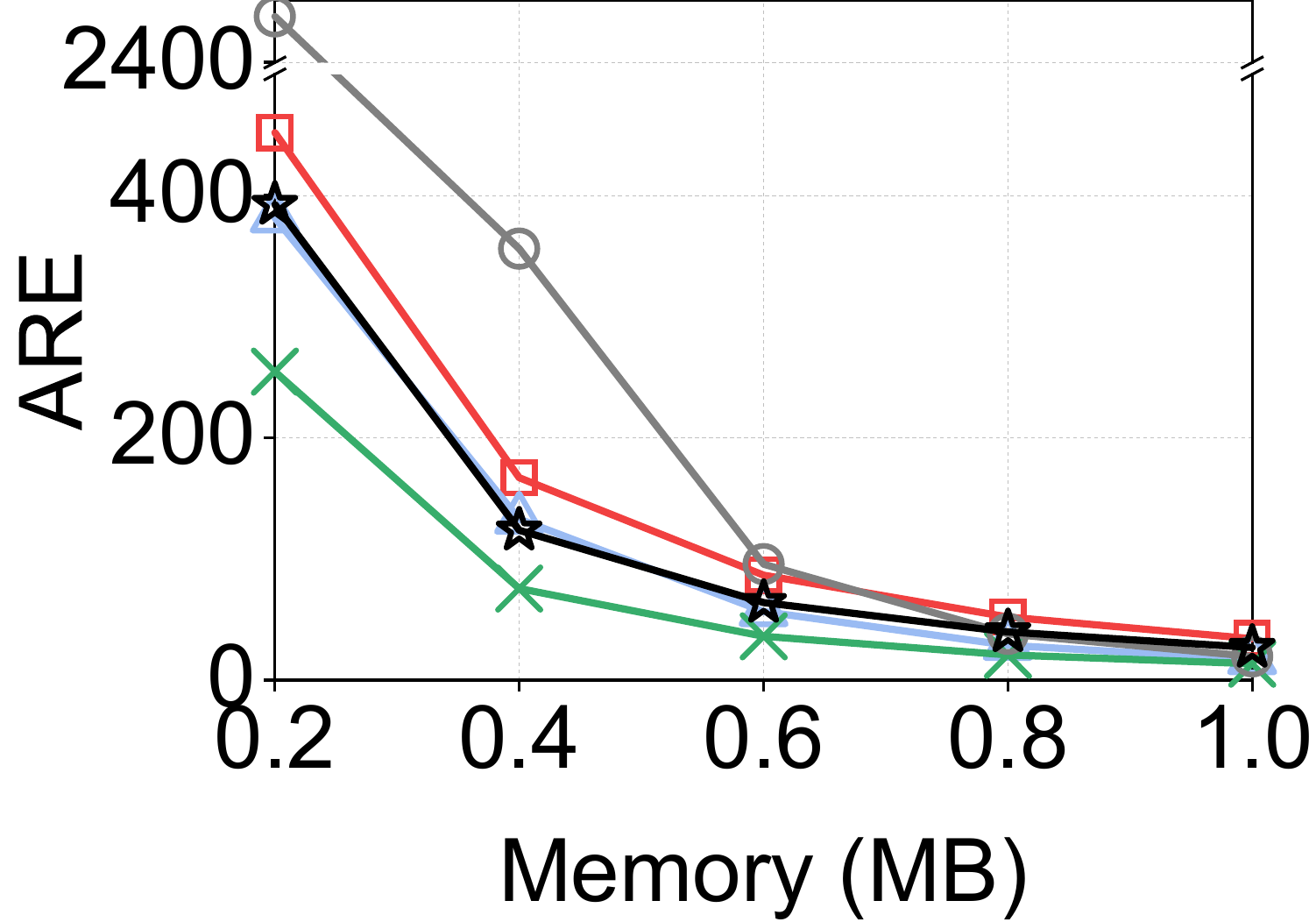}}
       \subfigure[Heavy hitter detection]
    {\includegraphics[width=0.19\textwidth]{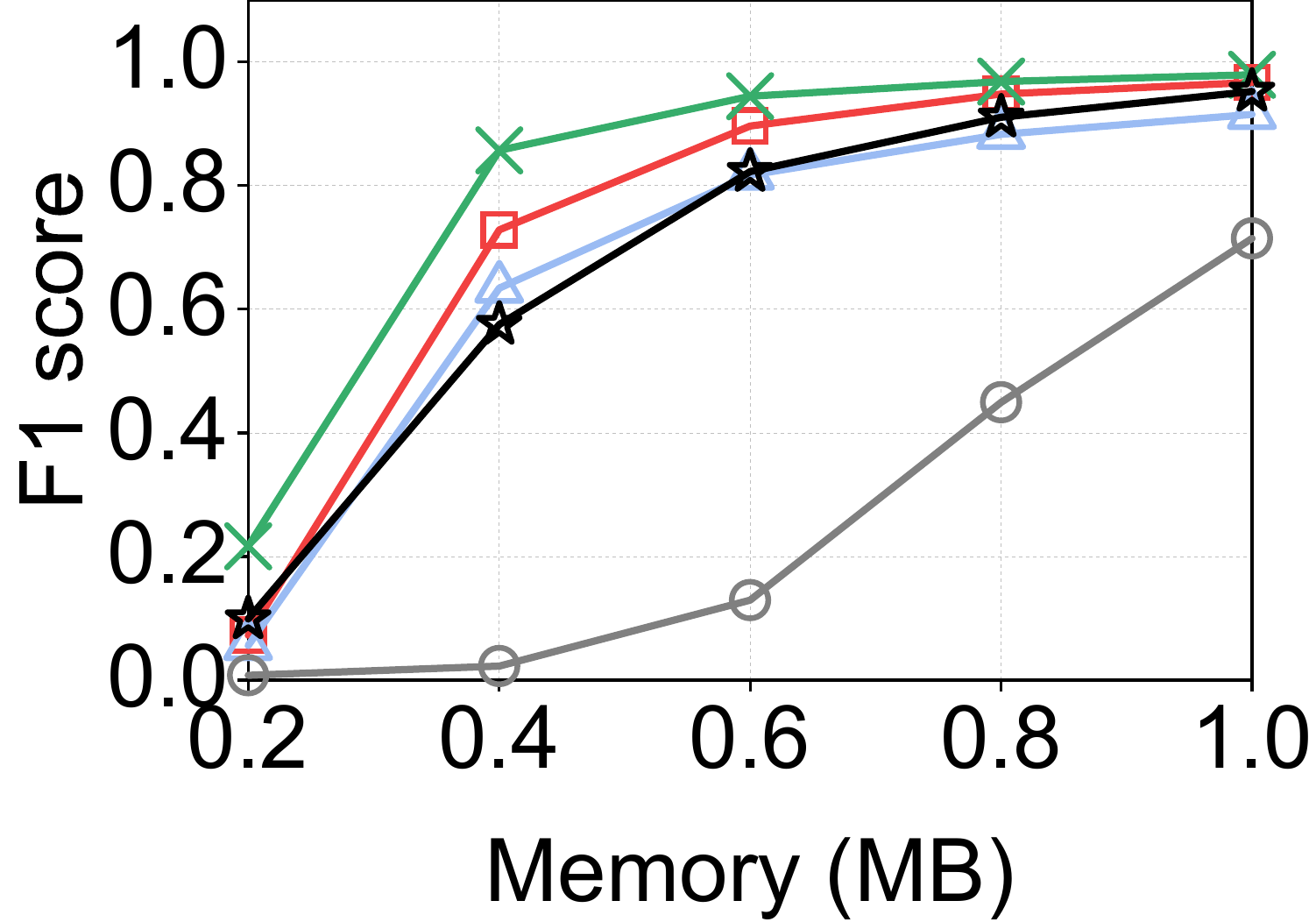}}
    \subfigure[Change detection]
    {\includegraphics[width=0.19\textwidth]{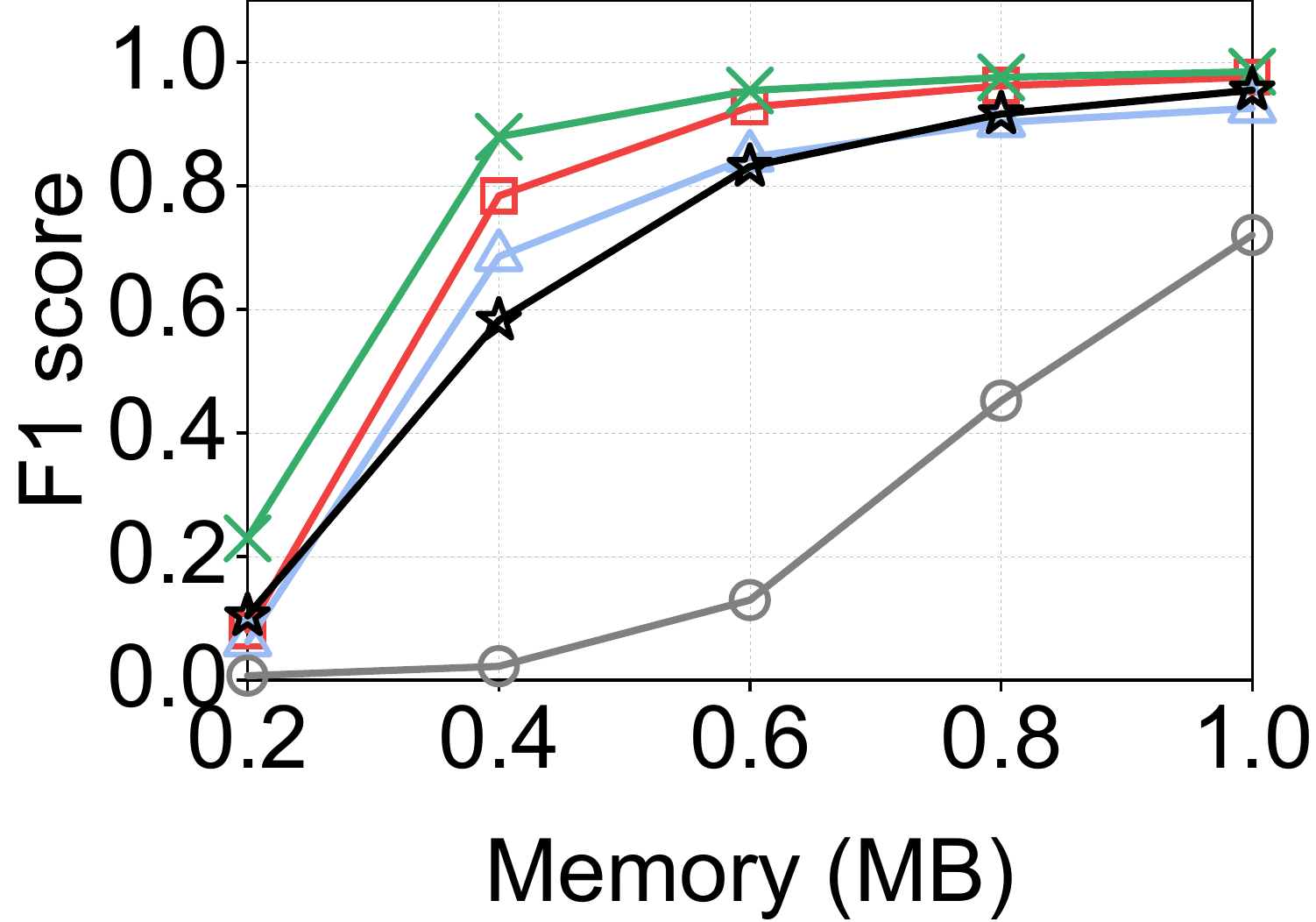}}
    \subfigure[Flow size distribution]
    {\includegraphics[width=0.19\textwidth]{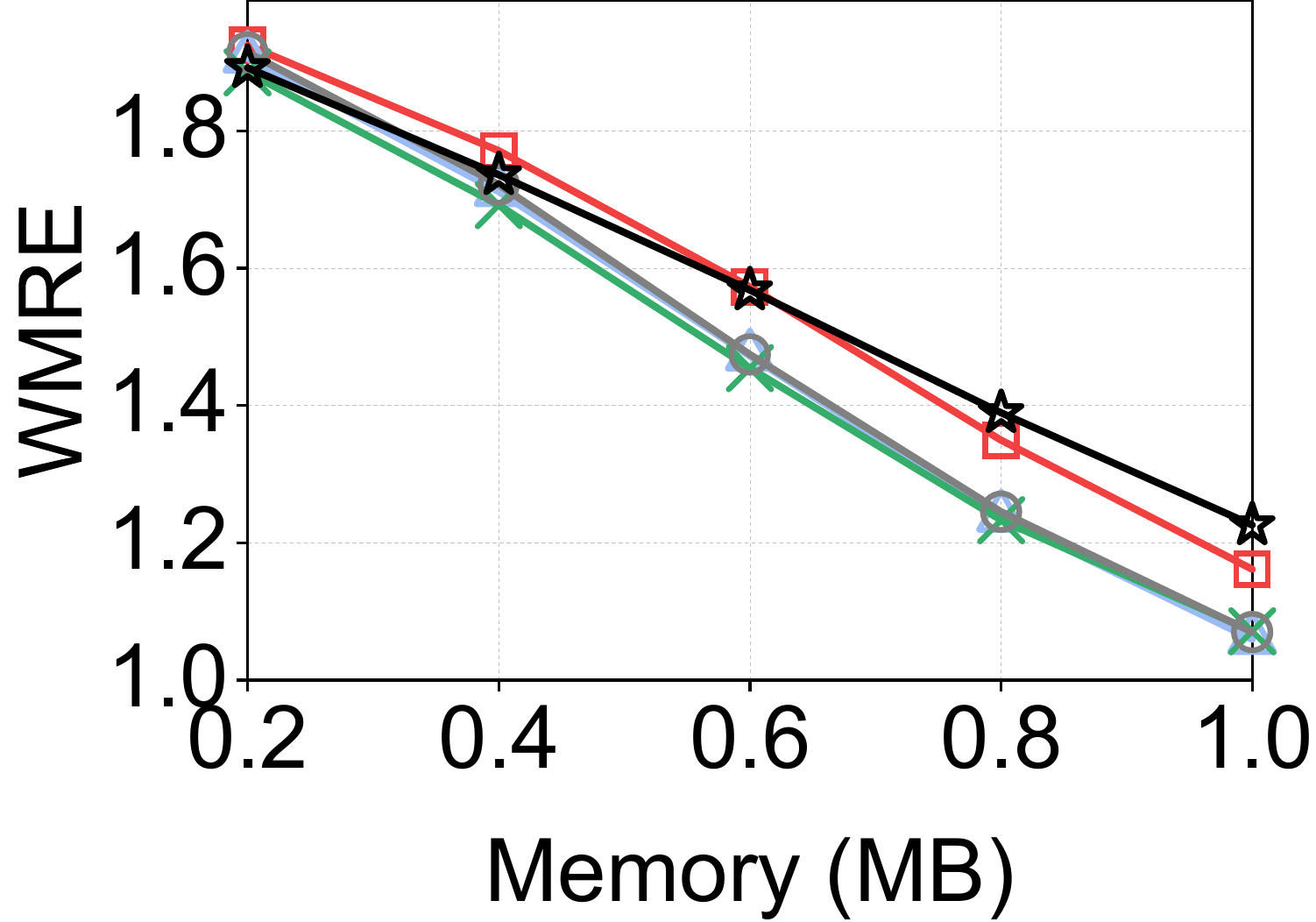}}
    \subfigure[Entropy]
    {\includegraphics[width=0.19\textwidth]{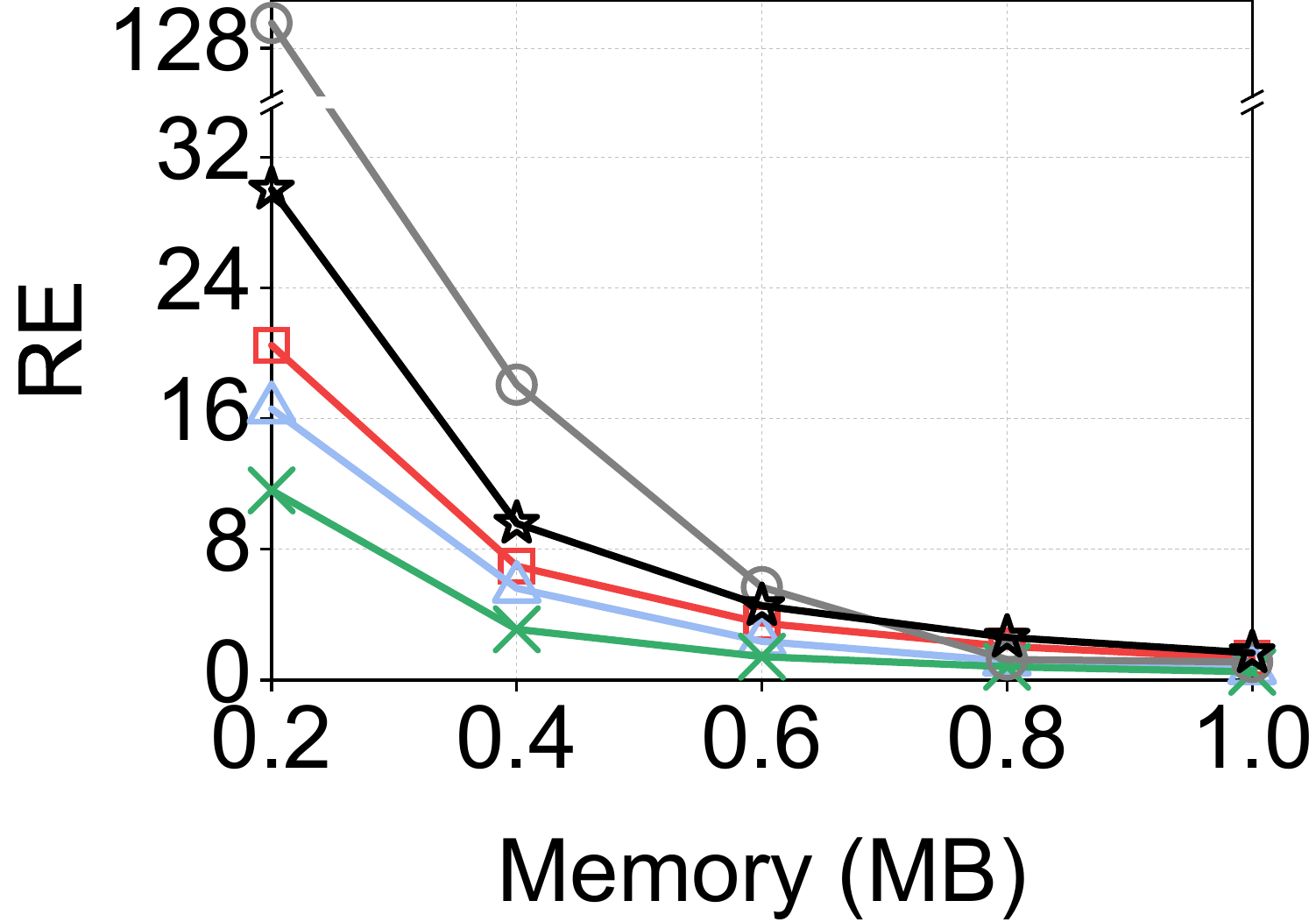}}
    \caption{Sketch-based security application under pollution attack varying memory: dynamic structures generally perform better than static structure sketches.
    }~\label{fig:sec5_eval_mem}
\end{figure*}

Similarly, as shown in Fig.~\ref{fig:sec5_eval_mem} (a), when varying the memory, dynamic merging SALSA outperforms FCM (the static counter extension) when the allocated memory is less than 0.6 MB by 50\% lower error on average. However, when the allocated memory exceeds 0.6 MB, FCM outperforms SALSA by 35\% less error, which suffers from plundering the neighboring counters. Across all schemes, \ours{} outperforms FCM, CountLess, SALSA, and ABC by 61\%, 43\%, 55\%, and 32\% on average.

\BfPara{Heavy Hitter Detection} Heavy hitter detection identifies network flows whose sizes exceed a predefined threshold $\phi$. This is particularly useful for detecting anomalies such as DDoS/DoS attacks~\cite{jaqen2021, yang2018elastic, zhou2023efficient} to identify suspicious flows or perform further analysis. We fixed the threshold as $0.004$\% of $N$, the total number of benign packets (equivalent to a threshold of 1500). As shown in Fig.~\ref{fig:sec5_eval} (b), dynamic schemes ABC and SALSA perform better across all attack scenarios due to their dynamic recycling design, compared to static extension approaches, which suffer from a high collision rate in the upper layer counters (e.g., 16-bit) where heavy flows reside. In this manner, the dynamic sketch SALSA and ABC outperform static extensions 4x times on average, and \ours{} outperforms static extension 6x and dynamic merging by 1.7x times on average. 
Fig.~\ref{fig:sec5_eval_mem} (b) shows the trend when varying the memory. As expected, dynamic counter merging consistently outperforms FCM, the static counter extension, across all memory settings. Moreover, CountLess, which uses the same static counter extension design but incorporates a dynamic operation design, demonstrates significant improvement compared to FCM with 41\% better accuracy on average.

\BfPara{Change Detection} Change detection identifies network flows with significant changes exceeding a threshold of $\phi$ over two continuous time windows. This is useful for detecting spikes or sudden drops in traffic, helping to identify flash crowds~\cite{jung2002flash} and stealthy DDoS~\cite{zhang2007low} attacks. Similar to heavy hitter detection, change detection exhibits an equivalent trend when $\phi$ is set to 0.004\% of $N$ (total number of benign packets), as depicted in Fig.~\ref{fig:sec5_eval} (c). In this manner, the dynamic sketch SALSA and ABC outperform static extensions FCM 4x times on average, and \ours{} outperforms static extension 6x and dynamic merging by 1.6x times on average.

Fig.~\ref{fig:sec5_eval_mem} (c) shows the trend when varying the memory, which is similar to heavy hitter detection. As shown, when the memory is set to as low as 0.2 MB, all schemes exhibit an F1 score of less than 0.06.  This is because static counter extension suffers from a high collision rate at upper layers and a high false positive rate, while dynamic counter merging suffers from plundering idle counters. In the same setting, even in such a low memory setting, \ours{} delivers 2.2x times more accurate performance on average, thanks to the late merging to enable long run sustainability of counters.

\BfPara{Flow Size Distribution} Flow size distribution measures the accuracy of the estimated distribution of flow size compared to the ground truth. This metric can be used to detect abnormal traffic distribution deviated from normality. As shown in Fig.~\ref{fig:sec5_eval} (d), we can infer that flow size distribution is more sensitive to the collision rate, as it directly impacts the estimation of the number of flows with size $i$. In this context, the static counter extension FCM consistently delivers better results, outperforming SALSA, the dynamic counter merging approach, by 4\% more accurate estimation on average. Despite suffering from a high collision rate in the upper layer, the results demonstrate that counter merging has a more negative impact compared to static counter extension. On the other hand, \ours{} outperforms all the schemes by 9\% less error on average.
Similarly, as shown in Fig.~\ref{fig:sec5_eval_mem} (d), we can observe that \ours{} consistently delivers 6\% more accurate performance across all memory settings.


\BfPara{Entropy} Entropy measures the randomness in the flow size distribution, which can be used for anomaly detection by identifying deviation from normal traffic patterns~\cite{da2020euclid}. It is being calculated using flow size distribution by $-\sum i*\frac{n_i}{N}*log(\frac{n_i}{N})$, where $N$ is the total number of flows and $n_i$ is the number of flows with size $i$. As shown in Fig.~\ref{fig:sec5_eval} (e), we can infer that dynamic merging delivers more accurate results in more aggressive attack scenarios when compared to a static counter extension. However, FCM shows resiliency (\ie better estimation) under light attack. In this manner, the static counter extension FCM outperforms SALSA by 20\% less error when the number of attack flows is less than 55k on average. Afterward, SALSA and ABC consistently outperform FCM by 63\% and 47\%, respectively. On the other hand, \ours{} outperforms all the schemes across all attack scenarios by reporting 53\% less error on average
Similarly, as shown in Fig.~\ref{fig:sec5_eval_mem} (e), we can observe that dynamic merging approaches consistently outperform static counter extensions. Notably, \ours{} unfailingly delivers 55\% more accurate estimation across all memory settings.

\BfPara{Takeaways} With its late merging design, \ours{} achieved superior performance compared to SOTA in five popular security applications, including flow size estimation, heavy hitter detection, change detection, and entropy estimation under varying levels of sketch pollution attacks and scenarios. As a result, \ours{} demonstrated a more robust approach to data stream estimation and took the first step toward security-aware sketching.

\subsection{In-depth Sketch Analysis}\label{sec:eval:in-depth}
In this section, we removed the handicap by excluding sketch attacks and focused on the sketch performance under normal traffic scenarios varying distributions.

\begin{figure*}[t]
    \centering
    \includegraphics[width=0.63\textwidth]{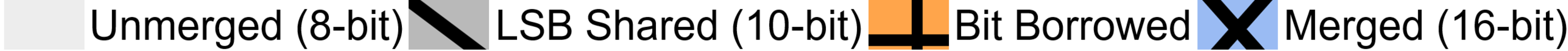}\vspace{-1mm}  \\ 
    \subfigure[CAIDA]
    {\includegraphics[width=0.19\textwidth]{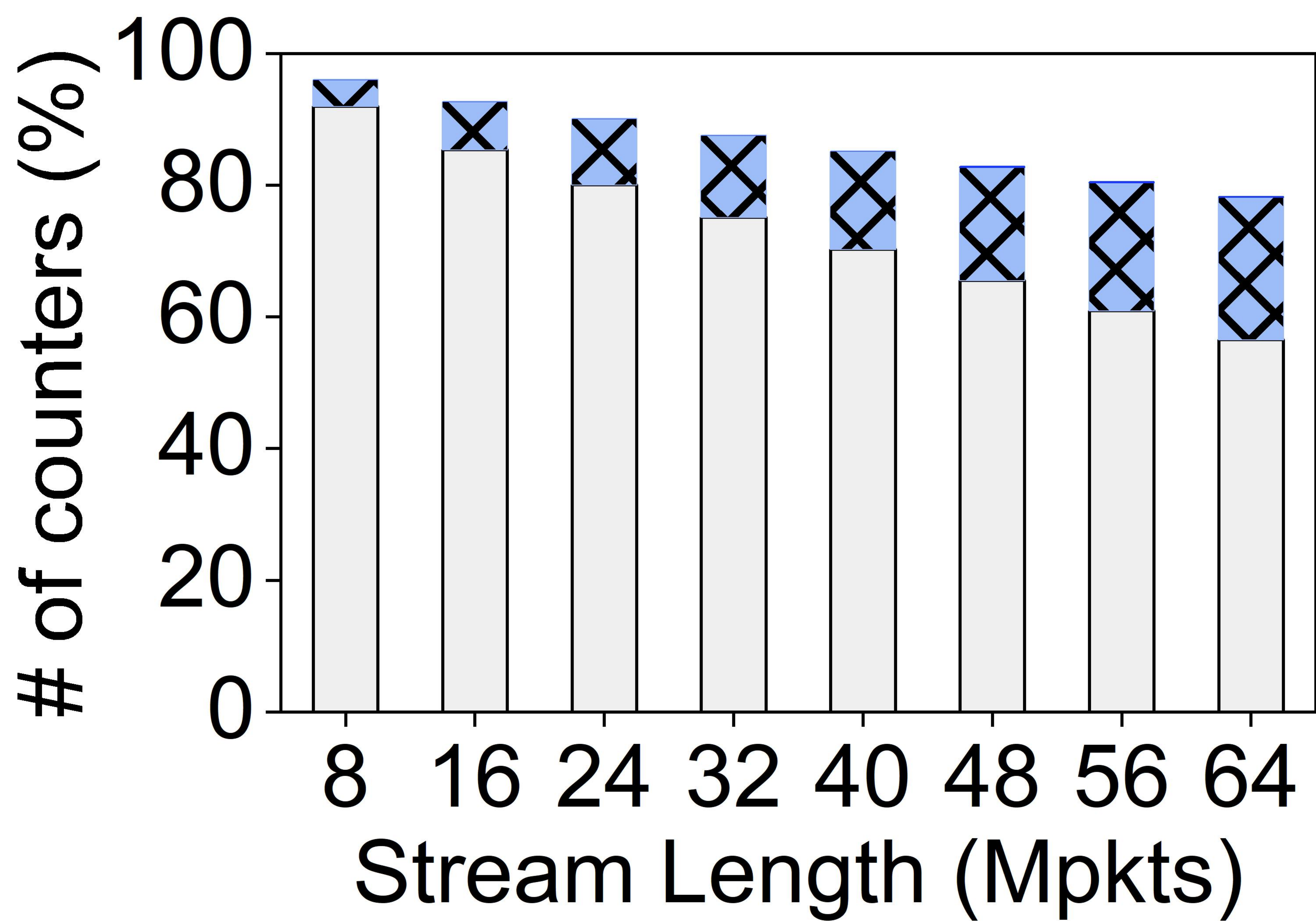}}
    \subfigure[MACCDC]
    {\includegraphics[width=0.19\textwidth]{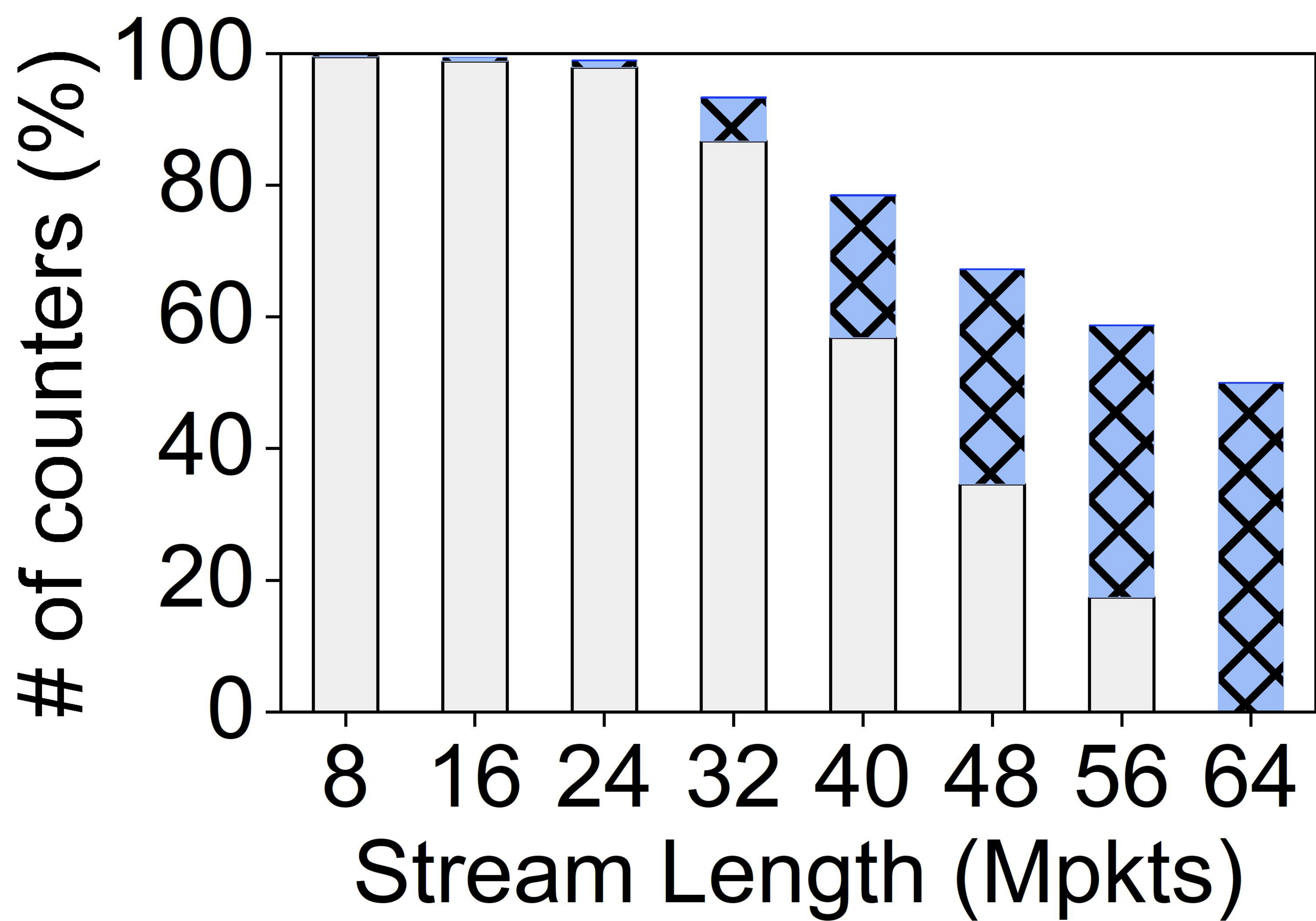}}
    \subfigure[CIC DDoS]
    {\includegraphics[width=0.19\textwidth]{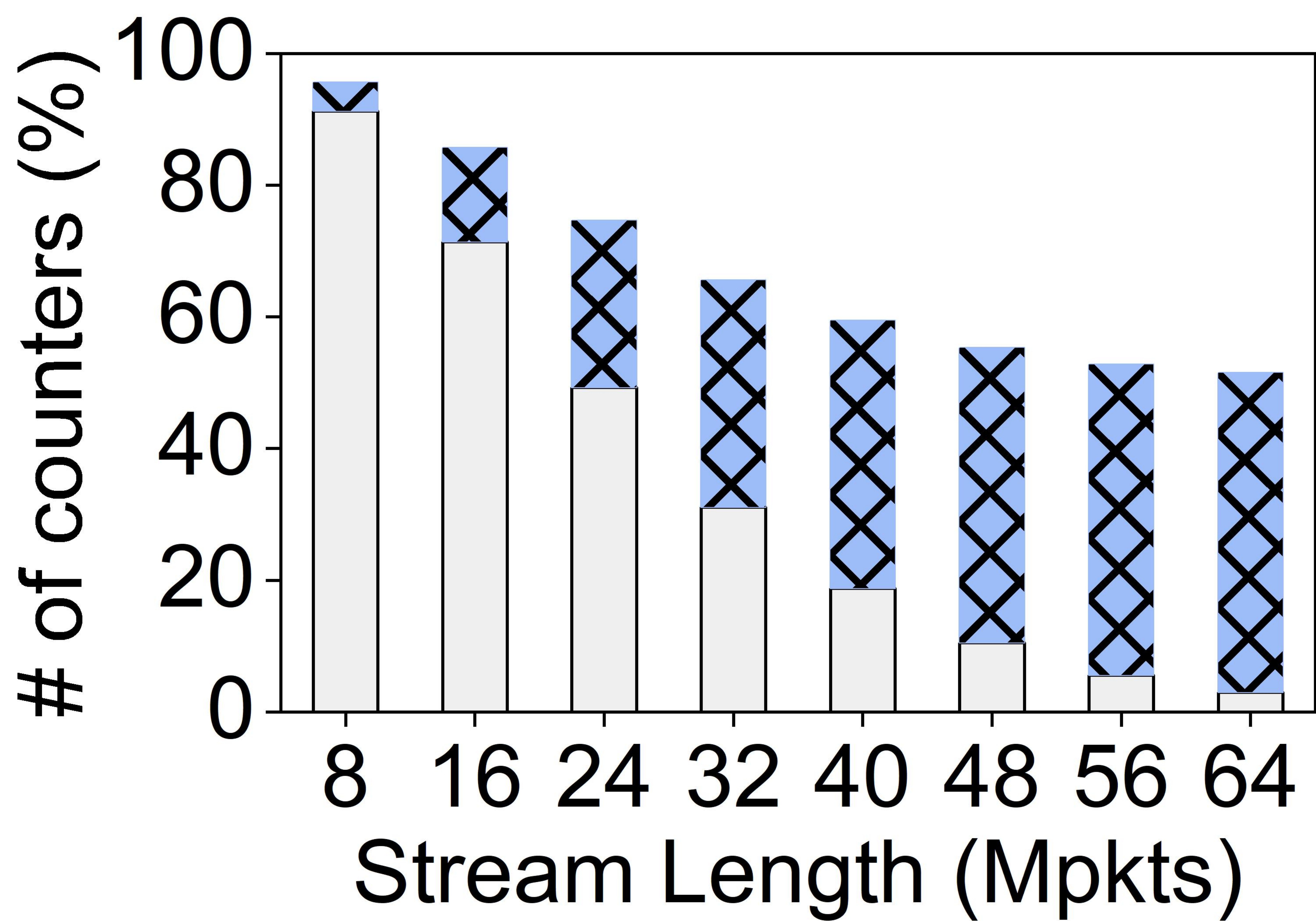}}
    \subfigure[CIC IDS]
    {\includegraphics[width=0.19\textwidth]{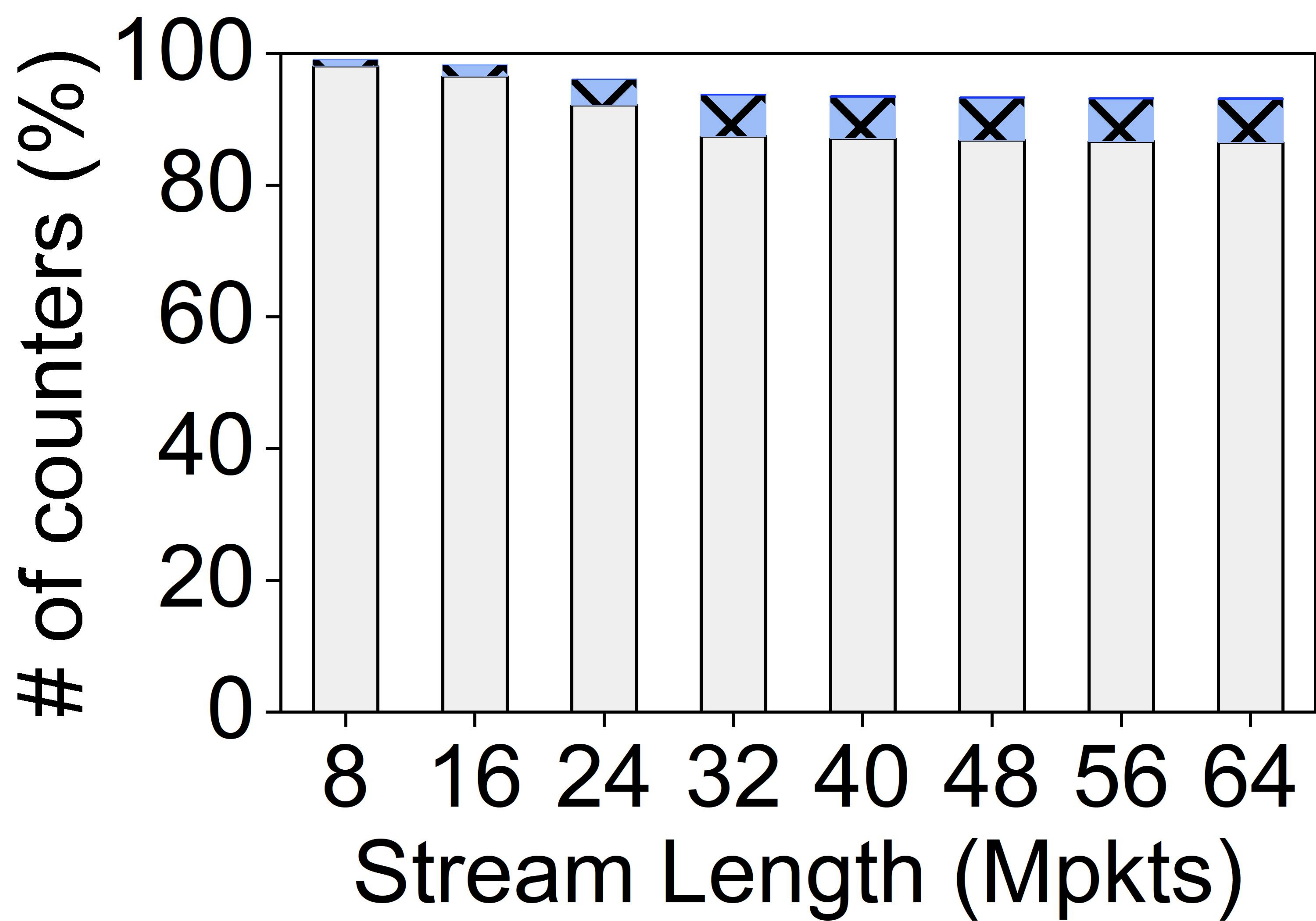}}
    \subfigure[UNSW DoS]
    {\includegraphics[width=0.19\textwidth]{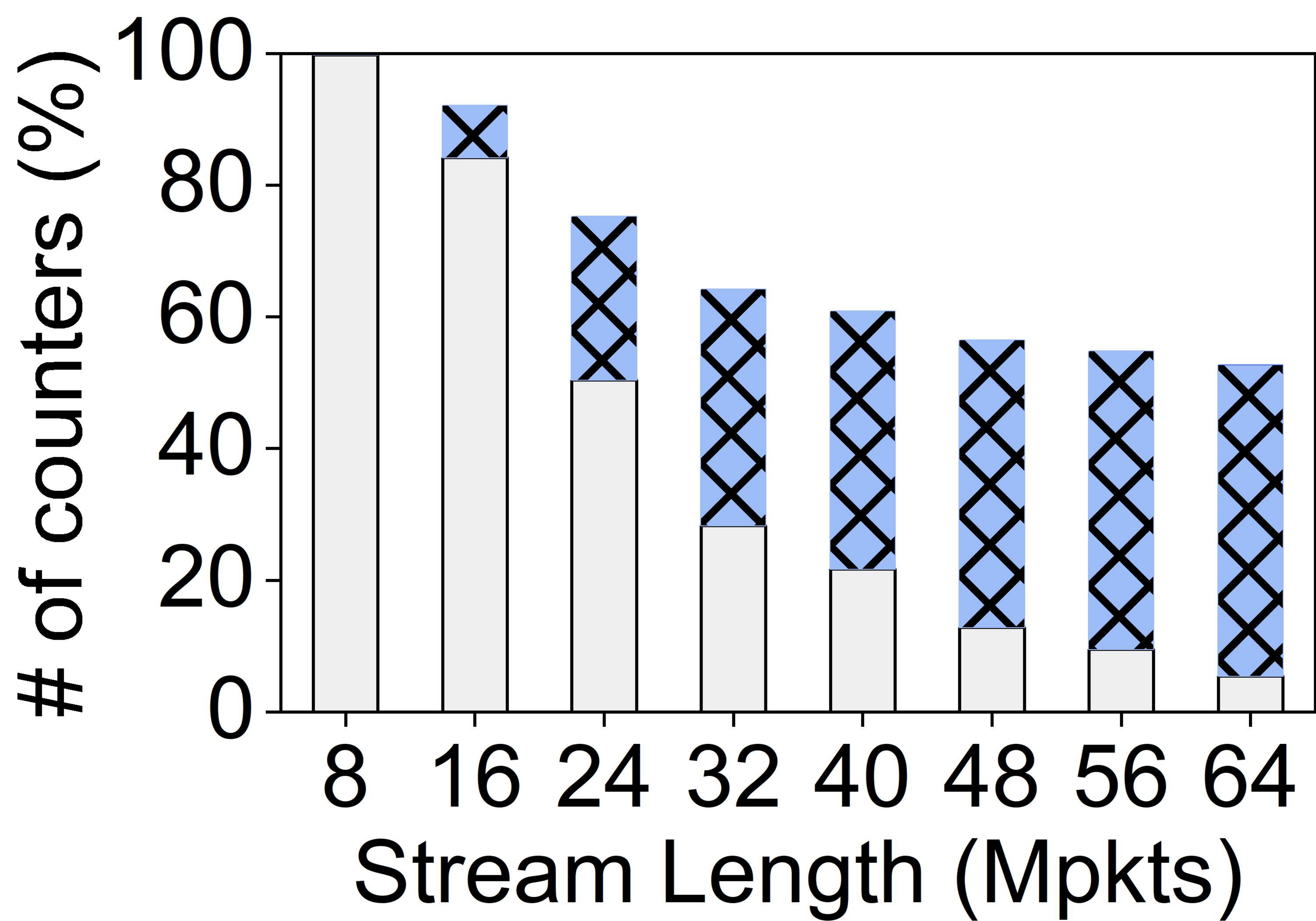}}\vspace{-2mm}
    \caption{Counter distribution: SALSA instantly merges the counters upon overflows reducing the number of counters rapidly.
    }~\label{fig:Counter_Dist0}\vspace{-4mm}
\end{figure*}

\begin{figure*}[t]
    \centering
    \subfigure[CAIDA]
    {\includegraphics[width=0.19\textwidth]{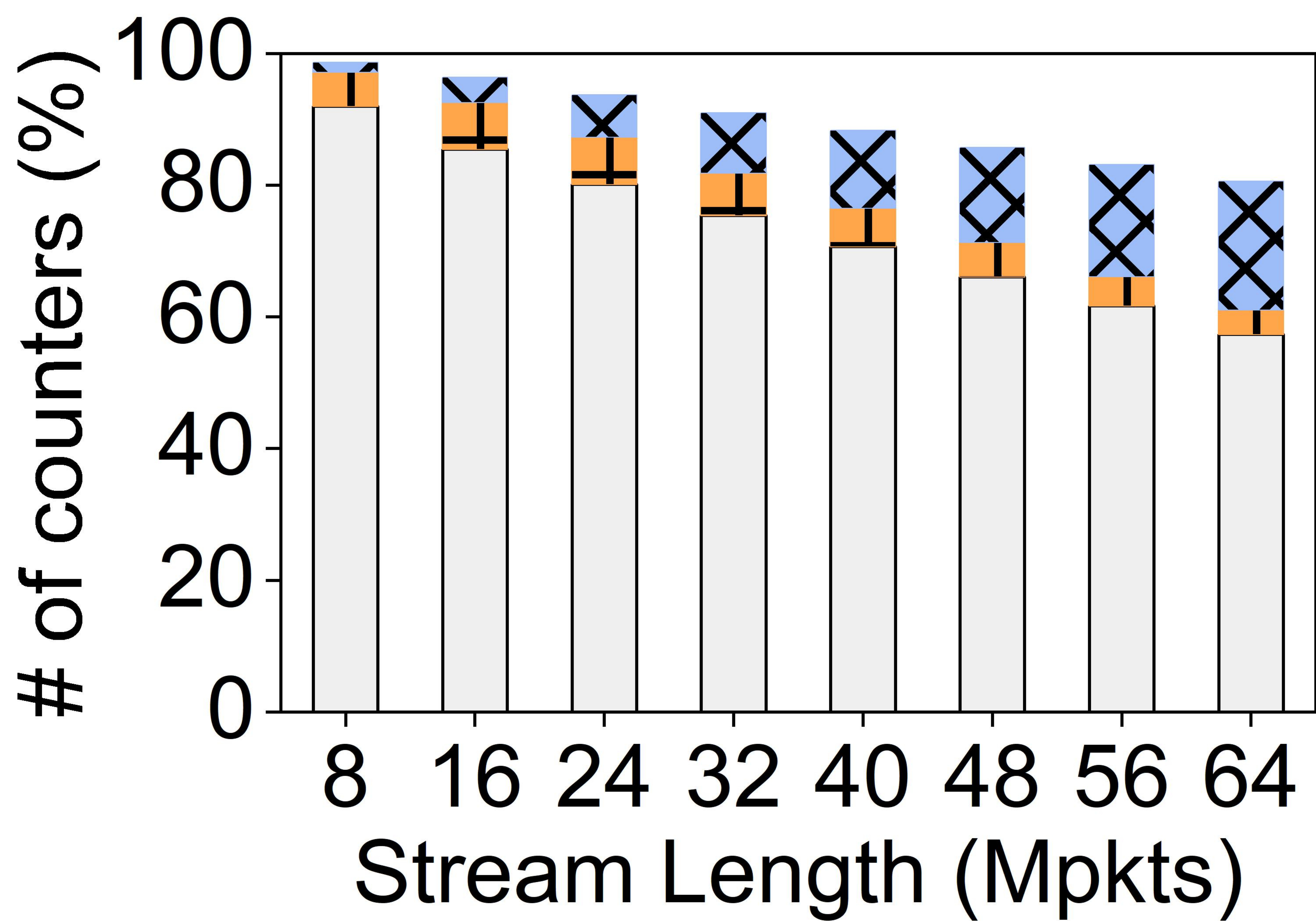}}
    \subfigure[MACCDC]
    {\includegraphics[width=0.19\textwidth]{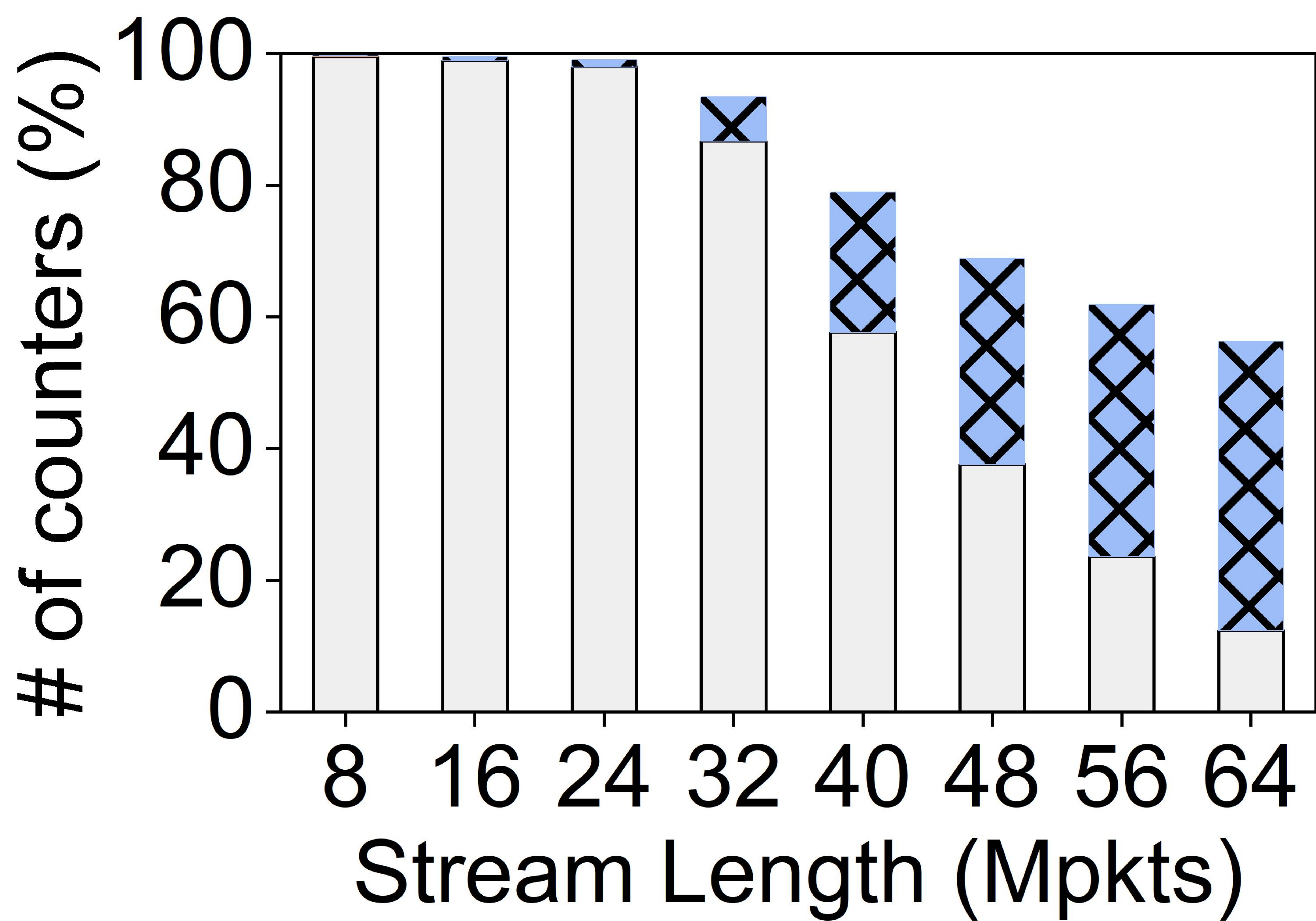}}
    \subfigure[CIC DDoS]
    {\includegraphics[width=0.19\textwidth]{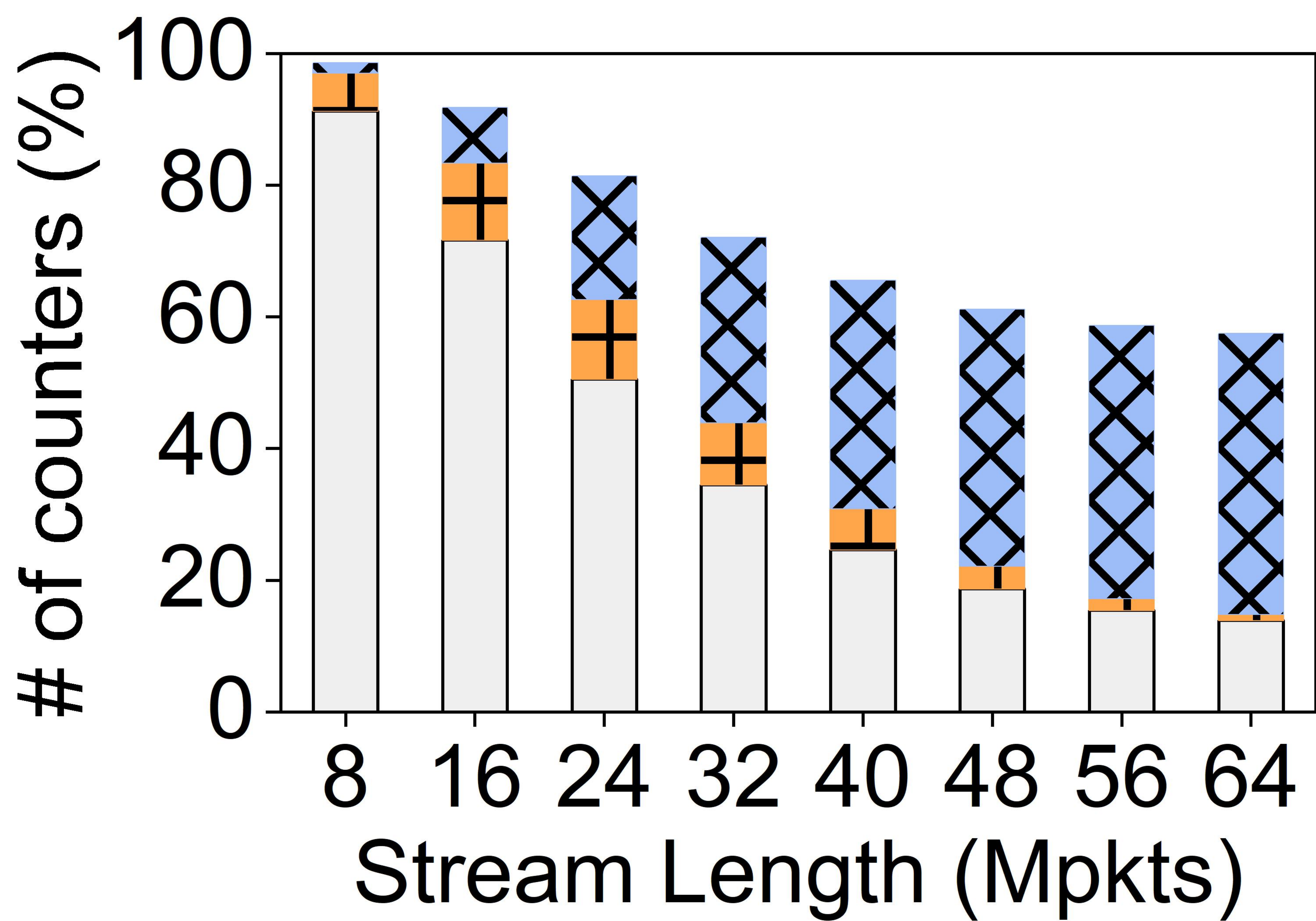}}
    \subfigure[CIC IDS]
    {\includegraphics[width=0.19\textwidth]{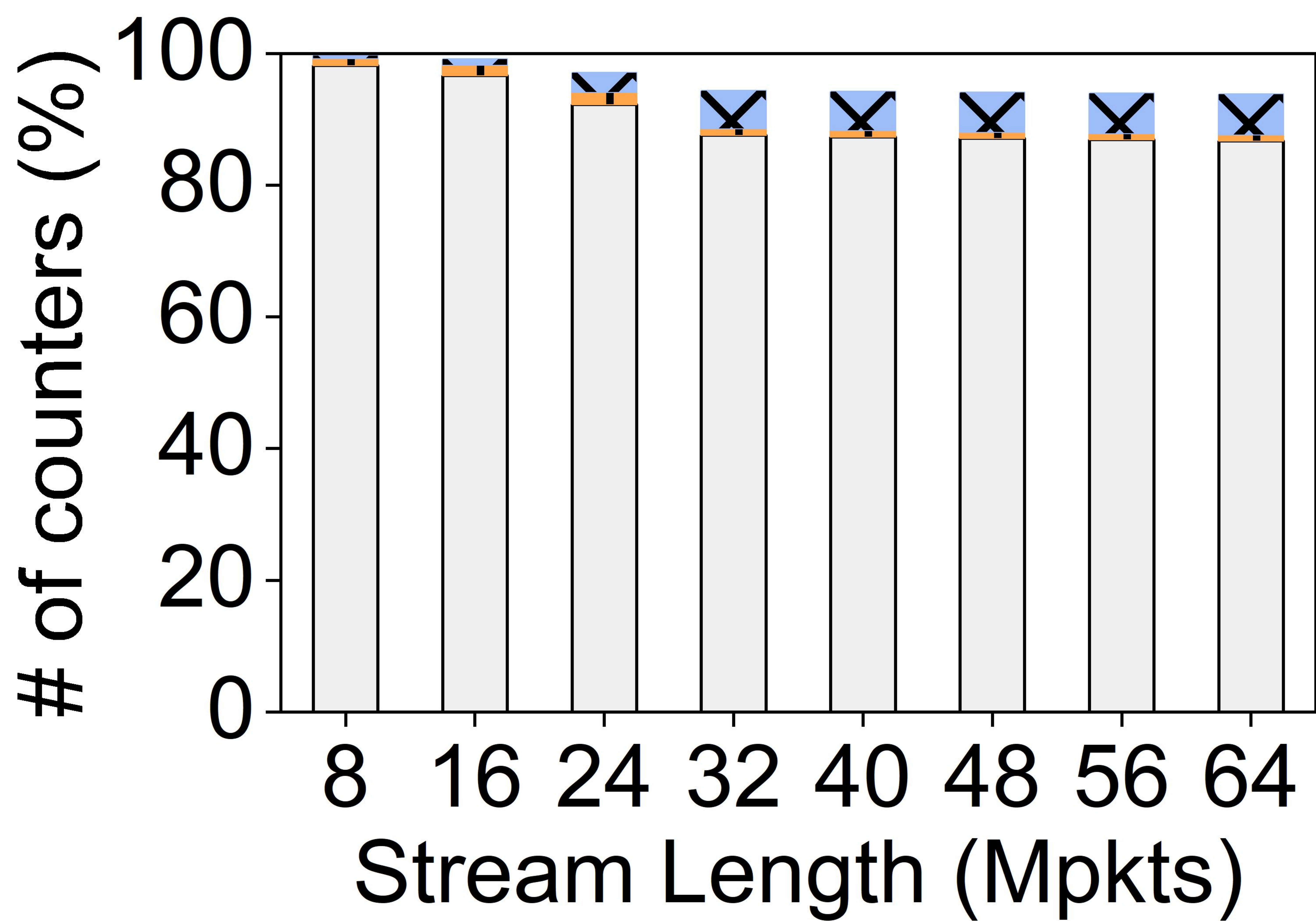}}
    \subfigure[UNSW DoS]
    {\includegraphics[width=0.19\textwidth]{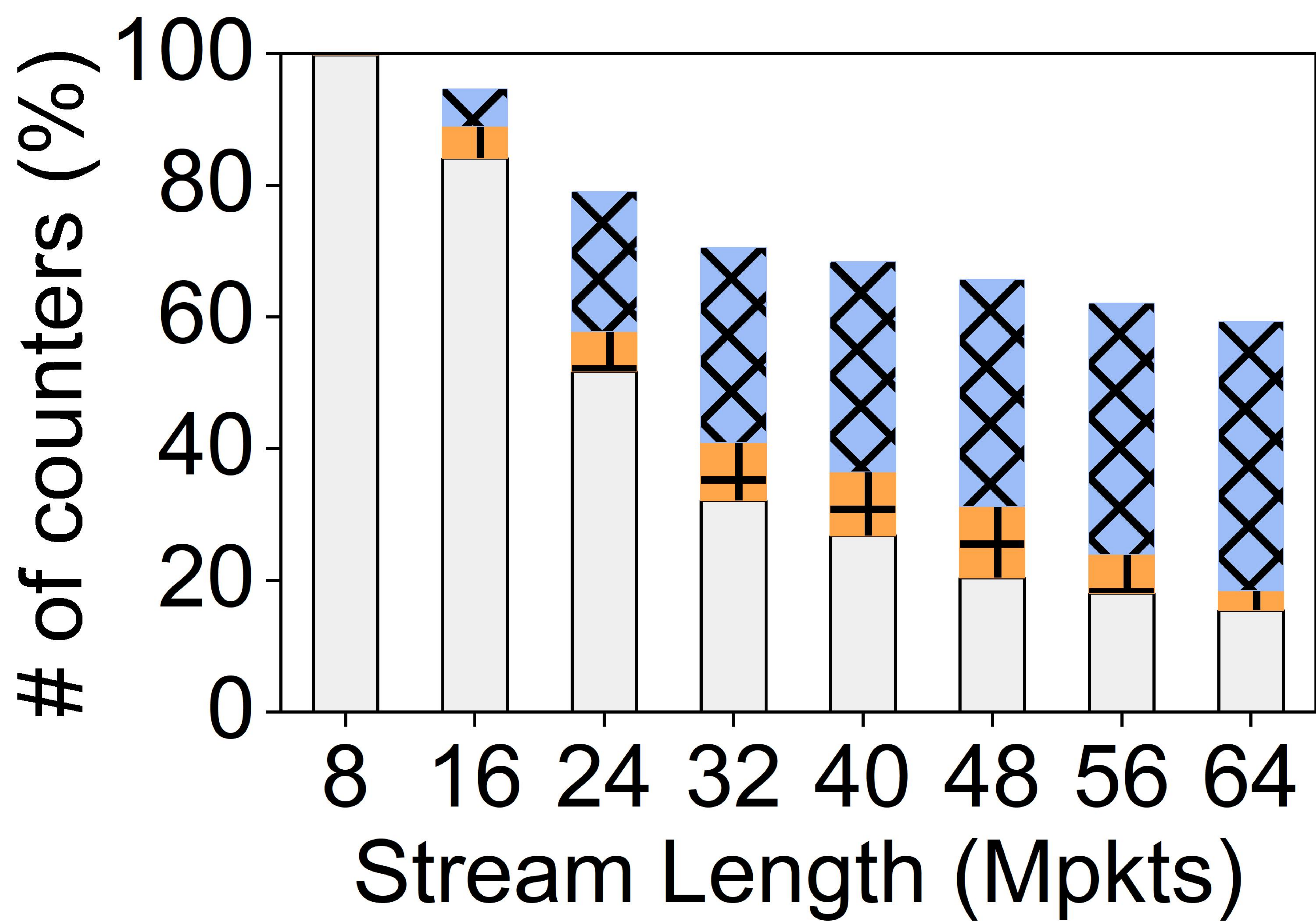}}\vspace{-2mm}
    \caption{Counter distribution: ABC's bit borrowing is ineffective and thus falls in the same category of instant merging.
    }~\label{fig:Counter_Dist1}\vspace{-4mm}
\end{figure*}

\begin{figure*}[t]
    \centering
    \subfigure[CAIDA]
    {\includegraphics[width=0.19\textwidth]{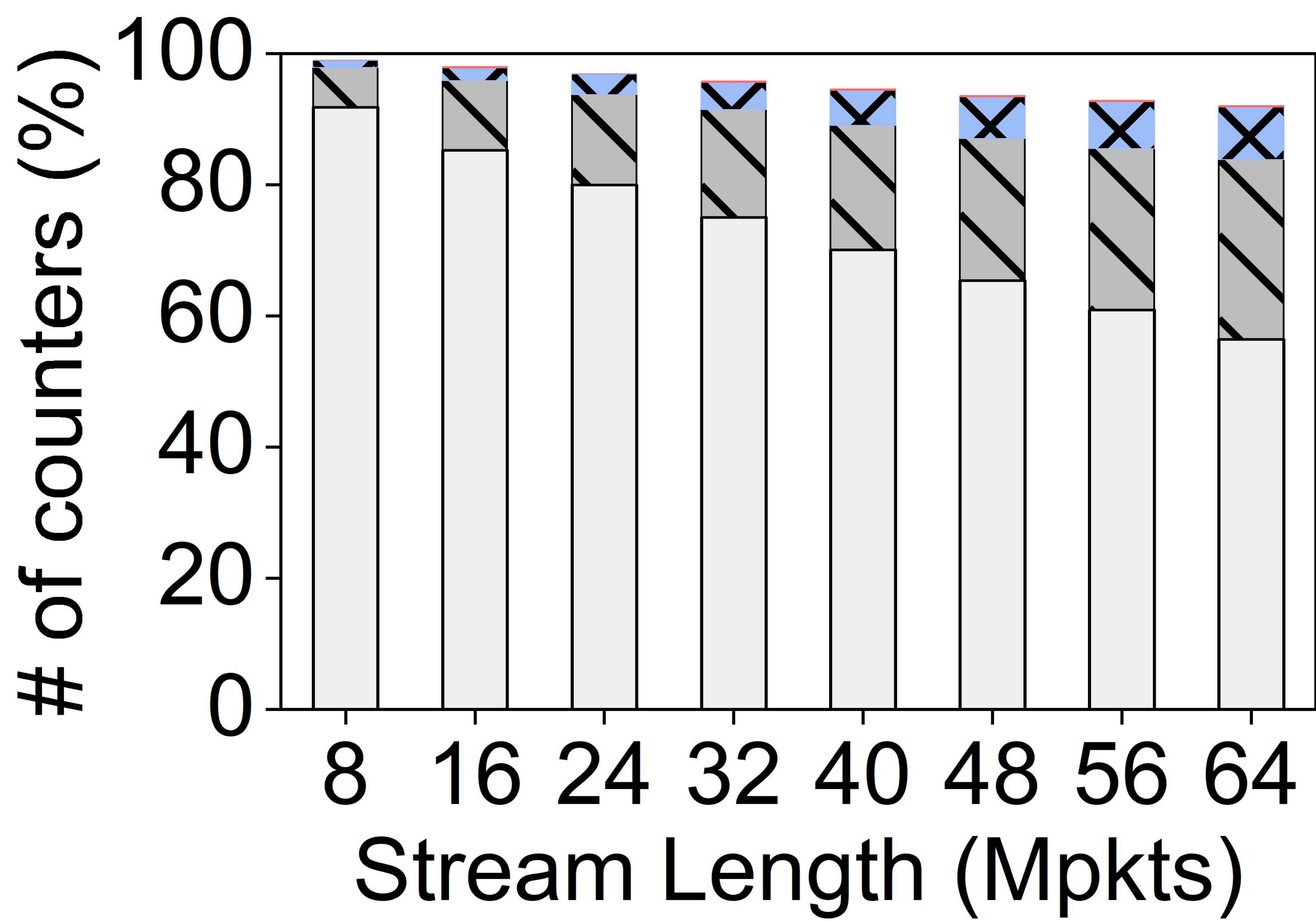}}
    \subfigure[MACCDC]
    {\includegraphics[width=0.19\textwidth]{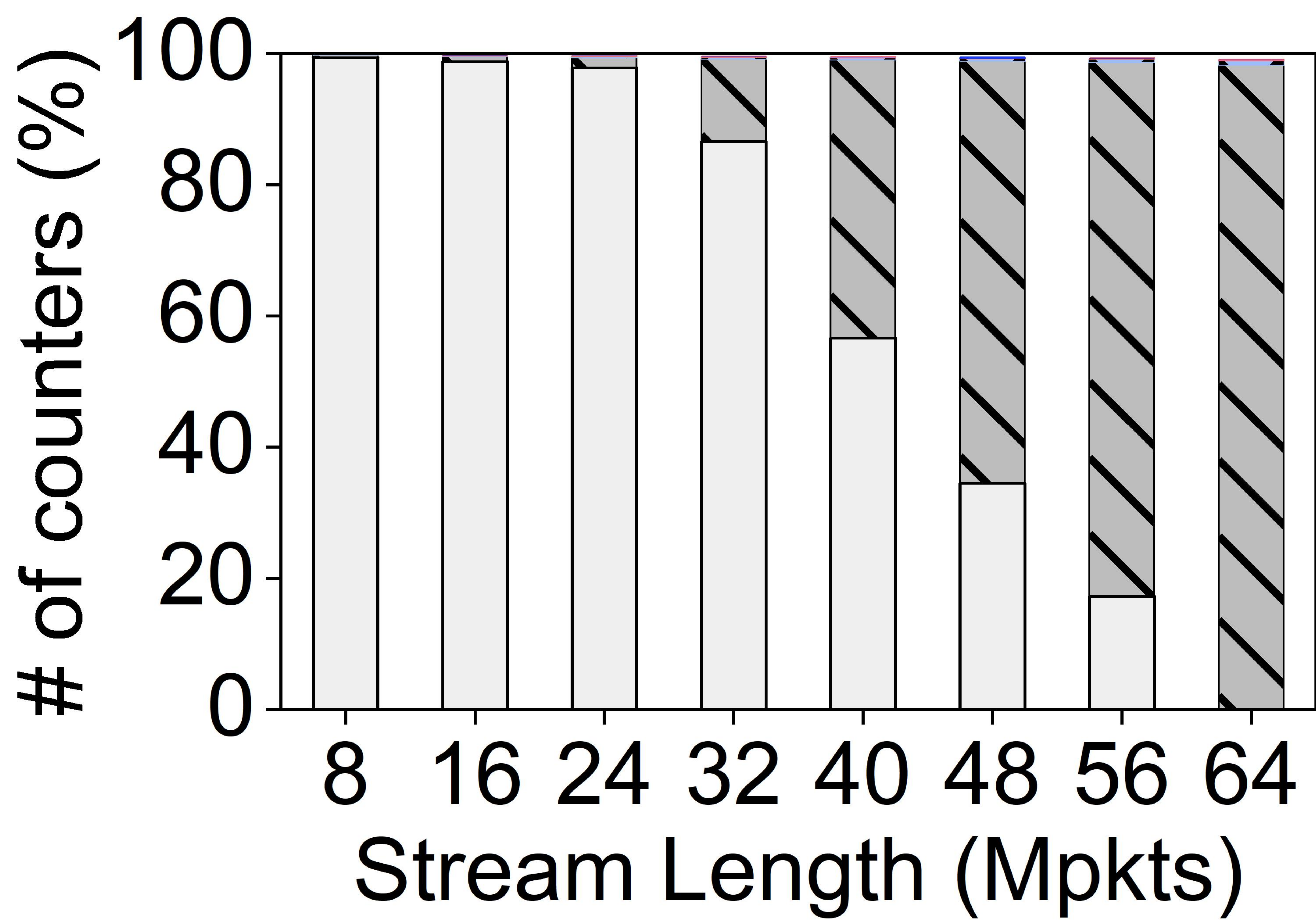}}
    \subfigure[CIC DDoS]
    {\includegraphics[width=0.19\textwidth]{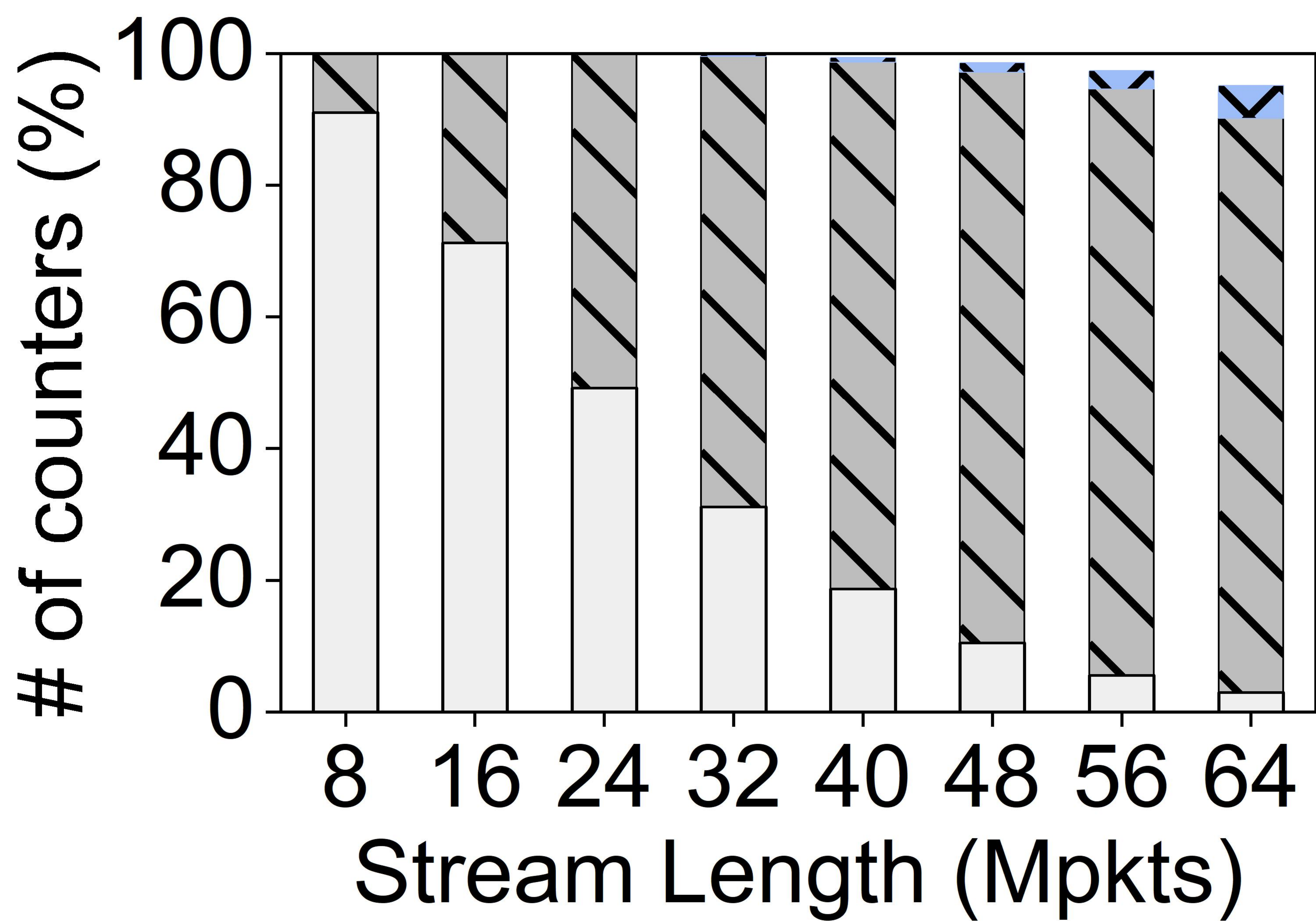}}
    \subfigure[CIC IDS]
    {\includegraphics[width=0.19\textwidth]{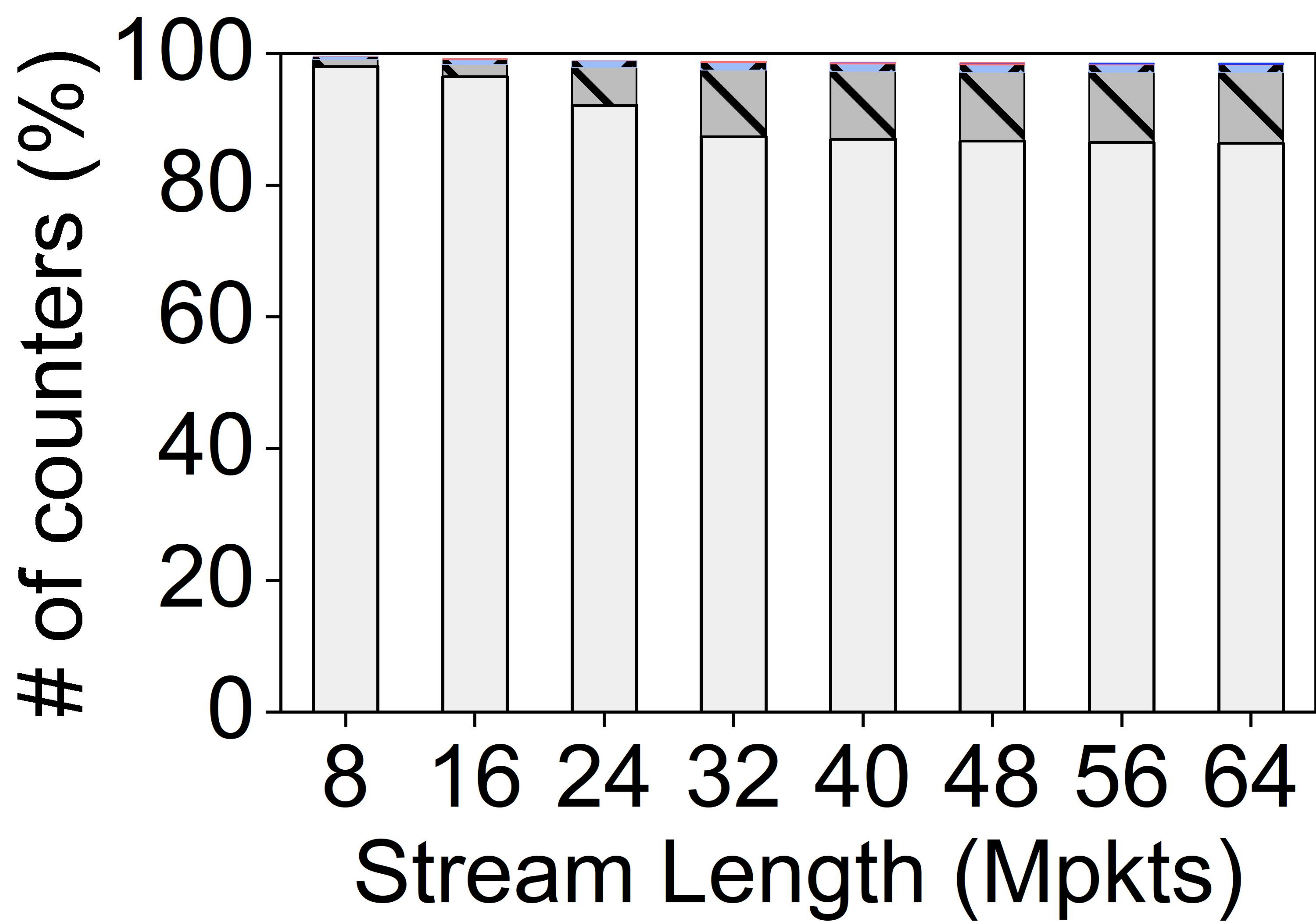}}
    \subfigure[UNSW DoS]
    {\includegraphics[width=0.19\textwidth]{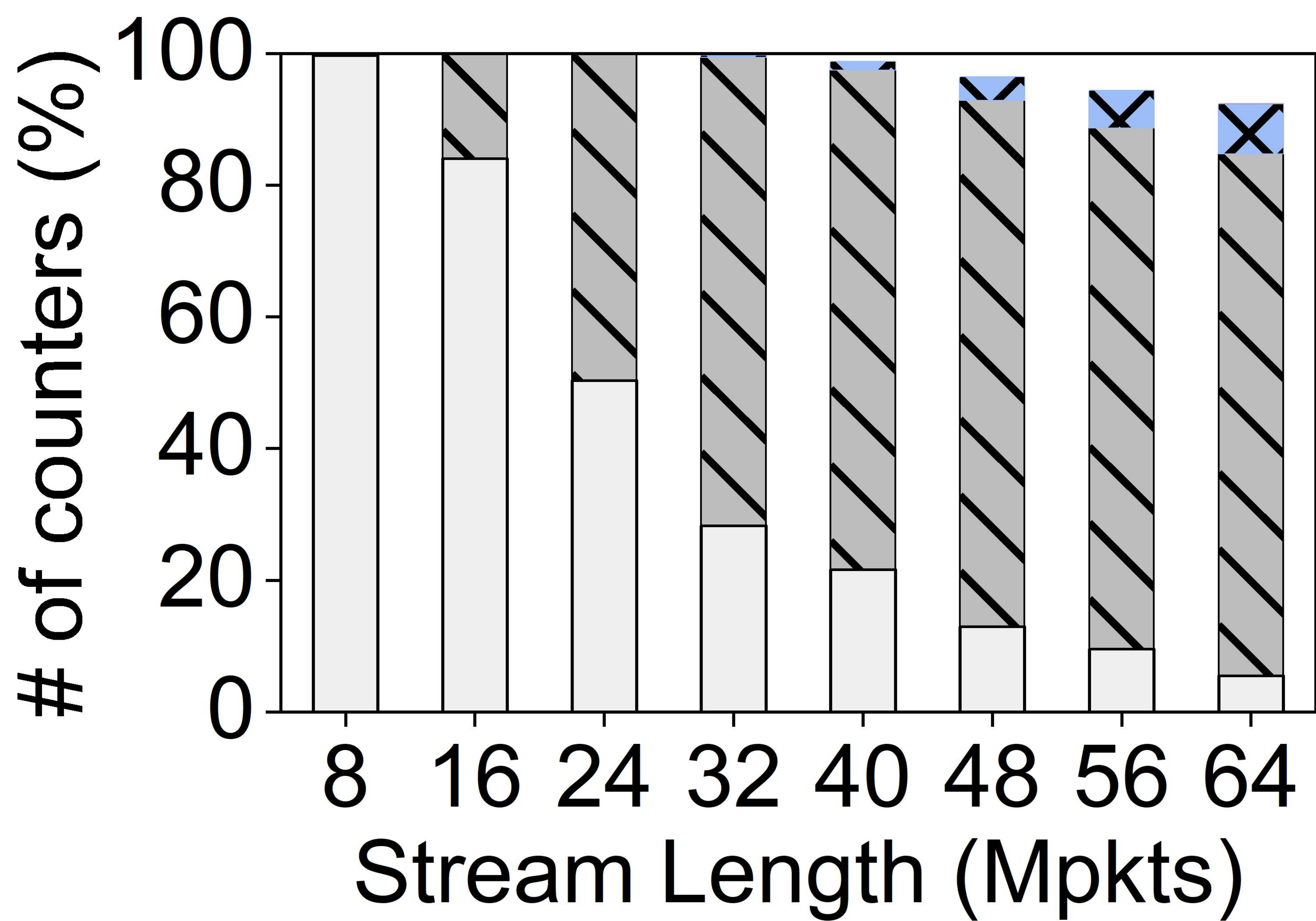}}\vspace{-2mm}
    \caption{Counter distribution: \ours{}'s delay merging event maximizes the number of counters over a long period.
    }~\label{fig:Counter_Dist2}
\end{figure*}

\BfPara{Number of Counters Over Time (Late Merging Effects)} 
We first demonstrate the impact of late merging on sketch performance over time when capturing sketch snapshots at different intervals. Fig.~\ref{fig:Counter_Dist0} to Fig.~\ref{fig:Counter_Dist2} demonstrate the number of counters over time for five real-world datasets when applying instant and late merging approaches. 
Interestingly, we observe that merging events triggered rapidly and extensively across the entire sketch over time, causing ABC’s bit-borrowing approach to be ineffective for retaining small counters due to the unavailability of idle bits from neighboring counters. As shown, among all the dataset distributions, the number of counters was preserved for a longer time with the late merging approach, resulting in a low hash collision rate. On the other hand, with the instant merging of SALSA~\cite{basat2021salsa}, the number of counters was about 1.5\%$\sim$ 31.7\% less than \ours{} on average as stream length increased from 8 M to 64 M packets, respectively, leading to a high rate of collision rate and higher estimation error. Similarly, the number of counters for ABC~\cite{gong2017abc}  was about 0.3\%$\sim$27.2\% less than \ours{} on average. Through these results, we verify how the later merging mechanism delivers long-term sustainability and independent operation of small counters, which is reflected in the improved estimation accuracy.

\begin{figure}[t]
    \centering
    \includegraphics[width=0.3\textwidth]{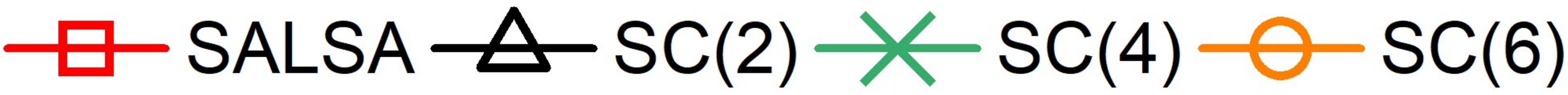}\\
    \subfigure[Z=0.6]
    {\includegraphics[width=0.15\textwidth]{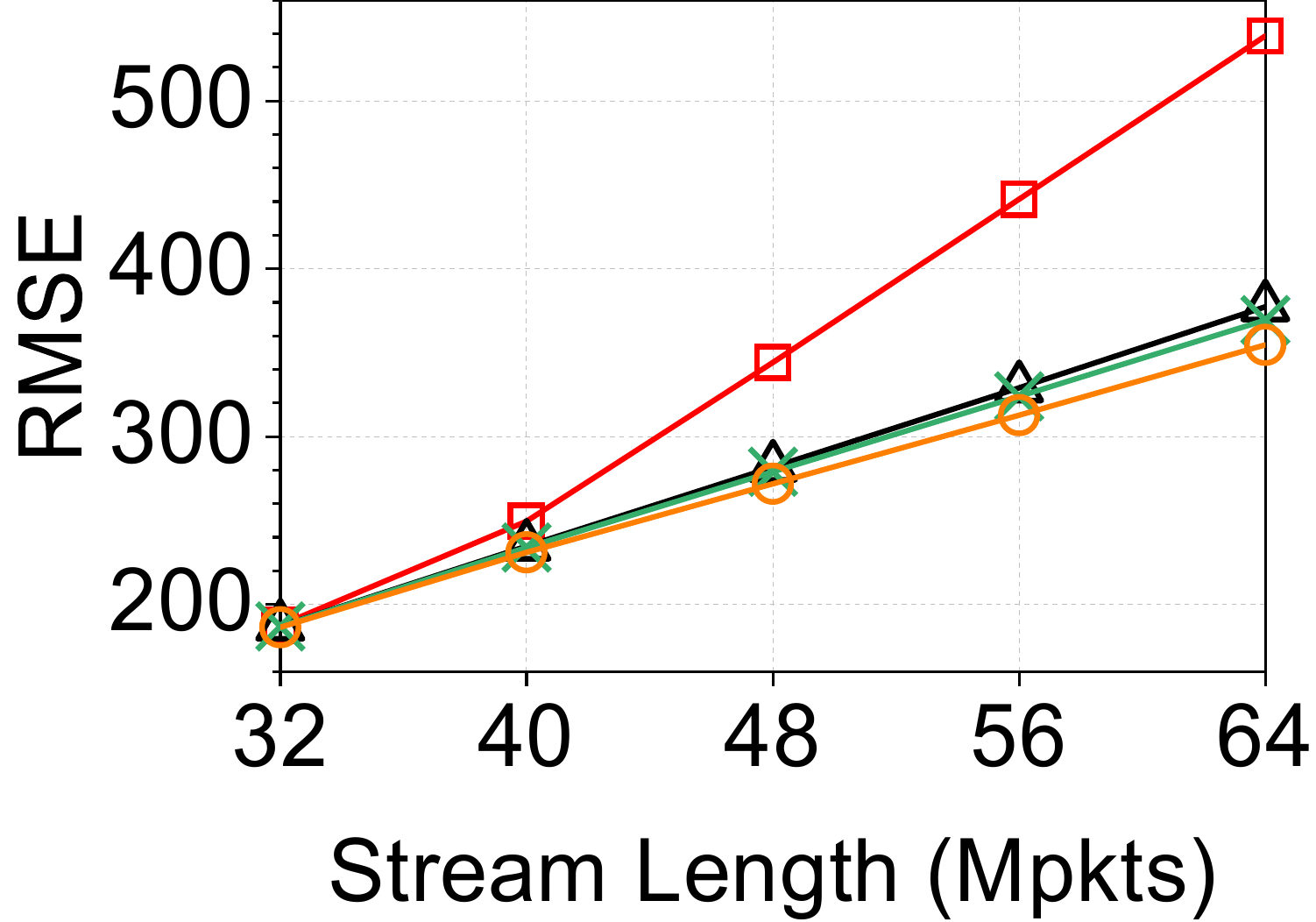}}
    \subfigure[Z=1.0]
    {\includegraphics[width=0.15\textwidth]{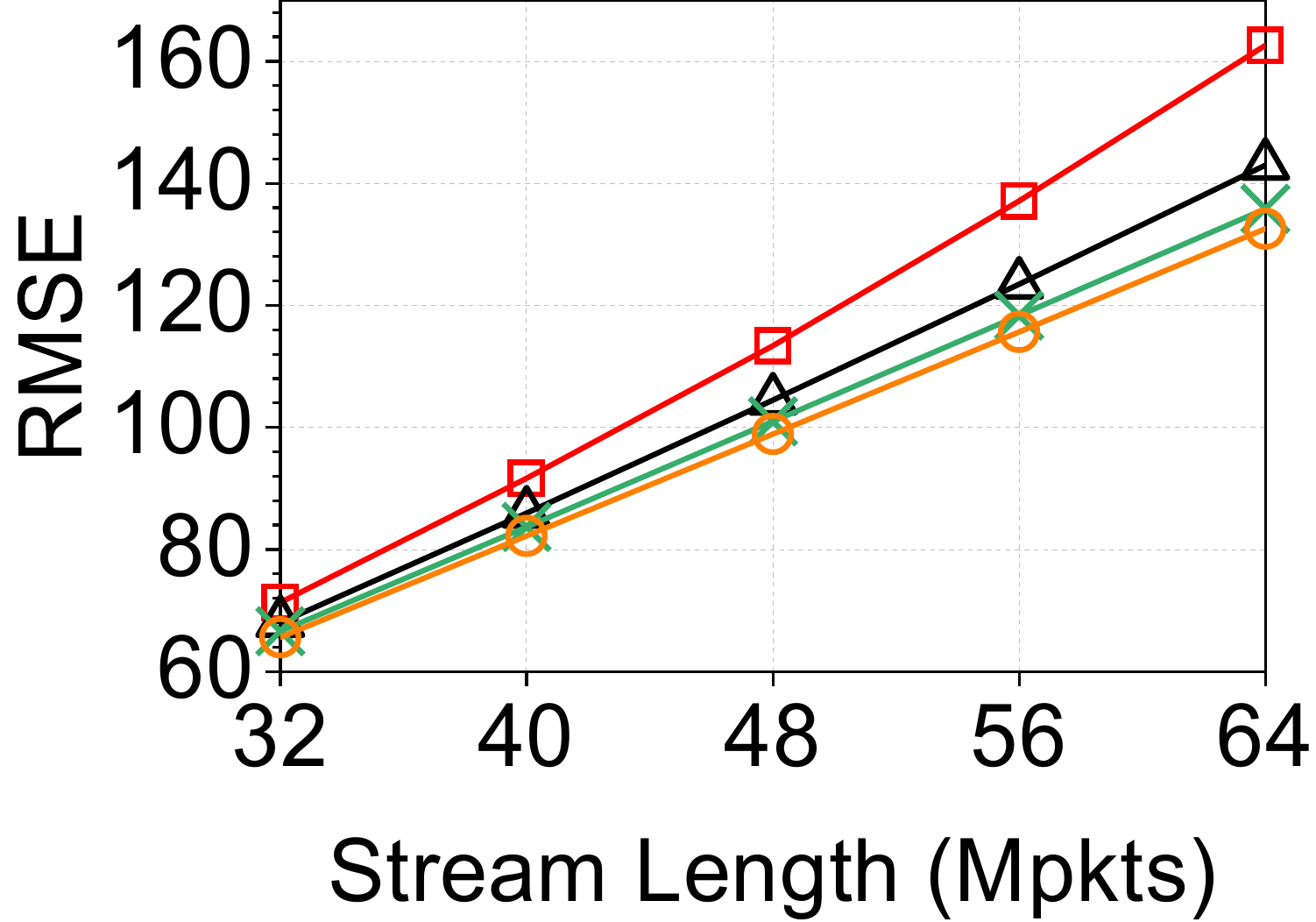}}
    \subfigure[Z=1.4]
    {\includegraphics[width=0.15\textwidth]{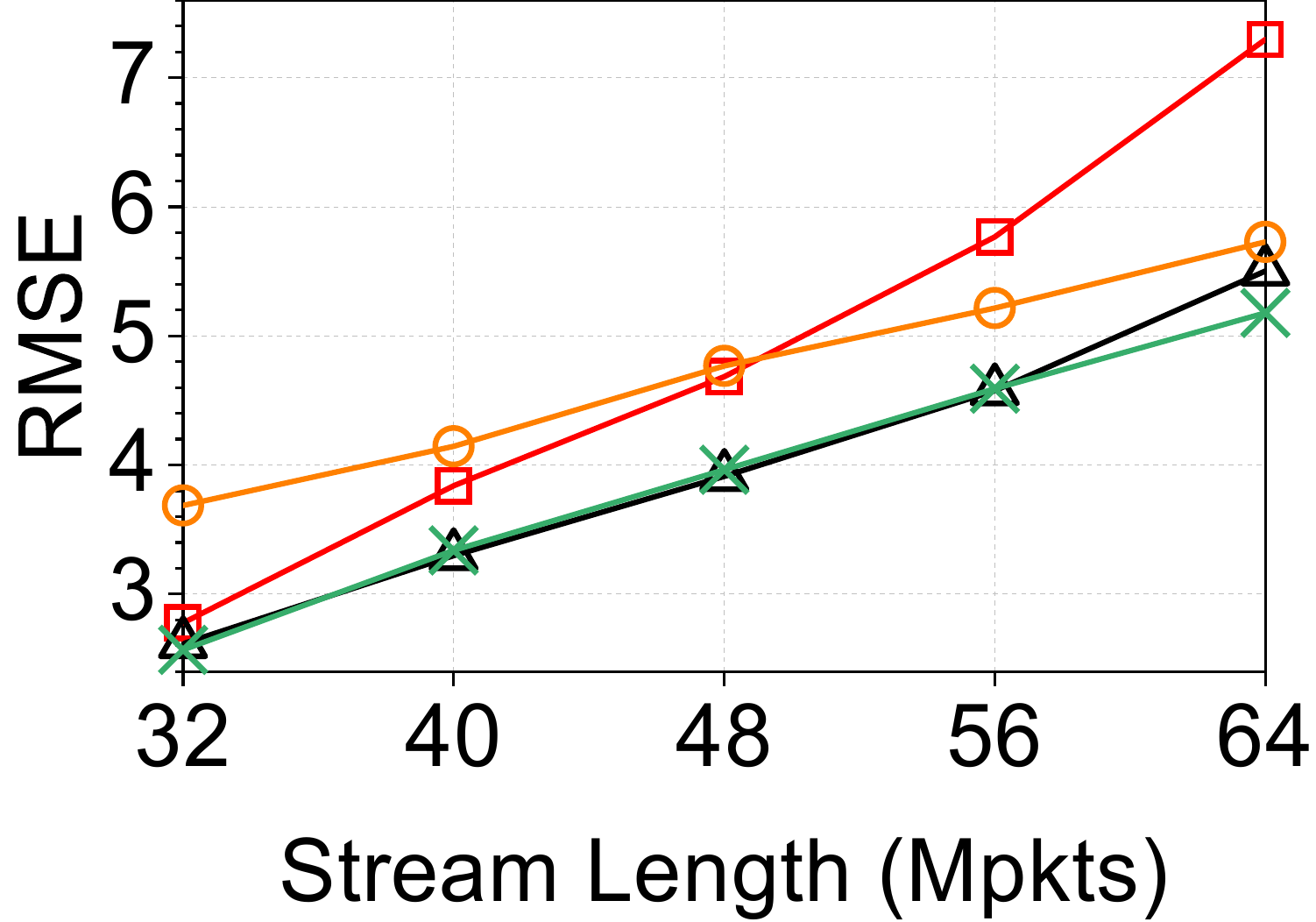}}     
    \caption{Flow size estimation with five synthetic datasets over different $K$: larger the $K$ values lead to slower counter merge and slower increase in error. }~\label{fig:Synthetic_CMP}
\end{figure}

\BfPara{Observations on variable $K$ using synthetic datasets}
We analyze \ours{} with $K$ varied from 2 to 6 to explore the optimal size of the shared LSB $K$. We used the state-of-the-art dynamic sketch, SALSA, as a baseline to demonstrate the performance trend of \ours{}. As shown, We define SC{($K$)} as the size of the shared LSB in \ours{}, where $K$ represents the amount of bits shared between two counters.
Fig.\ref{fig:Synthetic_CMP} demonstrates this through flow size estimation over time by varying skewness $Z$ from 0.6 to 1.4.
The overall error is reduced as skewness increases, since most packets belong to a smaller number of elephant flows, triggering fewer flow collisions and reducing error. For \ours{}, larger shared LSB $K$ lead to slower growth in error due to the delayed merging of counters, which can relax flow collisions significantly.
Therefore, when $Z=1.4$, although the initial error is relatively large with a large $K=6$, the error caused by the collision is significantly reduced as the flow size grows. We note that the sketch-based decoding is often required after the measurement window, thus we can expect \ours{} outperforms SALSA under pressure, as shown in Fig.\ref{fig:Synthetic_CMP} (c). On the other hand, \ours{} achieved better performance than SALSA consistently with $Z=0.6$ and $Z=1.0$, which proves its more robust estimation under various traffic patterns. Lastly, we found $K=4$ provides the most balanced performance varying skewness, thus, we used it as default in our work.

\begin{figure*}[t]
    \centering
    \includegraphics[width=0.45\textwidth]{figures/Legend.pdf} \\
    \subfigure[CAIDA]
    {\includegraphics[width=0.19\textwidth]{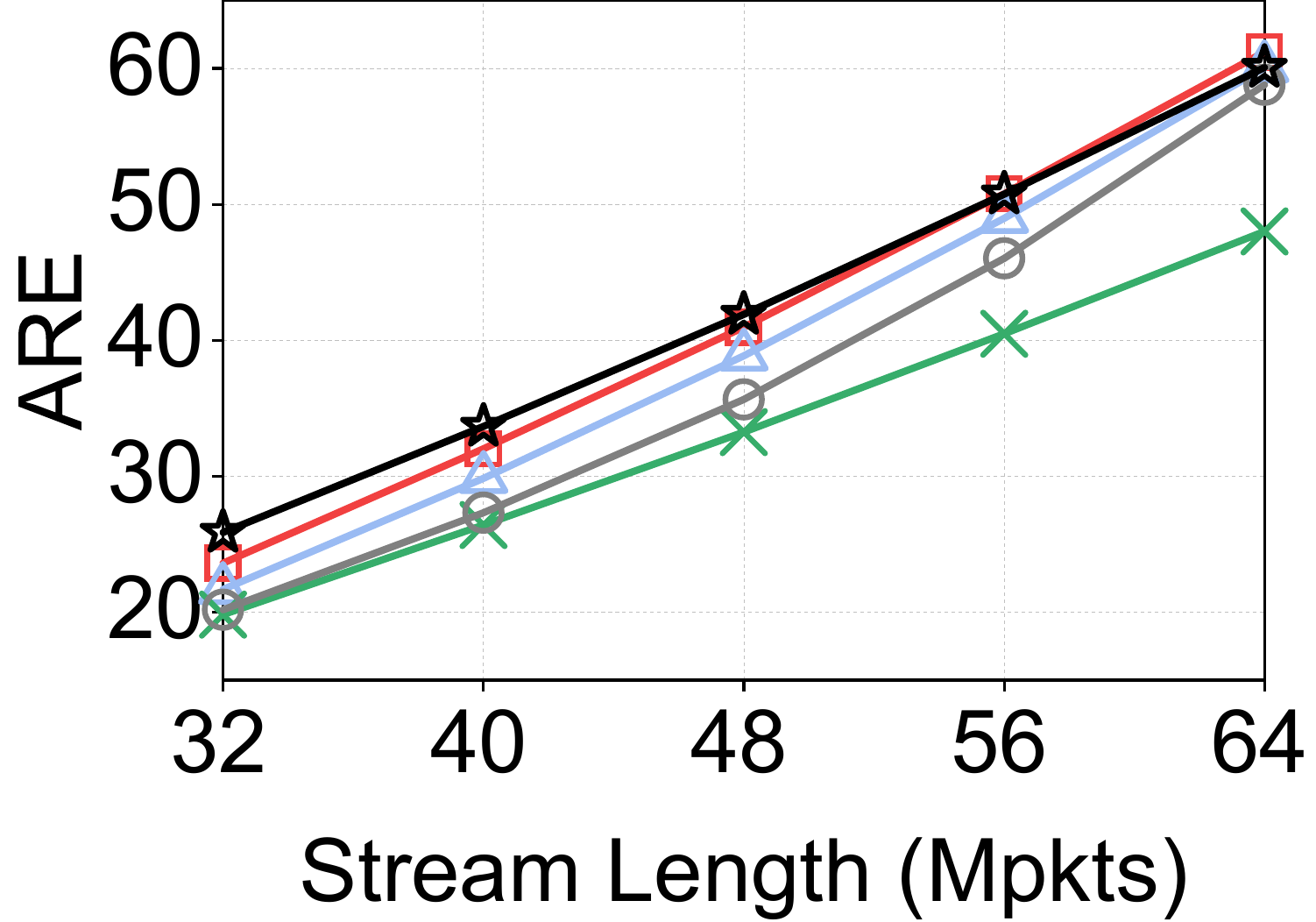}}
    \subfigure[MACCDC]
    {\includegraphics[width=0.19\textwidth]{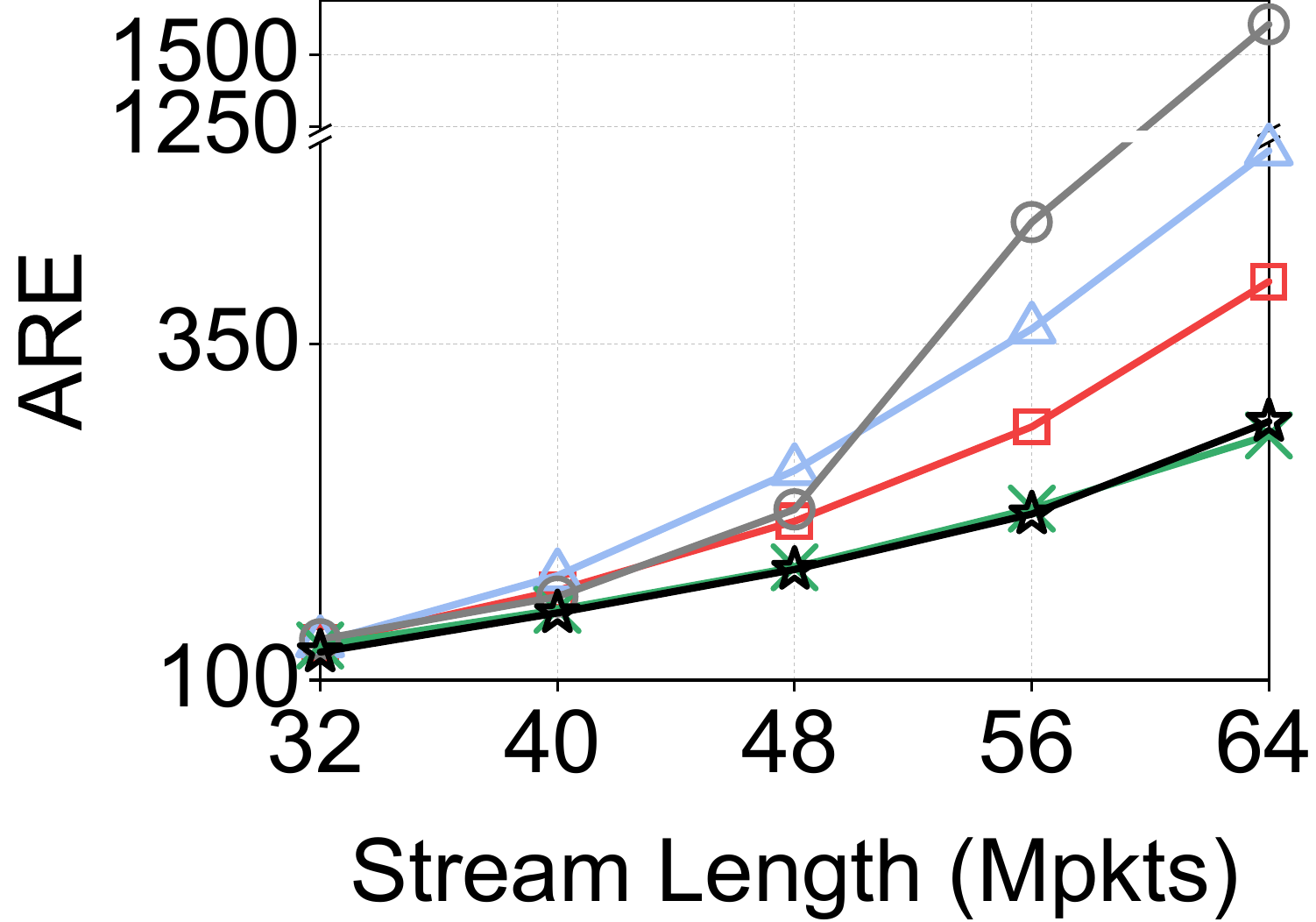}}
    \subfigure[CIC DDoS]
    {\includegraphics[width=0.19\textwidth]{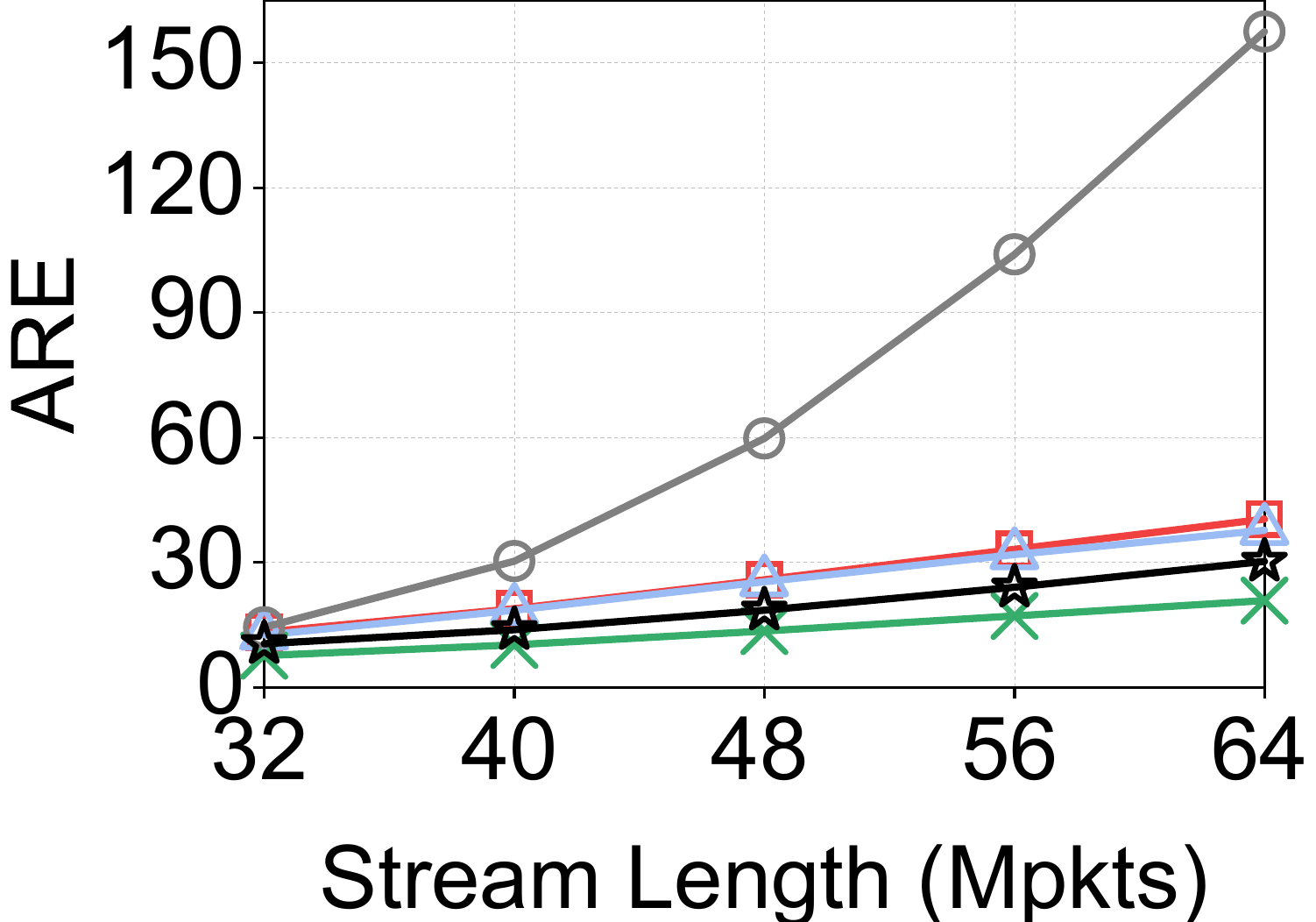}}
    \subfigure[CIC IDS]
    {\includegraphics[width=0.19\textwidth]{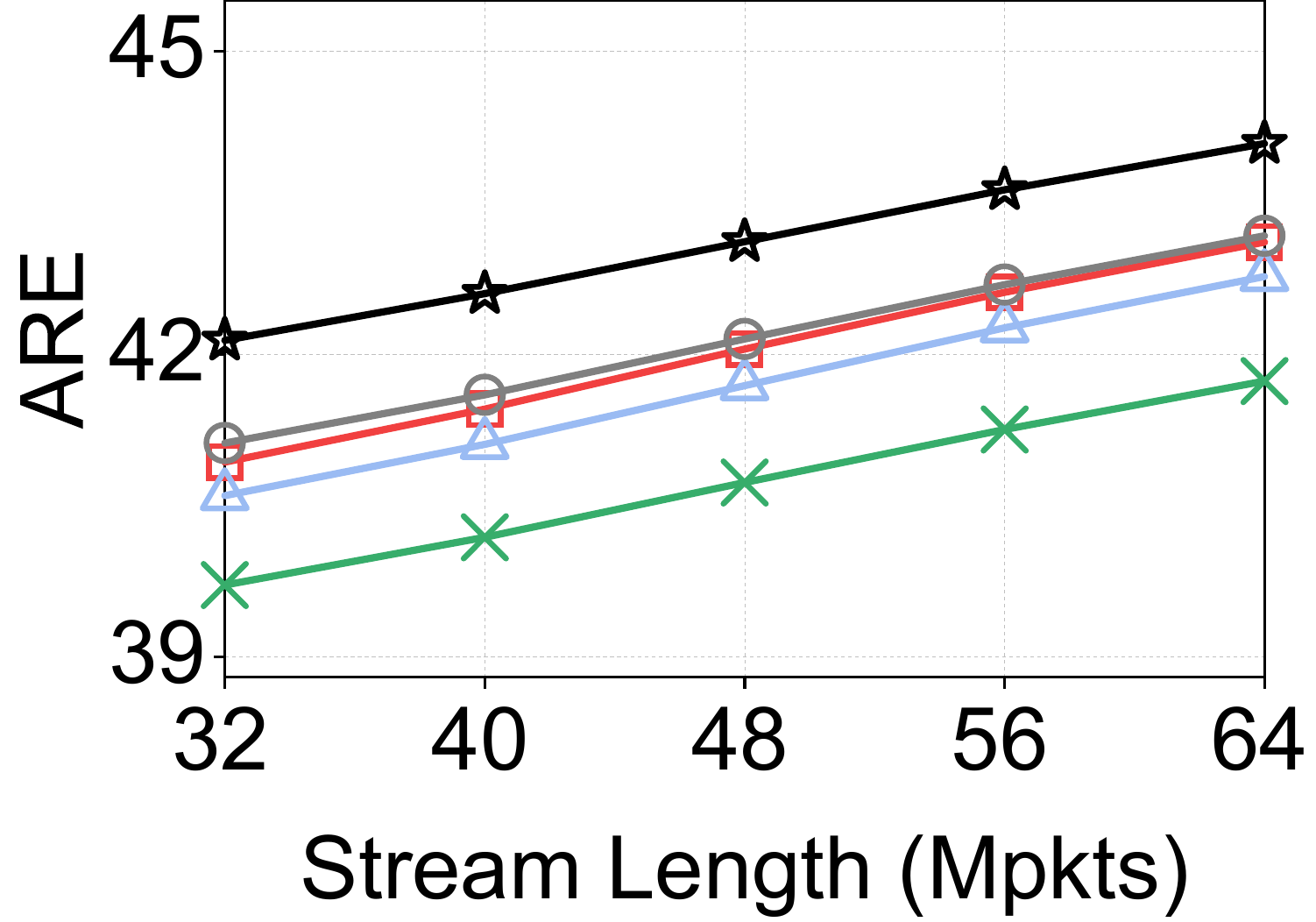}}
    \subfigure[UNSW DoS]
    {\includegraphics[width=0.19\textwidth]{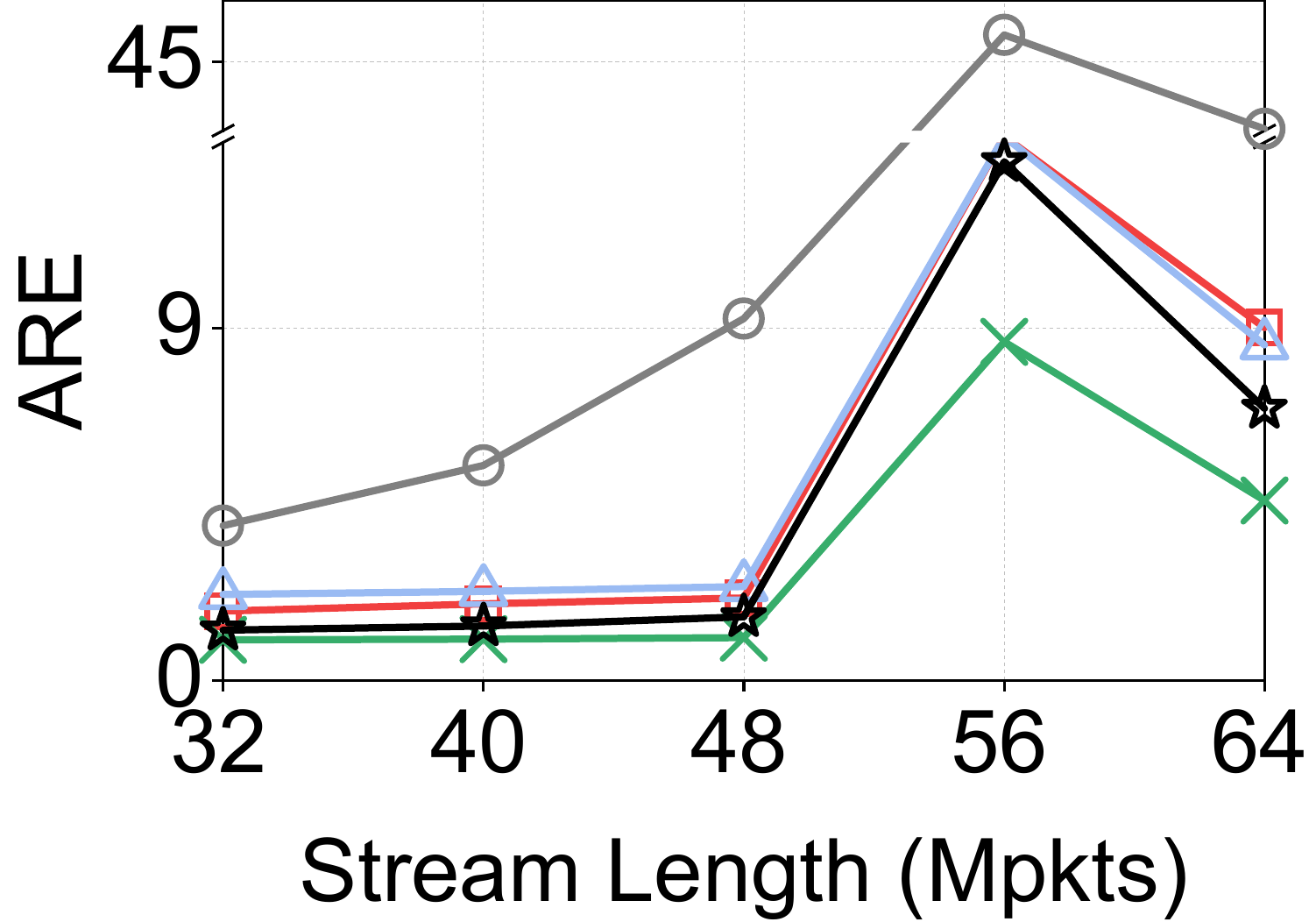}}
    \caption{Flow size estimation with five real-world datasets: for each experiment, the memory size is allocated to 0.4 MB.}~\label{fig:Real_FlowSize_0.4MB}\vspace{-4mm}
\end{figure*}

\BfPara{Varying Load with Fixed Memory Space} Next, we compare the accuracy of all the dynamic structure schemes for flow size estimation using five real-world datasets and five synthetic datasets. For fairness, we allocated 0.4 MB of memory to each scheme and varied the stream length from 8 M to 64 M packets from five different real-world datasets. As shown in Fig. \ref{fig:Real_FlowSize_0.4MB}, \ours{} showed superior performance in all the real-world datasets, thanks to the late merging of small counters. More specifically, \ours{} reduces error rates by about 2.0\% to 45.6\% compared to ABC~\cite{gong2017abc}, and by about 2.9\% to 46.1\% compared to SALSA~\cite{basat2021salsa}. When compared to FCM~\cite{song2020fcm}, a method using static counters, \ours{} shows a reduction in errors ranging from 3.3\% to 82\%. Although \ours{} generally outperforms static methods, FCM performs better by 11\% for the CAIDA dataset, which is direct evidence for our insight that counter isolation is beneficial for mouse flows. Additionally, CountLess, another method that combines static design with dynamic operations outperformed SALSA by 22\% and achieved similar performance with FCM.

\begin{figure*}[t]
    \centering
    \subfigure[CAIDA]
    {\includegraphics[width=0.19\textwidth]{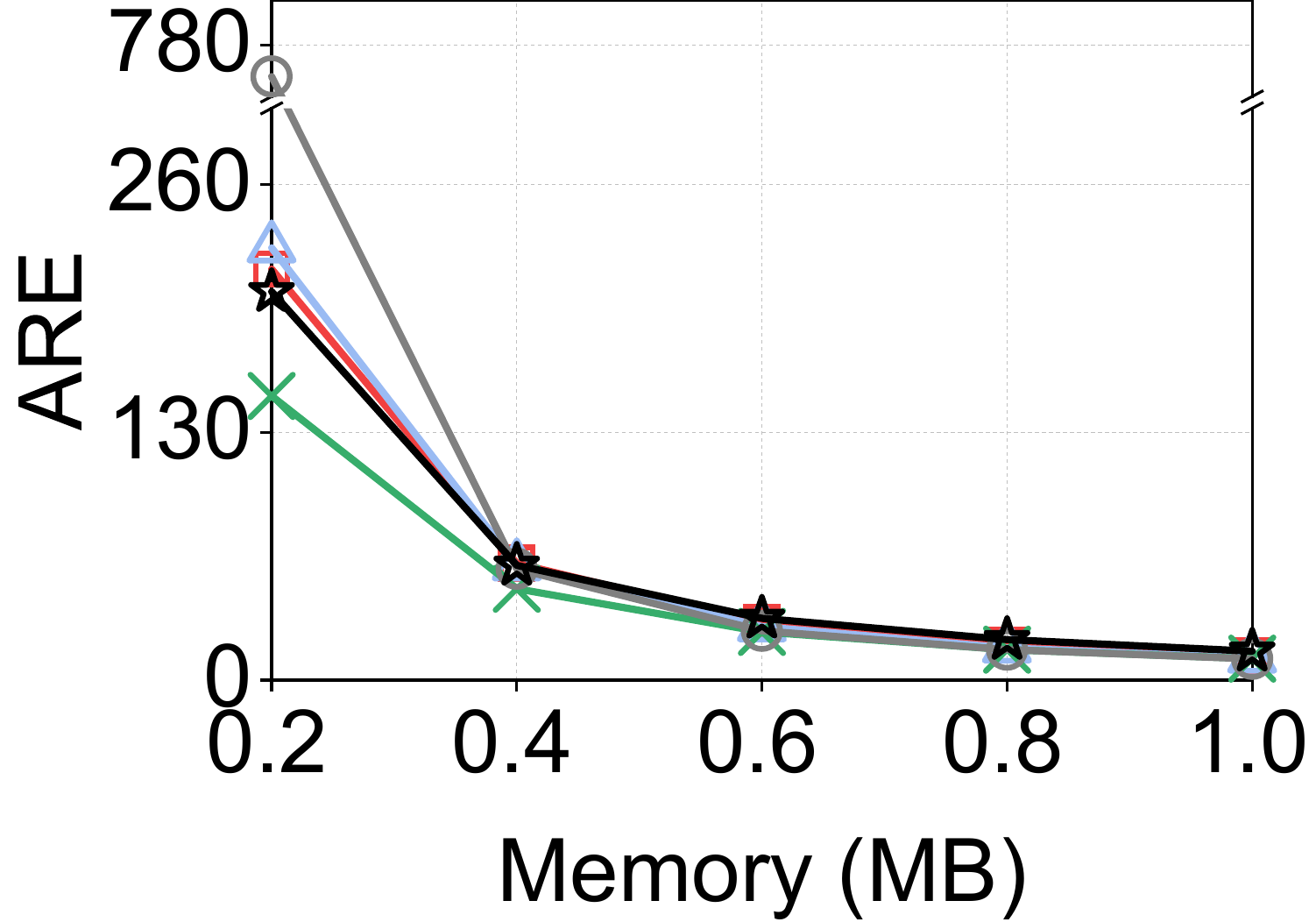}}
    \subfigure[MACCDC]
    {\includegraphics[width=0.19\textwidth]{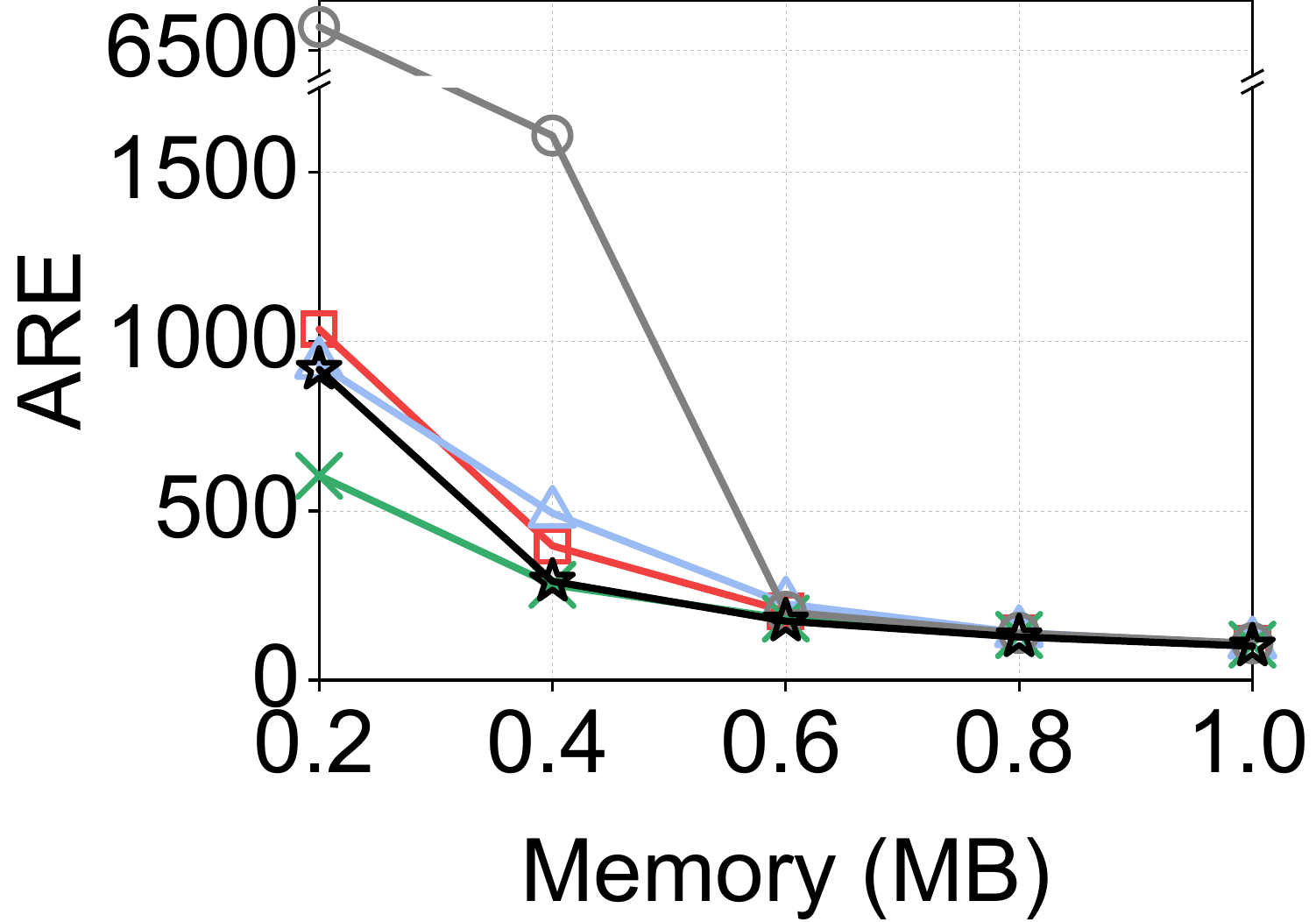}}
    \subfigure[CIC DDoS]
    {\includegraphics[width=0.19\textwidth]{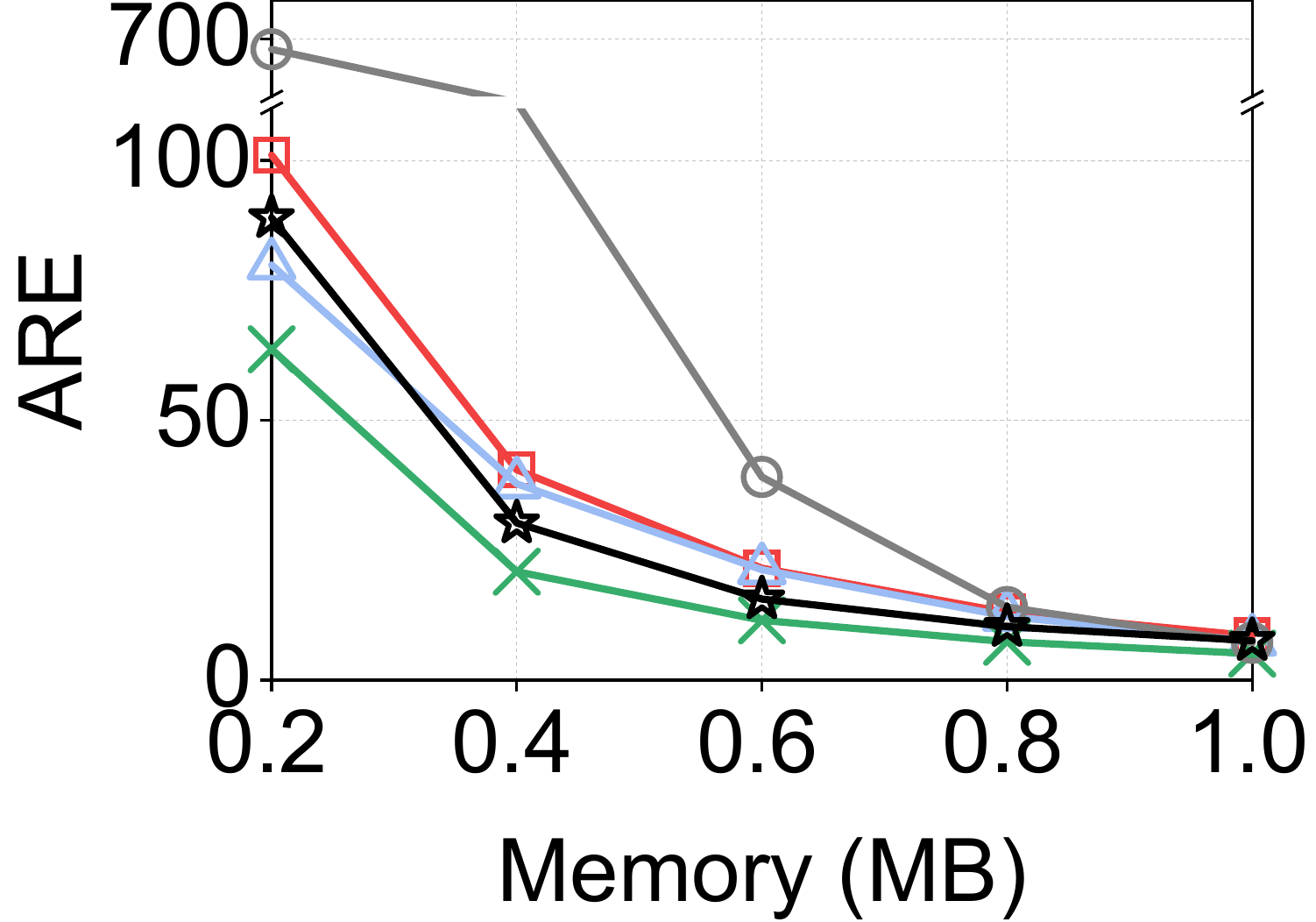}}
    \subfigure[CIC IDS]
    {\includegraphics[width=0.19\textwidth]{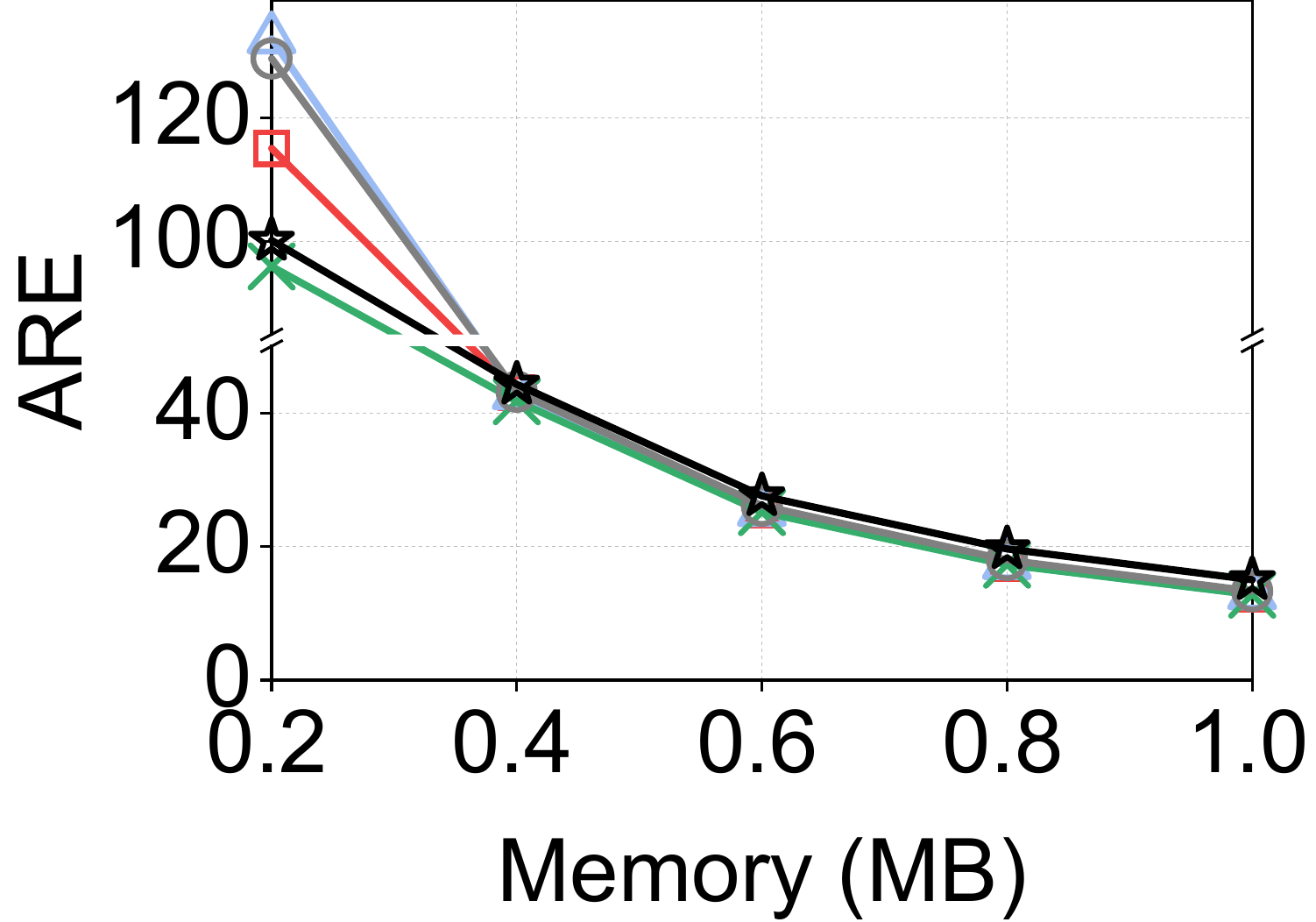}}
    \subfigure[UNSW DoS]
    {\includegraphics[width=0.19\textwidth]{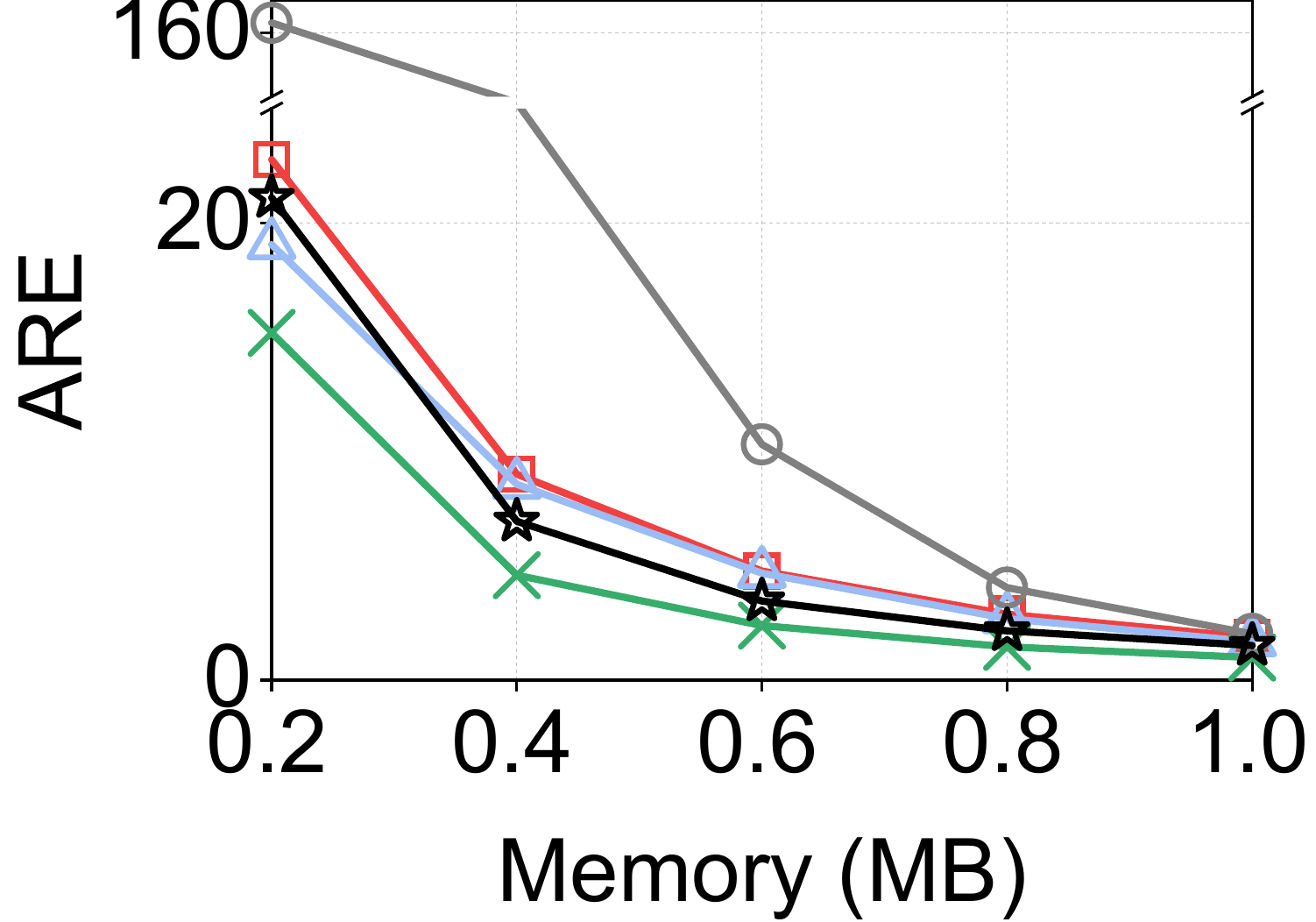}}
    \caption{Flow size estimation with five real-world datasets: the stream length of each real-world dataset is 64 M packets.}~\label{fig:Real_FlowSize_64Mpkts}\vspace{-4mm}
\end{figure*}
\begin{figure*}[t]
    \centering
    \subfigure[Z=0.6]
    {\includegraphics[width=0.19\textwidth]{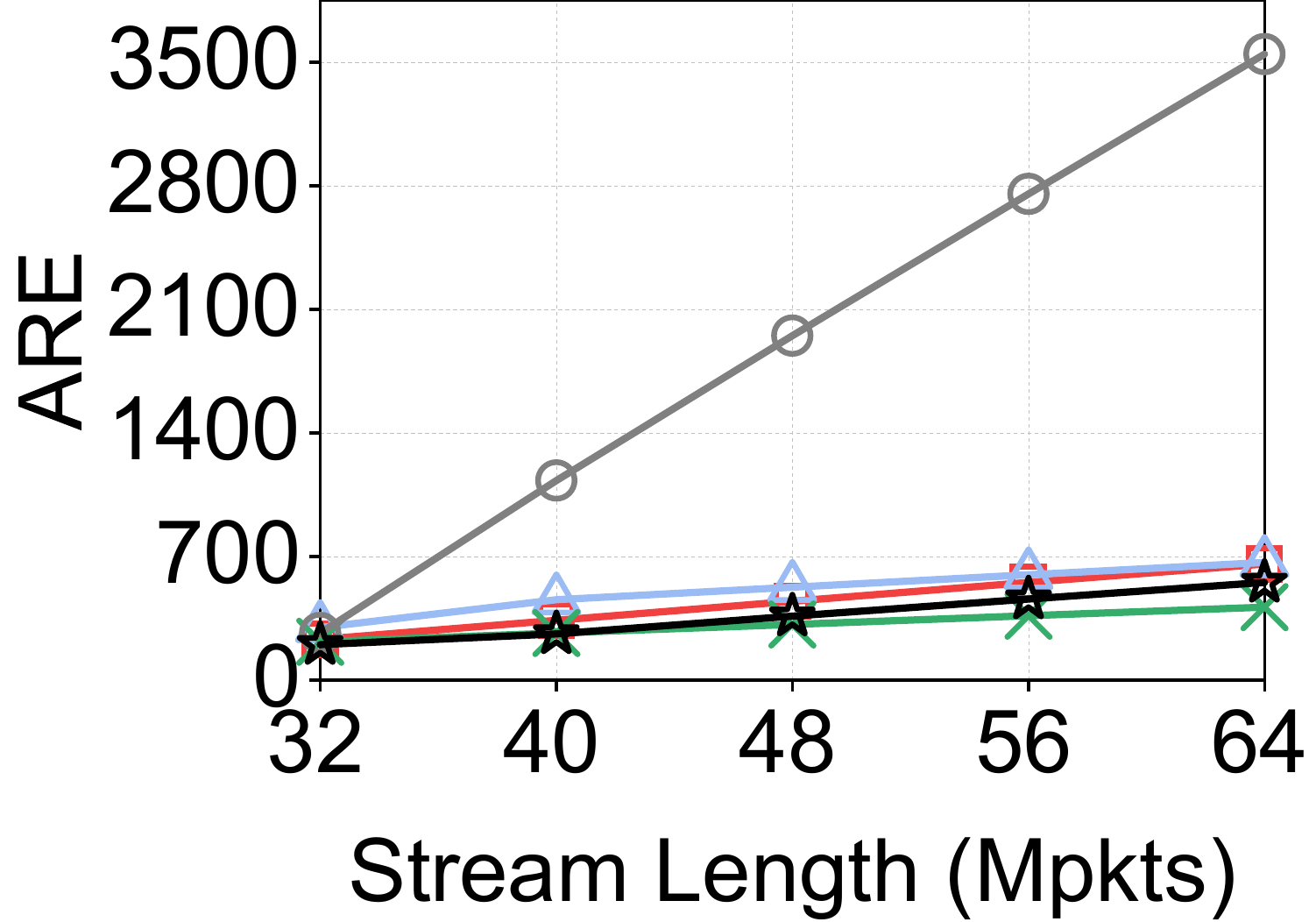}}
    \subfigure[Z=0.8]
    {\includegraphics[width=0.19\textwidth]{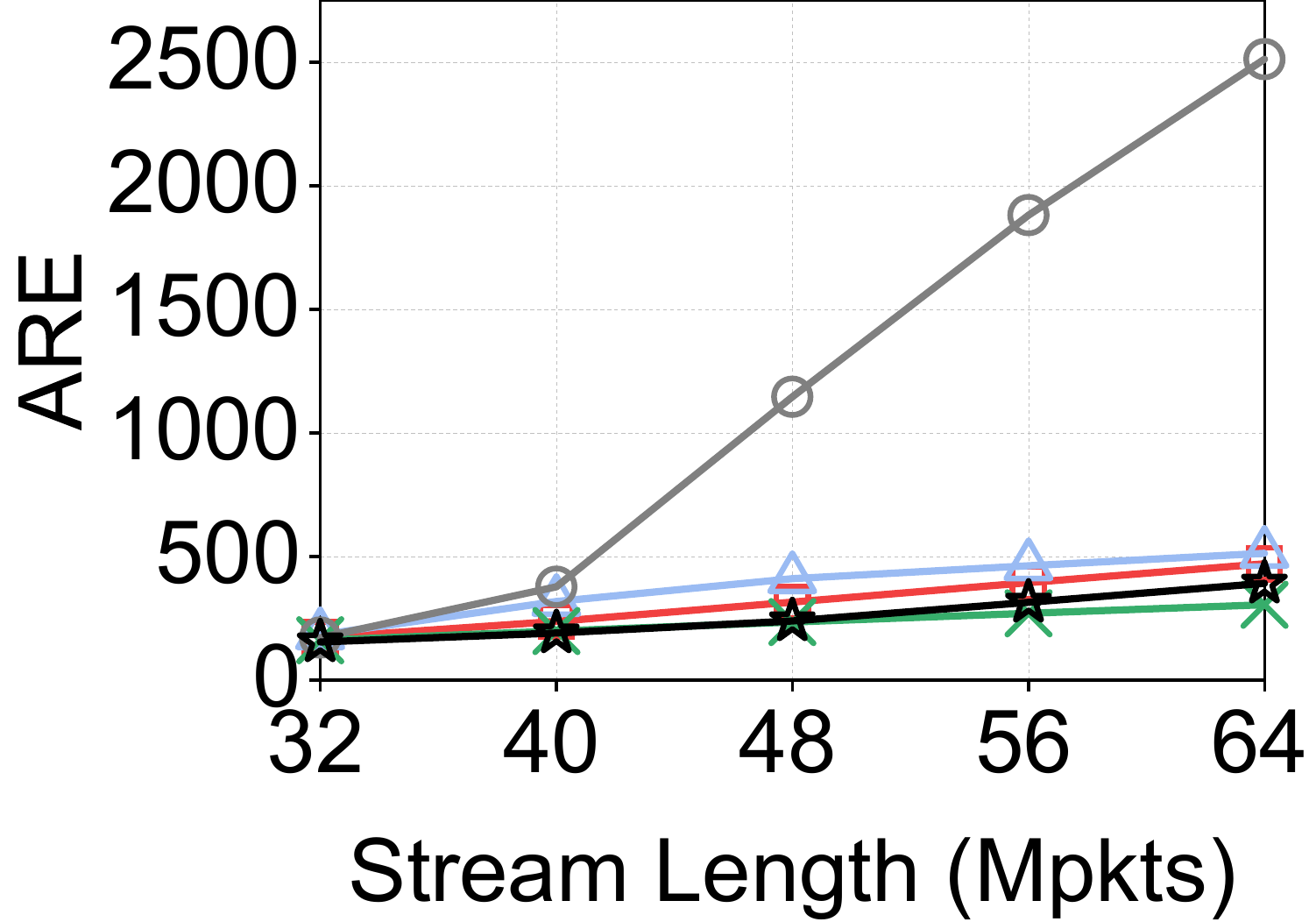}}
    \subfigure[Z=1.0]
    {\includegraphics[width=0.19\textwidth]{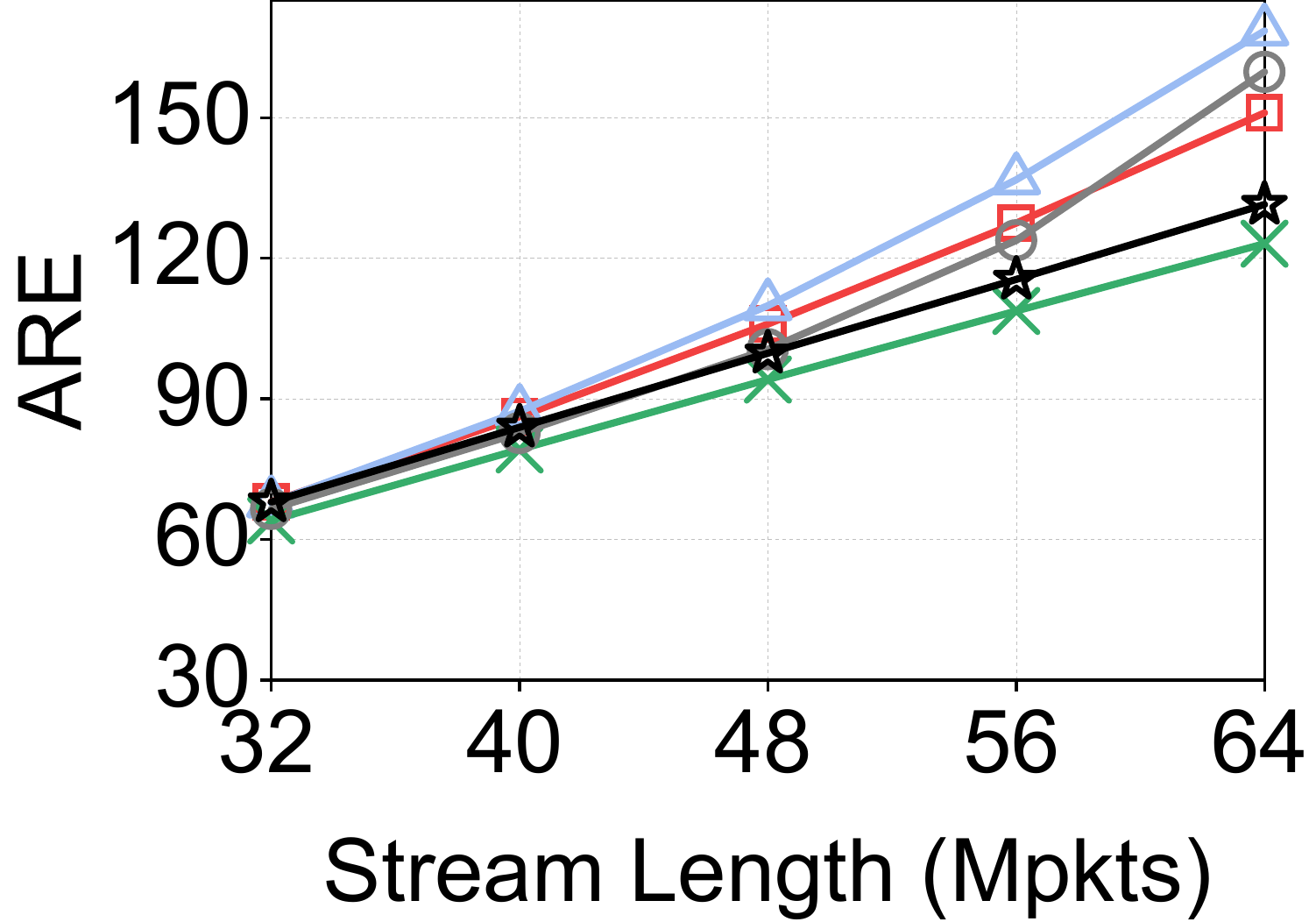}}
    \subfigure[Z=1.2]
    {\includegraphics[width=0.19\textwidth]{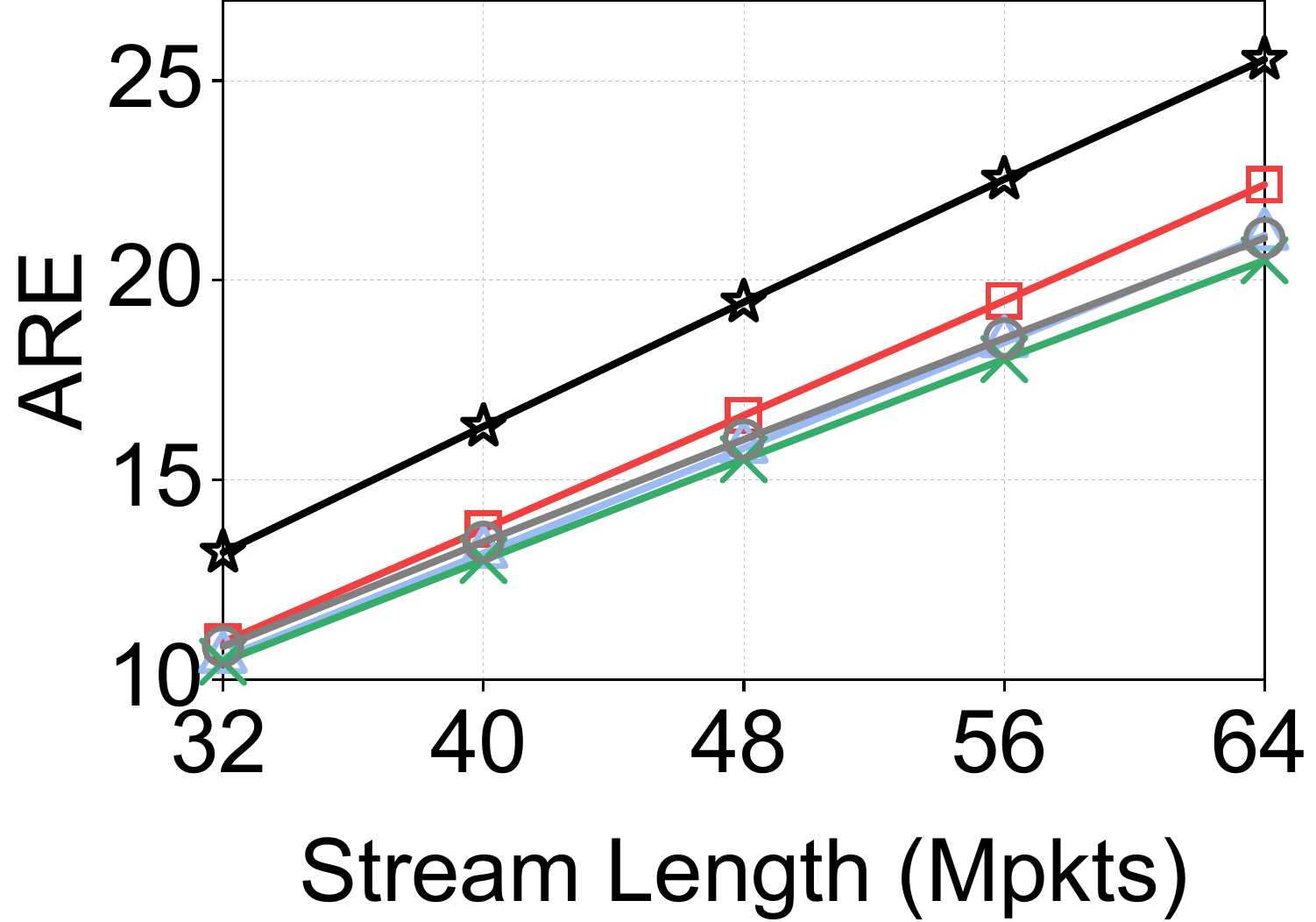}}
    \subfigure[Z=1.4]
    {\includegraphics[width=0.19\textwidth]{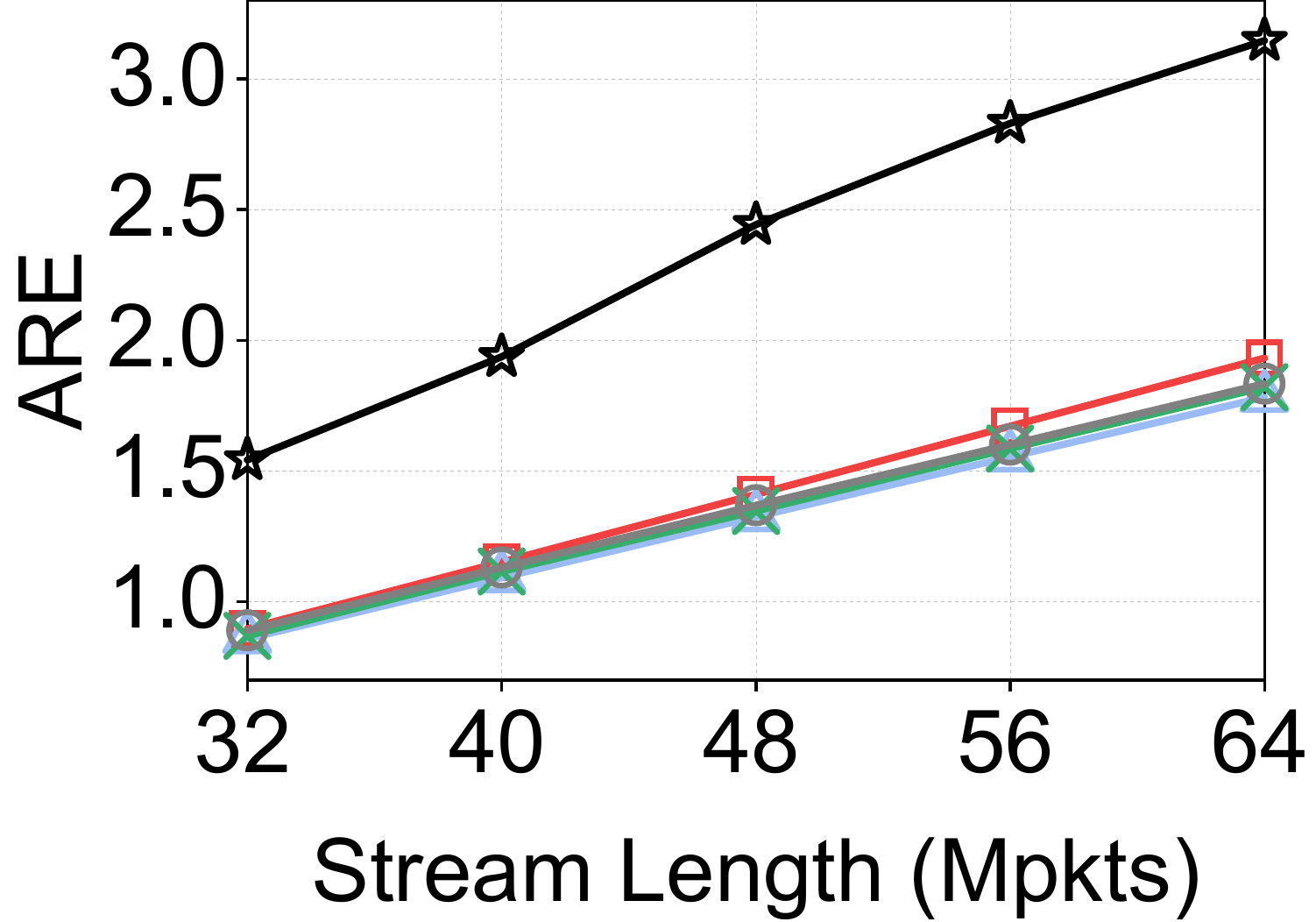}}
    \caption{Flow size estimation with five synthetic datasets:  for each experiment, the memory size is allocated to 0.4 MB.}~\label{fig:Synthetic_FlowSize_0.4MB}
\end{figure*}

\BfPara{Varying Memory Space with Fixed Load}
Next, we evaluate \ours{} by varying memory space. Here, we set the stream length to 32 M packets while varying the amount of memory from 0.2 MB to 1 MB. As shown in Fig.~\ref{fig:Real_FlowSize_64Mpkts}, \ours{} achieves higher accuracy than other methods across all datasets, especially when memory space is small, with more frequency counter expansions or merging. When the memory size is as small as 0.2 MB, \ours{} lowered errors from 24.6\% to 48.4\% compared to ABC. When compared with SALSA, \ours{} reduced errors from 15\% to 48.3\%. \ours{} also achieved about 8.1\% to 74\% lower errors compared to FCM. Lastly,  \ours{} showed  4.8\% to 32\% lower error compared to CountLess.
However, as more memory is allocated to the sketch, the collision rate decreases, and the counter merging rate. As a result, we would expect a closer gap between \ours{} and other schemes.  When the memory size increases and reaches 1.0 MB, the performance gap between \ours{} and other schemes is smaller compared to the small memory setting, but still shows clear advantages with the overall trend. 

\BfPara{Late Merging Impact on Synthetic Dataset}
To delve deeper into our analysis, we compared the flow size estimation using a synthetic dataset with varied traffic skewness.
Fig.~\ref{fig:Synthetic_FlowSize_0.4MB} shows the result when 0.4 MB of memory is allocated and the stream length is varied from 32 M to 64 M packets. Here, we observe that \ours{} outperforms SALSA, and the performance gap is greater when the skewness is small, which indicates LSB sharing can effectively delay the counter merging without instantly merging them upon saturation. Also, \ours{} improved the accuracy from 1.9\% (Z=1.2) to 37.0\% (Z=0.6), when compared to ABC~\cite{gong2017abc}. Moreover, \ours{} outperformed SALSA by lowering errors from 4.6\% (Z=1.4) to 25.6\% (Z=0.6). However, with the highly skewed traffic of Z = 1.4, ABC~\cite{gong2017abc} performs 2.2\% better than \ours{}; we observed that bit borrowing in this highly biased data set gives ABC the advantage of avoiding counter-merging, leading to slight improvement.
Lastly, \ours{} outperforms FCM on average by 30\% and CountLess by 17\%.

\BfPara{Takeaways} \ours{} demonstrates superiority not only under sketch pollution attacks but also in standard traffic measurement across various datasets. We showcase its effectiveness in real-world and synthetic traffic distributions, highlighting its counter sustainability and consistently outperforming state-of-the-art schemes.

\subsection{Data Processing Complexity: Throughput}
\begin{figure}[t]
    \centering
    \includegraphics[width=0.28\textwidth]{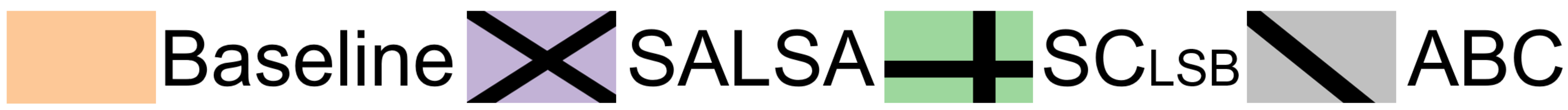} \\
    \includegraphics[width=0.4\textwidth]{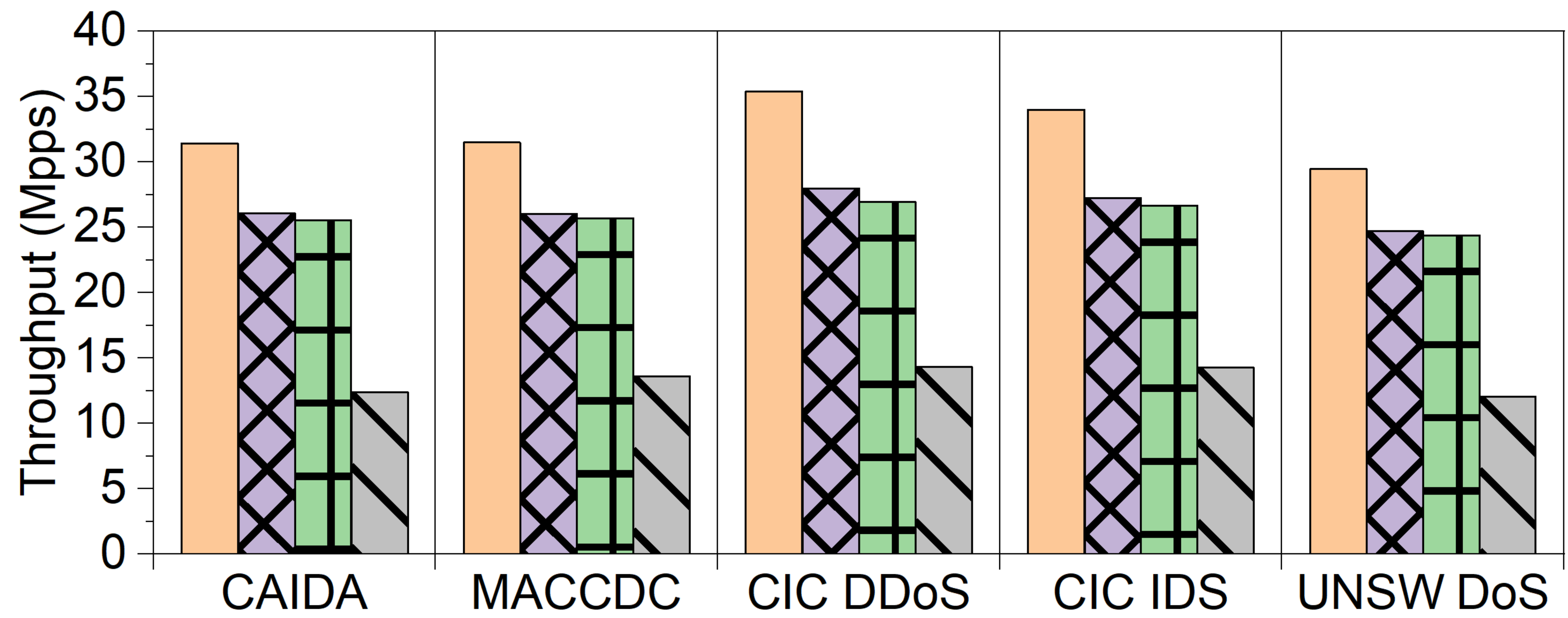}
    \caption{Throughput comparison among baseline, \ours{}, and dynamic counter merging schemes.}
    \label{fig:Throughput_CAIDA}
\end{figure}

Lastly, we evaluate the throughput of dynamic counter merging and compare it with the baseline. Here, we seek to answer the question of how much overhead the late merging effect introduces to provide improvement over frequency estimation. We ran the experiment 50 times and reported the average speed.
Fig.~\ref{fig:Throughput_CAIDA} shows the throughput of the baseline Count-Min sketch, SALSA, ABC, and \ours{} when measuring five real-world datasets. As shown, the Count-Min sketch is the fastest algorithm (baseline), followed by SALSA, \ours{}, and ABC. More specifically, the throughput of ABC is less than 15 Mpps, the lowest among all the schemes due to the complex bit-level counter operations. Although \ours{} showed 24.9\% lower throughput compared to the baseline, but achieved comparable performance with SALSA with 2.2\% overhead, mainly used for the LSB sharing operations. Therefore, we conclude that \ours{} achieves more robust measurement compared to state-of-the-art dynamic sketch~\cite{basat2021salsa} with negligible computational overhead.

\section{Discussions and Future Works} \label{sec:disc}
\BfPara{Sketch in practice} Due to the limitations of current hardware switches~\cite{tofino}, dynamic structure sketches cannot be deployed easily in such environments because they require complex operations such as multiple memory access for counter state tracking and merging. However, they are well-suited for a more flexible environment equipped with general-purpose CPUs and FPGAs. With their ability to adapt to arbitrary data streams, they are ideal for deployment in cloud environments for traffic monitoring~\cite{cheng2024trustsketch, liu2019nitrosketch, huang2017sketchvisor, yang2018elastic}, databases~\cite{redis, cmjulia} and large-scale data processing engines~\cite{Apachespark}. Therefore, one of the future directions is to enable dynamic sketch in the programmable data plane with a sketch approximation design~\cite{kim2023count} to adapt to pattern changes of the network traffic.

\BfPara{Limitation of LSB Sharing}
Note that the LSB sharing is designed based on the winner-take-all strategy, which aligns with the concept that flows appear in bursts. However, in cases where two flows compete in bursts, the LSB shared part can introduce errors and underestimations. However, our observations suggest that this error is negligible as the LSB shared part is small, minimizing the impact of the error. We also plan to apply LSB sharing concept to various primitive sketches, such as Count Sketch~\cite{charikar2002finding} and Count-Min with conservative update~\cite{estan2002new} to pursue advanced design of dynamic sketch.

\section{Conclusion} \label{sec:conclusion}
In this paper, we propose a novel sketch pollution attack targeting small counter designs in existing advanced data sketching, specifically static counter extensions and dynamic merging. Our findings indicate that dynamic merging is essential for counter recycling and memory efficiency, whereas static counter extensions benefit from independent counter operation.
Then, we introduced \ours{}, a novel dynamic structure sketch for resiliency against sketch pollution attacks, inspired by counter isolation design in static counter extension. Unlike the current strategies of resizing that instantly merge and reduce the number of counters, resulting in higher collision rates and estimation errors, \ours{}'s design incorporates late merging for long run counter sustainability. \ours{} shares the Least Significant Bit (LSB) among adjacent counters to expand the counter size while preserving counter quantity and collision rate, thereby enhancing estimation accuracy. Empirical and theoretical analyses show that \ours{} significantly improves accuracy under sketch pollution attacks and normal measurements across various real-world datasets, with minimal impact on throughput.
\label{last-page}

\bibliographystyle{IEEEtran} 
\bibliography{reference}

\end{document}